\documentclass[twocolumn, twocolappendix]{aastex701}

\usepackage{amsmath}
\usepackage{graphicx}
\usepackage{enumitem}
\usepackage{booktabs}
\usepackage{hyperref}
\usepackage{multirow}
\hypersetup{colorlinks=true, linkcolor=blue, citecolor=blue, urlcolor=blue}

\newcommand{\spherex}{SPHEREx}

\newcommand{\eazypy}{\texttt{eazy-py}}
\newcommand{\tractor}{\texttt{The Tractor}}
\newcommand{\galsim}{\texttt{GalSim}}
\newcommand{\simulator}{\texttt{SPHEREx Sky Simulator}}

\shorttitle{SPHEREx View of Galaxy Clusters}
\shortauthors{Bahk et al.}

\begin{document}
    \title{The SPHEREx View of Galaxy Clusters: A Simulation-based Validation of
    the Forced Photometry Pipeline for Extended Sources}

    \author[0009-0002-9878-1126]{Hyeonguk Bahk}
    \affiliation{Astronomy Program, Department of Physics and Astronomy, Seoul National University, 1 Gwanak-ro, Gwanak-gu, Seoul 08826, Republic of Korea}
    \email[show]{bahkhyeonguk@gmail.com}

    \author[0000-0003-3428-7612]{Ho Seong Hwang}
    \affiliation{Astronomy Program, Department of Physics and Astronomy, Seoul National University, 1 Gwanak-ro, Gwanak-gu, Seoul 08826, Republic of Korea}
    \affiliation{SNU Astronomy Research Center, Seoul National University, 1 Gwanak-ro, Gwanak-gu, Seoul 08826, Republic of Korea}
    \affiliation{Institute for Data Innovation in Science, Seoul National University, Seoul 08826, Republic of Korea}
    \email[show]{hhwang@astro.snu.ac.kr}

    \author[0000-0001-7665-5079]{Lindsey Bleem}
    \affiliation{High-Energy Physics Division, Argonne National Laboratory, 9700 South Cass Avenue., Lemont, IL, 60439, USA}
    \affiliation{Kavli Institute for Cosmological Physics, University of Chicago, 5640 South Ellis Avenue, Chicago, IL, 60637, USA}
    \affiliation{Department of Astronomy and Astrophysics, University of Chicago, 5640 South Ellis Avenue, Chicago, IL, 60637, USA}
    \email{lbleem@anl.gov}

    \author[0000-0003-3078-2763]{Yujin Yang}
    \affiliation{Korea Astronomy and Space Science Institute (KASI), 776 Daedeok-daero, Yuseong-gu, Daejeon 34055, Republic of Korea}
    \email{yyang@kasi.re.kr}

    \author[0000-0002-2618-1124]{Yoonsoo P. Bach}
    \affiliation{Korea Astronomy and Space Science Institute (KASI), 776 Daedeok-daero, Yuseong-gu, Daejeon 34055, Republic of Korea}
    \email{ysbach93@gmail.com}

    \author[0000-0002-5437-0504]{Yun-Ting Cheng}
    \affiliation{Department of Physics, California Institute of Technology, 1200 East California Boulevard, Pasadena, CA 91125, USA}
    \affiliation{Jet Propulsion Laboratory, California Institute of Technology, 4800 Oak Grove Drive, Pasadena, CA 91109, USA}
    \email{ycheng3@caltech.edu}

    \author[0000-0002-4650-8518]{Brendan P. Crill}
    \affiliation{Department of Physics, California Institute of Technology, 1200 East California Boulevard, Pasadena, CA 91125, USA}
    \affiliation{Jet Propulsion Laboratory, California Institute of Technology, 4800 Oak Grove Drive, Pasadena, CA 91109, USA}
    \email{brendan.p.crill@jpl.nasa.gov}

    \author[0000-0001-7432-2932]{Olivier Doré}
    \affiliation{Jet Propulsion Laboratory, California Institute of Technology, 4800 Oak Grove Drive, Pasadena, CA 91109, USA}
    \affiliation{Department of Physics, California Institute of Technology, 1200 East California Boulevard, Pasadena, CA 91125, USA}
    \email{odore@caltech.edu}

    \author[0000-0002-9382-9832]{Andreas L. Faisst}
    \affiliation{Department of Physics, California Institute of Technology, 1200 East California Boulevard, Pasadena, CA 91125, USA}
    \affiliation{IPAC, California Institute of Technology, 770 S. Wilson Ave, Pasadena, CA 91125, USA}
    \email{afaisst@ipac.caltech.edu}

    \author[0009-0009-1219-5128]{Zhaoyu Huai}
    \affiliation{Department of Physics, California Institute of Technology, 1200 East California Boulevard, Pasadena, CA 91125, USA}
    \affiliation{Jet Propulsion Laboratory, California Institute of Technology, 4800 Oak Grove Drive, Pasadena, CA 91109, USA}
    \email{zhuai@caltech.edu}

    \author[0000-0002-2770-808X]{Woong-Seob Jeong}
    \affiliation{Korea Astronomy and Space Science Institute (KASI), 776 Daedeok-daero, Yuseong-gu, Daejeon 34055, Republic of Korea}
    \email{jeongws@kasi.re.kr}

    \author[0000-0003-1954-5046]{Bomee Lee}
    \affiliation{Korea Astronomy and Space Science Institute (KASI), 776 Daedeok-daero, Yuseong-gu, Daejeon 34055, Republic of Korea}
    \email{bomee@kasi.re.kr}

    \author[0000-0003-3301-759X]{Jeong Hwan Lee}
    \affiliation{Astronomy Program, Department of Physics and Astronomy, Seoul National University, 1 Gwanak-ro, Gwanak-gu, Seoul 08826, Republic of Korea}
    \affiliation{Research Institute of Basic Sciences, Seoul National University, Seoul 08826, Republic of Korea}
    \email{joungh93@gmail.com}

    \author[0000-0001-9937-8270]{Jeonghyun Pyo}
    \affiliation{Korea Astronomy and Space Science Institute (KASI), 776 Daedeok-daero, Yuseong-gu, Daejeon 34055, Republic of Korea}
    \email{jhpyo@kasi.re.kr}

    \author[0000-0001-8253-1451]{Michael Zemcov}
    \affiliation{School of Physics and Astronomy, Rochester Institute of Technology, 1 Lomb Memorial Drive, Rochester, NY 14623, USA}
    \affiliation{Jet Propulsion Laboratory, California Institute of Technology, 4800 Oak Grove Drive, Pasadena, CA 91109, USA}
    \email{mbzsps@rit.edu}


    \begin{abstract}
        We present a simulation-driven assessment of the performance of the \spherex{}
        pipeline for galaxy cluster science, focusing on photometry, source
        blending, survey depth, and photometric redshift accuracy. To do that,
        we compile a sample of eight galaxy clusters spanning a wide redshift
        range ($z \approx 0.02$--$1.1$) and develop an end-to-end pipeline. We
        use the ancillary data from the DESI Legacy Survey and COSMOS survey, and
        generate realistic mock SPHEREx observations with the \simulator. By
        performing forced photometry on these images with \tractor, we quantify the
        characteristic biases and uncertainties relevant to cluster science. We
        find that the photometry is generally unbiased, but source blending is the
        primary driver of catastrophic outliers, particularly when the combined
        flux of neighbors is comparable to the flux of targets. Measuring the effective
        survey depth, we find that SPHEREx detects members down to
        $K_{s}\approx 20$~AB ($5\sigma$), 7--9 mag fainter than the brightest
        cluster galaxy (BCG) in nearby clusters but only 1--2 mag for clusters at
        $z \sim 1$, where the BCG itself has faded close to this depth. Despite these
        challenges, we demonstrate that SPHEREx can achieve a photometric
        redshift precision of $\sigma_{\rm NMAD}\approx 0.003$--$0.01$ for cluster
        galaxies with an appropriate sample selection based on brightness or signal-to-noise.
        Combining the redshifts of quality-selected members, we recover cluster redshifts
        with a bias of $|\Delta z|/(1+z) < 0.002$ and a scatter of $\sigma \approx
        0.002$ at $z \lesssim 0.5$, meeting the precision required for cluster
        cosmology.
    \end{abstract}


    \section{Introduction}
    \label{sec:intro}

    Clusters of galaxies, as the most massive gravitationally bound structures
    in the Universe, serve as powerful probes for both cosmology and galaxy evolution.
    Originating from the highest peaks of the primordial density field within
    the framework of hierarchical structure formation \citep{peebles80, kravtsov12},
    the abundance and clustering of these objects are sensitive to the growth of
    cosmic structure. This sensitivity allows their population statistics to place
    powerful constraints on key cosmological parameters, including the
    properties of dark energy \citep{allen11, huterer18}.

    Simultaneously, their deep gravitational potential wells and dense
    populations of galaxies make them unique laboratories for studying the
    environmental processes that shape galaxy evolution. Physical mechanisms such
    as ram-pressure stripping and galaxy mergers are prevalent in these environments,
    providing important insights into the quenching of star formation \citep{gunn72, toomre72, dressler80, park09, peng10, haines13, wetzel14, lee22}.
    To achieve both scientific goals with galaxy clusters, it is necessary to have
    a large, homogeneously selected samples with accurately measured properties
    for their member galaxies, most notably their redshifts and spectral energy distributions
    (SEDs).

    Over the past two decades, a variety of surveys across different wavelengths
    successfully identified large samples of galaxy clusters. Optical surveys primarily
    use red-sequence methods, which exploit the tight color--magnitude relation of
    passively evolving cluster galaxies, to locate galaxy overdensities \citep[e.g.,][]{rykoff14, rykoff16, oguri18, wenhan24},
    often in combination with gravitational lensing studies \citep[e.g.,][]{abbott20}.
    Large and dense spectroscopic surveys further extend cluster identification
    through three-dimensional galaxy distributions \citep[e.g.,][]{huchra82,eke04,berlind06,tempel14,sohn21}.
    Meanwhile, X-ray surveys exploit the thermal bremsstrahlung emission from
    the intracluster medium \citep[ICM; e.g.,][]{rasscl, mcxc, erass}, while the
    Sunyaev--Zel'dovich (SZ) effect enables all-sky cluster catalogs from Planck
    \citep{psz2, bahk24} and high-resolution surveys from the South Pole
    Telescope \citep[SPT;][]{bleem15, bleem24, kornoelje25} and the Atacama Cosmology
    Telescope \citep[ACT;][]{hilton21, actdr6}.

    Obtaining accurate redshifts is an essential step in constructing and fully
    utilizing these cluster catalogs. For optically selected samples, red-sequence-based
    methods \citep{gladders00, koester07} have been widely adopted and provide reliable
    cluster redshift estimates, while dedicated tools have been developed to
    confirm ICM-selected cluster candidates and assign cluster redshifts via
    their optical member galaxies \citep[e.g.,][]{klein18, kluge24}. Nevertheless,
    obtaining spectroscopic redshifts for the vast number of individual member
    galaxies remains observationally prohibitive due to the immense telescope
    time required, even in the era of Stage-IV spectroscopic surveys \citep{desi}.

    This information is incomplete precisely for the large ICM- and optically selected
    samples that drive cluster cosmology, including those from eROSITA \citep{erass},
    SPT, and the upcoming Euclid \citep{sartoris16} and Rubin/LSST \citep{lsst09}
    surveys, which contain far more clusters than can be characterized
    spectroscopically at the member level. Even where a spectroscopic cluster redshift
    is available, it typically rests on the Brightest Cluster Galaxy (BCG) or a
    few members rather than a spectroscopic census of the membership: in the eRASS1
    catalog only a minority of clusters have a spectroscopic redshift (3{,}210
    of the $\sim$12{,}000 systems; \citealt{kluge24}), and even for SZ-selected massive
    clusters the fraction, while higher (e.g., $\sim$80\% of the all-sky UPCluster-SZ
    sample; \citealt{bahk24}), reflects cluster-level rather than member-level spectroscopy.

    Photometric redshifts \citep[][for review]{salvato19} provide a natural solution
    to this problem, allowing efficient redshift estimation for vast samples of
    galaxies. Although less precise than spectroscopy, photometric redshifts are
    indispensable for wide-field surveys \citep[e.g.,][]{connolly95, ilbert06, lsst09, newman15, kim25}.
    Previous studies show that cosmological constraints from cluster number counts
    demand stringent control over photometric redshift uncertainties. For
    instance, to avoid significant degradation ($< 10 \%$) of dark energy constraints,
    the systematic bias and scatter in photometric redshift must be controlled
    to be less than $\lesssim 0.003$ and $\sim 0.03$, respectively, with the
    bias roughly an order of magnitude more important than the scatter for
    cluster number counts \citep{huterer04, lima07}.

    In this context, the Spectro-Photometer for the History of the Universe, Epoch
    of Reionization, and Ices Explorer \citep[\spherex{};][]{spherex, dore14}, an
    all-sky near-infrared (NIR) spectrophotometric survey, offers a promising
    opportunity by providing dense spectrophotometric sampling of rest-frame
    optical and NIR features across a wide redshift range. Recent simulations of
    the SPHEREx survey predict a photometric redshift precision of $\sigma_{z}< 0
    .003 (1+z)$ for $>10^{7}$ bright galaxies \citep{feder24}, well within the scatter
    requirement noted above. Combined with its all-sky coverage, this makes SPHEREx
    well suited to assembling large, well-characterized cluster samples for such
    analyses.

    Moreover, the scientific utility of SPHEREx data extends beyond precise redshifts.
    The continuous near-infrared spectra will enable the study of key spectral
    features for galaxy evolution, such as the 1.6~$\mu$m stellar bump (a tracer
    of intermediate- and old-age stellar populations) and Polycyclic Aromatic Hydrocarbon
    (PAH) emission features, which are powerful diagnostics of dust and star
    formation \citep{zhang25}.

    Despite its potential for cluster science, the instrumental characteristics
    of \spherex{} introduce several observational challenges that need to be addressed
    to enable robust scientific use of its data. Primarily, the large pixel scale
    of $6\farcs15 \times 6\farcs15$ \citep{korngut18} is expected to cause
    significant source blending, particularly in the dense cores of galaxy clusters.
    The planned mitigation strategy for this issue is forced photometry, which relies
    on high-resolution prior catalogs to deblend sources \citep{feder24, huai25, akeson25},
    together with accurate modeling of the undersampled SPHEREx Point Spread Function
    \citep[PSF;][]{symons21}.

    The challenge of performing accurate photometry with SPHEREx has been
    previously explored. \citet{stickley16} conducted an end-to-end simulation
    using a COSMOS-based catalog \citep{cosmos} and demonstrated that contamination
    from neighboring sources produces significant systematic biases in the extracted
    fluxes. These biases grow toward longer wavelengths, where the larger PSF blends
    light from a greater number of neighbors.

    More recent investigations have revisited this issue using complementary approaches.
    \citet{dachan24} examined the impact of blending on galaxies and galaxy luminosity
    functions by defining blended sources as those within the same SPHEREx pixel.
    Using this simple geometric criterion, they found that only $\sim$0.7\% of galaxies
    are classified as blended in the all-sky survey data, suggesting a minimal
    effect on overall photometric accuracy, although the fraction increases to $\sim$9\%
    in SPHEREx deep fields. We note that these estimates are derived from
    extragalactic fields at high Galactic latitudes and do not capture the
    substantially higher source density expected at low Galactic latitudes or in
    dense cluster environments. In a separate effort, \citet{huai25} addressed the
    problem at the image level by performing forced photometry on simulated SPHEREx
    images to assess the impact of blending and spectral confusion on flux recovery
    and photometric redshifts. They found that unbiased fluxes can be obtained (with
    only minor blending-induced noise) when the covariance between neighboring
    sources is properly modeled, while spectral confusion becomes significant primarily
    in deep survey regions. Together, these efforts demonstrate the feasibility of
    the forced-photometry approach adopted by the official SPHEREx Level~3 pipeline
    \citep{akeson25}.

    However, two gaps persist when this approach is applied to galaxy cluster fields,
    which motivate our study. First, while \citet{stickley16} modeled contamination
    in a general field and \citet{huai25} characterized blending and spectral
    confusion in extragalactic settings, the performance of forced photometry in
    the extreme overdensities characteristic of galaxy cluster cores, where
    source blending is most severe, remains unexplored. Second, the photometry extraction
    in \citet{stickley16} and \citet{huai25} was performed with a PSF model,
    effectively treating all sources as unresolved. This simplification does not
    address how well model-based forced photometry can reproduce the true total flux
    of intrinsically extended sources such as typical cluster member galaxies,
    which are significantly undersampled by the large SPHEREx pixels. Furthermore,
    the effective depth to which SPHEREx can characterize faint cluster members
    within its nominal 2-year survey requires a dedicated investigation. Addressing
    these specific challenges is therefore a necessary prerequisite essential to
    fully exploiting SPHEREx for galaxy cluster science.

    To address these prerequisites, we present an end-to-end simulation designed
    to quantitatively assess the performance of the SPHEREx pipeline for galaxy
    cluster science. We generate realistic mock SPHEREx observations for a sample
    of eight galaxy clusters spanning $z \approx 0.02$--$1$. We then apply a forced
    photometry pipeline to these simulated images to evaluate its performance in
    several key aspects. First, we test the fundamental performance of model-based
    forced photometry using \tractor{} \citep{tractor, hogg13, lang20} to measure
    the flux of intrinsically extended sources, which are undersampled by the
    SPHEREx pixels. Second, we quantify how this photometric accuracy is
    affected by severe source blending, a prevalent issue in the dense cores of these
    clusters. Third, we determine the effective survey depth for reliably characterizing
    cluster members in the nominal 2-year mission. Finally, we assess the resulting
    fidelity of the photometric redshifts derived from the recovered spectrophotometry.

    This paper is structured as follows. In Section~\ref{sec:sample}, we describe
    the cluster sample and the setup of our simulations. Section~\ref{sec:end2end}
    introduces the photometry pipeline and validation procedures. Section~\ref{sec:phot}
    presents results on photometric measurement bias, blending, and depth.
    Section~\ref{sec:photoz} shows results on testing photometric redshift
    performance. We discuss implications for cluster science in Section~\ref{sec:discussion}
    and summarize our conclusions in Section~\ref{sec:conclusion}. Throughout
    this work, we adopt a flat $\Lambda$CDM cosmology with
    $H_{0}= 70 \, \mathrm{km \, s^{-1} \, Mpc^{-1}}$, $\Omega_{m}= 0.3$, and
    $\Omega_{\Lambda}= 0.7$ and use the AB magnitude system.

    \section{Cluster Sample and Ancillary Data}
    \label{sec:sample}

\begin{figure}
        \centering
        \includegraphics[width=0.48\textwidth]{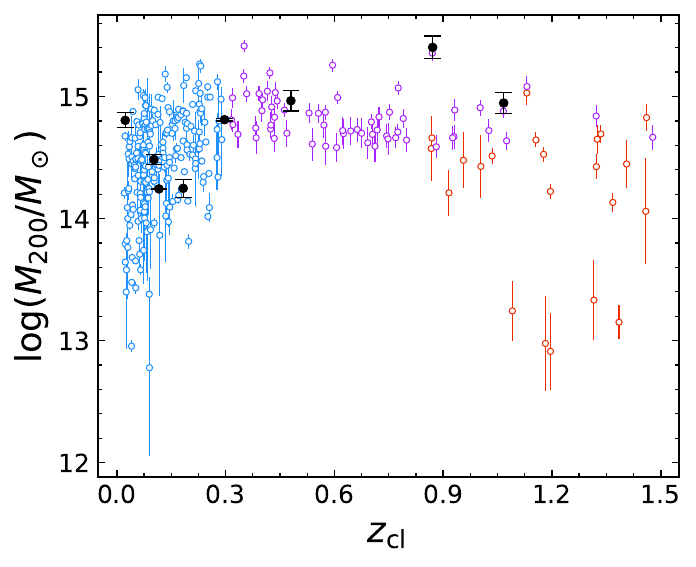}
        \caption{ Cluster redshift versus mass distribution. Different colors
        indicate clusters from the three cluster catalogs: HeCS-omnibus at low redshift,
        SPT at intermediate redshift, and GOGREEN at high redshift. Black filled
        symbols mark the subsample of clusters selected for our analysis. }
        \label{fig:zmass}
    \end{figure}

    We base our analysis on three spectroscopic cluster samples that span a wide
    redshift range. The HeCS-omnibus catalog \citep{sohn20} provides a compilation
    of spectroscopic data for 227 massive clusters at low redshift ($0 .02 < z <
    0.3$). At intermediate redshifts, the SPT-SZ sample \citep{ruel14, bayliss17}
    consists of galaxy clusters discovered via the SZ effect with the South Pole
    Telescope (SPT), with spectroscopic follow-up providing redshifts for 61 clusters
    and velocity dispersions for 48 systems. At higher redshift ($0.8 < z < 1.5$),
    the GOGREEN survey \citep{balogh21} targets 26 galaxy groups and clusters
    with deep spectroscopy and multi-wavelength imaging, providing critical constraints
    on galaxy populations and environments in the early universe. Together, these
    catalogs span complementary redshift and mass ranges (see Figure
    \ref{fig:zmass}).

    To construct our working sample, we randomly selected seven clusters from
    the HeCS-omnibus, SPT-SZ, and GOGREEN catalogs: Abell 2055, Abell 1361, Abell
    2187, Abell 2537 (HeCS-omnibus), SPT-CL J2145$-$5644, SPT-CL J0546$-$5345 (SPT-SZ),
    and SpARCS J1613+5649 (GOGREEN). The randomization ensures that the
    resulting set spans the mass distribution of the parent catalogs and the
    redshift range from the local universe to $z \sim 1$. No additional secondary
    cuts were applied before this random draw. We did not supplement the draw
    with $z \gtrsim 1.5$ clusters, where SPHEREx is too shallow for their members
    (Sections~\ref{sec:depth}, \ref{sec:photoz}). We note that the parent catalogs
    are already biased toward massive systems by their selection methods, and
    the resulting sample happens to favor the massive end within each catalog, particularly
    at high redshift. Because more massive clusters reside in more crowded
    environments with more severe blending, they represent a challenging test case
    for photometric performance, and our results can therefore be regarded as
    conservative relative to typical clusters at these redshifts.

    In addition, we deliberately include the Coma cluster as a benchmark system.
    Coma represents one of the most massive and well-studied clusters in the
    local universe, and its proximity makes it a natural reference point. Based
    on caustic mass profile measurements (e.g., \citealt{rines03, sohn17, kang25}),
    weak-lensing results (e.g., \citealt{kubo07, gavazzi09, okabe14, hyeonghan24}),
    and galaxy distribution analyses (e.g., \citealt{ho22}), $M_{200}$ estimates
    for Coma span $(0.5\textrm{--}1.9 )\times 10^{15}\,M_{\odot}$. In this study,
    we adopt a value of $M_{200}=8\times 10^{14}\,M_{\odot}$ \citep{rines03, hyeonghan24},
    corresponding to $R_{200}\approx1.9\,\mathrm{Mpc}$.

    \begin{figure}
        \centering
        \includegraphics[width=0.48\textwidth]{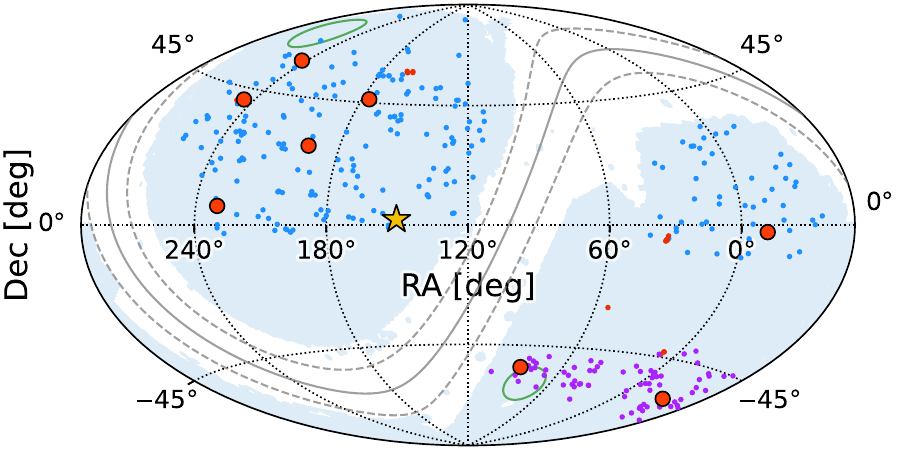}
        \caption{ Sky distribution of the cluster samples used in this study.
        Small points represent clusters from HeCS-omnibus, SPT, and GOGREEN. Large
        red circles mark the cluster fields selected for our analysis. The
        COSMOS field is shown with a yellow star, and the approximate locations of
        the SPHEREx NEP and SEP deep fields are outlined in green. The blue
        shaded regions indicate the footprint of the DESI Legacy Survey. }
        \label{fig:aitoff}
    \end{figure}

    \begin{deluxetable}
        {ccccccccc} \tablecaption{Cluster Sample in This Work \label{tab:clusters}}
        \tablehead{ \colhead{Name} & \colhead{RA} & \colhead{Dec} & \colhead{$z$} & \colhead{$R_{200}$} & \colhead{$N_{\rm mem}$} & \colhead{$N(K_{s}<19)$} & \colhead{$N(N_{\rm ch}>50)$} & \colhead{$N(N_{\rm ch}=102)$} \\ & \colhead{(deg)} & \colhead{(deg)} & & \colhead{(deg)} & & & & }
        \startdata Coma & 194.9531 & 27.9807 & 0.023 & 1.1454 & 657 (632) & 577 &
        576 & 174 \\
        Abell 2055 & 229.7063 & 6.2318 & 0.1023 & 0.1978 & 127 (120) & 115 & 116
        & 22 \\
        Abell 1361 & 175.925 & 46.3296 & 0.1159 & 0.1468 & 52 (52) & 52 & 52 & 5
        \\
        Abell 2187 & 246.0668 & 41.24 & 0.1829 & 0.0984 & 50 (50) & 50 & 50 & 4 \\
        Abell 2537 & 347.0926 & $-$2.1922 & 0.2969 & 0.1011 & 112 (111) & 86 &
        69 & 2 \\
        SPT-CL J2145-5644 & 326.4694 & $-$56.7477 & 0.48 & 0.0787 & 23 (23) & 16
        & 14 & 1 \\
        SpARCS J1613+5649 & 243.311 & 56.825 & 0.871 & 0.073 & 32 (26) & 4 & 4 &
        0 \\
        SPT-CL J0546-5345 & 86.6403 & $-$53.761 & 1.0669 & 0.0453 & 33 (18) & 1
        & 0 & 0 \\
        COSMOS & 150.1192 & 2.2058 & -- & -- & -- & -- & -- & -- \\
        \enddata \tablecomments{$N_{\rm mem}$ is the total number of spectroscopic members within the analyzed radius; the value in parentheses counts members that retain at least one valid SPHEREx observation after the sensitivity pre-selection (Section~\ref{sec:mockimg}). The remaining columns give the number of these members passing the $K_{s}<19$, $N_{\rm ch}(\mathrm{S/N}>5)>50$, and $N_{\rm ch}(\mathrm{S/N}>5)=102$ selections used in Section~\ref{sec:photoz}, where $N_{\rm ch}(\mathrm{S/N}>5)$ denotes the number of the 102 Secondary-Catalog spectral channels (Section~\ref{sec:end2end}) in which a member is detected at $\mathrm{S/N}>5$. COSMOS is listed as a field-galaxy reference and has no member definition.}
    \end{deluxetable}

    The selection outcome is illustrated in Figure~\ref{fig:zmass} and \ref{fig:aitoff}.
    Figure~\ref{fig:zmass} compares their redshift and mass distributions. Figure~\ref{fig:aitoff}
    shows the sky distribution of the parent catalogs and the chosen clusters.
    For the SPT-SZ clusters, the catalog provides $M_{500}$ estimates; we
    converted these to $M_{200}$ by assuming an NFW profile with a concentration
    factor of $c=4$ \citep{duffy08}. The final set covers a broad redshift
    baseline ($0<z<1$) with a preference for high-mass systems. A summary of the
    adopted cluster properties is provided in Table~\ref{tab:clusters}. In
    addition to the basic cluster properties, Table~\ref{tab:clusters} also lists
    the number of spectroscopic members that pass the brightness ($K_{s}<19$) and
    data-quality ($N_{\rm ch}(\mathrm{S/N}>5)>50$ and $=102$) selections used later
    to define high-fidelity subsamples for the photometric-redshift performance analysis
    (Section~\ref{sec:photoz}). Here $N_{\rm ch}(\mathrm{S/N}>5)$ is the number
    of SPHEREx spectral channels in which a source is detected above
    $\mathrm{S/N}=5$, out of the 102 channels of the Secondary Catalog (Section~\ref{sec:end2end});
    the two cuts therefore select members detected in more than half and in all
    channels, respectively. These counts illustrate how the number of reliably
    characterized members decreases with cluster redshift.

    To obtain $R_{200}$ values, we adopt different approaches depending on the available
    information in each catalog. For the HeCS-omnibus clusters, we directly use the
    $R_{200}$ values provided in the catalog, which are derived from caustic
    mass profiles. For the SPT-SZ sample, we determine $R_{200}$ from the
    converted $M_{200}$ estimates and the critical density at the cluster
    redshift. For the GOGREEN survey, no direct mass estimates are available; instead,
    we use the published velocity dispersions. We apply the calibration of \citet{saro13}
    between velocity dispersion and dynamical mass to infer $M_{200}$, and compute
    $R_{200}$ as for the SPT-SZ sample.

    We use photometry and shape parameters from the DESI Legacy Imaging Survey DR10
    Tractor catalog \citep{dey19}. For each cluster, we extract all sources
    within $R_{200}$, except for the Coma cluster where we adopt a smaller radius
    of $0.5R_{200}$ due to its large extent. For all objects, we use fluxes in the
    $gri zW1W2$ bands, corrected for Galactic extinction using the Corrected SFD
    map \citep[CSFD;][]{chiang23}. In addition, we extract source types (PSF, REX,
    EXP, DEV, SER, DUP)\footnote{ PSF: point sources; REX: round exponential
    galaxies; EXP: exponential; DEV: de Vaucouleurs; SER: S\'ersic profiles; DUP:
    Gaia sources fitted with a extended model and retained for catalog
    completeness. See \url{https://www.legacysurvey.org/dr10/description/\#morphological-classification}}
    and shape parameters including the S\'ersic index, half-light radius (effective
    radius; $r_{e}$), and ellipticities \citep[$e_{1}$, $e_{2}$; see ][for
    reference]{kitching08}. We exclude DUP sources, which are assigned to Gaia-matched
    objects without independent flux measurements in the Legacy Survey. These quantities
    are later used for generating simulated source images.

    To compare our results for cluster galaxies with those for field galaxies,
    we include galaxies from the COSMOS region \citep{cosmos}. Unlike the
    cluster fields, COSMOS offers extensive multi-wavelength coverage, making it
    a valuable complementary dataset. We use the COSMOS2025 catalog \citep{cosmos2025},
    which combines photometric measurements in the COSMOS2020 \citep{weaver22}
    and COSMOS-Web \citep{cosmosweb}. In COSMOS2025, we use CFHT $u^{*}$, HSC
    $grizy$, Subaru medium- and narrow-band (IB427, IA484, IB505, IA527, IB574,
    IA624, IA679, IB709, IA738, IA767, IB827, NB711, and NB816) filters, UltraVISTA
    $YJHK_{\mathrm{s}}$, \textit{Spitzer}/IRAC channels 1 and 2, \textit{HST}/ACS
    F814W, and \textit{JWST}/NIRCam filters (F115W, F150W, F277W, F444W), providing
    broad to medium-band coverage across the optical to NIR. To obtain
    morphological information, we cross-match COSMOS sources with the Legacy Survey
    catalog within $0.5^{\prime\prime}$, so that the COSMOS field-galaxy sample
    is characterized with the same morphological measurements as the cluster
    fields, keeping the simulation inputs homogeneous across all fields. This
    cross-match is also expected to filter out most artifacts, as genuine sources
    should be detected in both datasets. These ancillary data provide high-quality
    SED coverage that we use in field-galaxy modeling and source image
    simulations. The COSMOS sample further serves as the field reference throughout
    this work: it provides the field-galaxy comparison baseline for the
    photometric bias and scatter analysis (Section~\ref{sec:phot-bias}), the crowding
    environment characterization and radial blending trends (Section~\ref{sec:blend}),
    and the photometric redshift performance benchmark (Section~\ref{sec:photoz}).

    We further cross-match our Legacy Survey sources with the SPHEREx Reference
    Catalog (version~v0.5; Y.~Yang et al.\ in prep.; see also \citealt{crill25})
    using a $0.5^{\prime\prime}$ tolerance. The Reference Catalog provides all-sky
    photometric measurements compiled from Gaia \citep{gaia}, Pan-STARRS \citep{panstarrs1},
    DESI Legacy Imaging Survey, 2MASS \citep{2mass}, ALLWISE \citep{wise, neowise, allwise},
    and CatWISE \citep{catwise}, serving as the reference source list for the SPHEREx
    forced photometry pipeline and catalog construction. By performing this
    cross-match, we aim to reproduce the selection criteria and content of the forthcoming
    SPHEREx Level~3 products. Beyond the DUP exclusion, no additional magnitude or
    flag cuts are applied, as the cross-match with the multi-survey Reference
    Catalog is expected to exclude most spurious detections and imaging
    artifacts.

    For spectroscopic redshifts, we compile data from multiple catalogs,
    including DESI DR1, the SPT-SZ redshift catalog, the compilation of \citet{sifon16},
    the GOGREEN survey, and SDSS DR17 \citep{sdssdr17}. In addition, to improve
    coverage for faint galaxies, we supplement with literature data from the Hectospec
    series \citep{hecs13, hecs-sz, hecs-red, hectomap_dr2} and other
    compilations \citep{hwang10, hwang14}. For the COSMOS field, we use the spectroscopic
    redshift compilation of \citet{cosmos-specz}, retaining only high-confidence
    measurements with confidence levels of 80\% or higher. These redshifts are used
    both to fix galaxy redshifts during SED modeling and to identify cluster member
    galaxies (see Section~\ref{sec:source}).


    \section{End-to-End Pipeline}
    \label{sec:end2end}

    \begin{figure*}
        \centering
        \includegraphics[width=\textwidth]{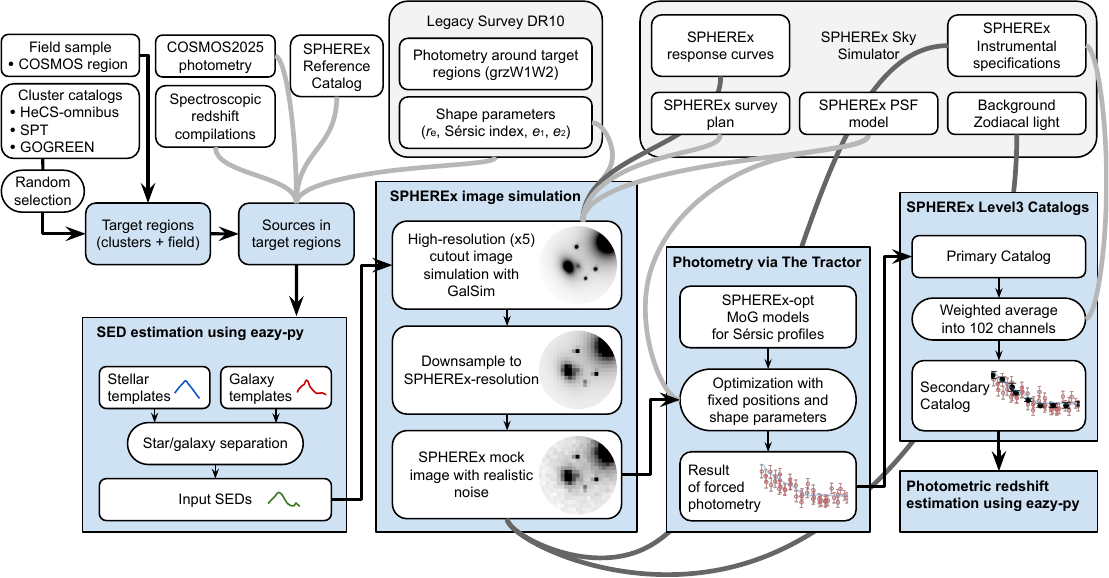}
        \caption{Schematic overview of the end-to-end pipeline used in this work.
        Cluster and field samples (HeCS-omnibus, SPT, GOGREEN, and COSMOS)
        provide the input target regions. Legacy Survey photometry and COSMOS data
        are used to estimate stellar and galaxy SEDs with \eazypy, forming the
        spectral and morphological inputs for the SPHEREx image simulations. Using
        the \simulator, high-resolution \galsim{} cutouts are generated,
        convolved with the SPHEREx PSF, downsampled to the SPHEREx pixel scale,
        and combined with realistic instrumental effects and noise. Forced photometry
        is then performed with \tractor, which fits SPHEREx-optimized Mixture-of-Gaussians
        (MoG) S\'ersic models to the images while fixing source positions and
        shapes. The resulting measurements are compiled into Level~3 Primary Catalog,
        containing individual measurements, and Secondary Catalog, containing
        binned spectrophotometry averaged over multiple measurements per channel.
        Finally, \eazypy{} is applied again to the simulated SPHEREx spectrophotometry
        to derive photometric redshifts, enabling quantitative validation of the
        SPHEREx forced-photometry pipeline for extended sources in dense cluster
        environments. }
        \label{fig:pipeline}
    \end{figure*}

    Our goal is to perform a first end–to–end assessment of redshift and
    photometry performance of \spherex{} for clusters. We simulate \spherex{} observations
    of cluster fields and then carry out forced photometry on the resulting images,
    emulating the \spherex{} data products. With these controlled experiments, we
    test how characteristic features of \spherex{} data impact the recovery of
    physical quantities for member galaxies—most notably (i) photometry of
    extended sources under the large pixel scale, (ii) source blending in crowded
    cluster environments, and (iii) survey depth across channels.

    The end-to-end pipeline developed for this work is schematically summarized
    in Figure~\ref{fig:pipeline}. The process begins with the selection of targets
    from the cluster and field catalogs. We then use ancillary photometry from
    the Legacy Survey (or COSMOS) to perform an initial SED estimation for each
    source with \eazypy{} \citep{brammer08,eazypy}, a template-fitting code for photometric
    redshift and SED estimation. These initial SEDs, along with morphological information,
    serve as inputs to the \simulator{} to generate realistic mock images. We
    then perform forced photometry on these images using \tractor{} to measure the
    fluxes. The resulting measurements are compiled into Level~3-like Primary
    and Secondary catalogs. The final photometric redshifts, which are one of
    the main focus of our performance analysis, are subsequently derived by applying
    \eazypy{} to the binned spectrophotometry of the Secondary Catalog. The subsequent
    sections will describe each of these major steps in detail.

    \subsection{SPHEREx Survey Strategy}
    \label{sec:survey} To understand the data products we emulate, it is
    essential to first consider the SPHEREx survey strategy. \spherex{} conducts
    an all-sky spectrophotometric survey over $0.75$--$5\,\mu$m with a spectral
    resolution of $R \approx 35$--$130$. It does not obtain a spectrum in a
    single observation; instead, it builds up spectrophotometric information for
    every pixel on the sky from a series of images taken over its 2-year mission
    \citep{dore14, crill20, spherex}. The telescope utilizes Linear Variable
    Filters (LVFs), where the transmitted wavelength varies continuously along one
    direction of the detector arrays \citep{korngut18, spherex}. Over the course
    of four complete all-sky surveys, a series of small and large slews repoints
    the telescope, causing any given celestial object to be measured at different
    detector positions, and thus at different wavelengths \citep{spangelo15, bryan25}.
    These individual exposures are processed into Level 2 Calibrated Spectral
    Images, and the Level 3 catalogs are derived from these images \citep{crill20, crill25, akeson25}.
    The Level 3 data consist of two main types: Primary Catalog and Secondary
    Catalog.\footnote{The publicly released SPHEREx Level~3 products correspond
    to the ``High Reliability'' Primary and Secondary catalogs, a quality-filtered
    subset of the full catalogs. As we construct our own Level~3-like catalogs
    from forced photometry, this distinction does not affect our analysis.} The
    Primary Catalog contains every individual photometric measurement for each
    source at its unique observed wavelength. The Secondary Catalog is created by
    binning and averaging the measurements from the Primary Catalog onto a common
    grid of 102 predefined spectral channels. This process simplifies the data
    into a single, consistently-sampled spectrum per object and increases the effective
    depth in each channel \citep{crill25}.

    The first step in our pipeline is to generate realistic mock observations
    using the \simulator{} \citep{crill25}. This simulator integrates three core
    components to reproduce SPHEREx observations with high fidelity: (i) a
    comprehensive sky model incorporating astrophysical emission from various
    sources, including Zodiacal light, diffuse Galactic light, and stars/galaxies
    seeded from the Reference Catalog; (ii) a survey plan that dictates the
    precise pointing, timing, and orientation of the spacecraft for each
    observation; and (iii) a sophisticated instrument model based on pre-launch characterization
    data. The instrument model is particularly important for this work, as it implements
    detailed simulations of the PSF, the pixel-dependent bandpass of the LVFs, and
    a variety of detector-level noise and systematic effects. By combining these
    elements, the simulator can generate reliable data essential for testing the
    performance of our photometry pipeline in the dense and complex environments
    of galaxy clusters.

    \subsection{Source Characterization}
    \label{sec:source}

    \subsubsection{SED Modeling}
    \label{sec:sed}

    The primary goal of this stage is to generate a self-consistent SED for sources
    in our target fields. Using \eazypy, we fit the $grizW1W2$ photometry from the
    Legacy Survey for the cluster fields, and the multi-wavelength photometry (see
    Section~\ref{sec:sample}) for the COSMOS field.

    We perform two separate fits for each source: one with a library of galaxy templates
    and the other with stellar templates. First, each source is fit with a library
    of 160 galaxy templates compiled by \citet{feder24} from \citet{brown14} and
    \citet{ilbert09}, to determine its best-fit galaxy model and corresponding goodness-of-fit,
    $\chi^{2}_{\text{gal}}$. We use single-template fitting (i.e., no linear
    combination between templates) over a redshift grid of $z= 0.001$-$5.0$ where
    the grid step $\Delta z= 0.01$, without applying any priors. A flat 10\%
    systematic error is added to the photometric uncertainties to account for
    systematics not captured by the formal catalog errors \citep[e.g.,][]{serra11, lang16, weaver22}.
    The star--galaxy classification purity (see Section \ref{sec:sgsep}) is
    marginally higher at 10\% than at a smaller floor and changes negligibly across
    the 5--20\% range. More complex corrections, such as zero-point adjustments or
    template error functions, are not applied, as our goal is to create a reproducible
    input model rather than achieve the most precise SED fit. For sources with
    available spectroscopic redshifts, we fix the redshift to its known value during
    this galaxy template fitting process to ensure an accurate SED model for confirmed
    cluster members.

    Second, every source is also fit with a library of theoretical stellar
    templates (PHOENIX BT-Settl; \citealt{allard12}) to find its best-fit star
    model and corresponding $\chi^{2}_{\text{star}}$. This process results in every
    source in our catalog to have both best-fit galaxy and star models, the
    combination of which is used for classification in the next step.

    To assess the fidelity of the SED templates adopted as the ground truth for
    our simulation, we validate the fitting quality by comparing the photometric
    redshift estimates ($z_{\text{phot}}$) with the available spectroscopic redshifts
    ($z_{\text{true}}$) for sources in our sample. We quantify the performance
    using several standard statistical metrics. Letting $\Delta z = z_{\text{phot}}
    - z_{\text{true}}$, these are: the bias, defined as the mean of the
    normalized redshift error, $\langle \Delta z / (1+ z_{\text{true}}) \rangle$;
    the scatter, given by the normalized median absolute deviation,
    $\sigma_{\text{NMAD}}= 1.48 \times \text{median}[|\Delta z - \text{median}(\Delta
    z)| / (1+z_{\text{true}})]$; and the catastrophic outlier fraction ($\eta$),
    which we define using two common criteria: the fraction of sources with $|\Delta
    z| / (1+z_{\text{true}}) > 0. 15$ , $\eta_{0.15}$, and the fraction with
    $|\Delta z| / \hat{\sigma}> 3$, $\eta_{\hat{\sigma}}$, where $\hat{\sigma}$
    is the estimated error on the photometric redshift.

    For the subset of sources with available spectroscopy, running the fit without
    fixing their redshifts yields a bias of $\sim$0.05 and a scatter of
    $\sigma_{\text{NMAD}}\approx 0.07$. The corresponding outlier fractions are $\eta
    _{0.15}\approx 22\%$ and $\eta_{\hat{\sigma}}\approx 5.9\%$, respectively. We
    note that these statistics are heavily influenced by faint sources with lower
    signal-to-noise ratios. When restricting the comparison to bright sources
    with Legacy Survey $z$-band magnitude $m_{z, \text{LS}}< 19$, the fidelity improves:
    the scatter decreases to $\sigma_{\text{NMAD}}\approx 0.05$, and the catastrophic
    outlier fraction drops to $\eta_{0.15}\approx 9.9\%$ (with bias of $\sim$0.06
    and $\eta_{\hat{\sigma}}\approx 5.4\%$). Similarly, the fitting quality
    depends strongly on galaxy type: for passive galaxies (E, S0, and Sa
    templates), the scatter improves to $\sigma_{\rm NMAD}\approx 0.05$ with
    $\eta_{0.15}\approx 2.7\%$, compared to $\sigma_{\rm NMAD}\approx 0.14$ and $\eta
    _{0.15}\approx 25\%$ for star-forming galaxies, reflecting the smoother SEDs
    of quiescent systems. These results confirm that the template library provides
    a reliable characterization of the galaxy population. Therefore, for sources
    lacking spectroscopic confirmation, we adopt the best-fit SEDs from this
    process as the input for our simulation. However, as the photometric-redshift
    scatter from this input-SED fit ($\sigma_{\text{NMAD}}\approx 0.07$) is too
    large for reliable cluster member selection, we reinforce our decision to use
    spectroscopic redshifts exclusively for that purpose in the subsequent analysis.

    \subsubsection{Star--Galaxy Separation}
    \label{sec:sgsep}

    To apply the correct SED model and simulate sources realistically, we classify
    each object as either a star or a galaxy. An object is classified as a
    galaxy if it satisfies at least one of the following three criteria: (i) its
    morphological type in the Legacy Survey Tractor catalog is not `PSF'; (ii)
    it has a spectroscopic redshift $z_{\text{spec}}> 0.002$; or (iii) the goodness-of-fit
    for the galaxy template model is better than that for the stellar template model
    ($\chi^{2}_{\text{gal}}< \chi^{2}_{\text{star}}$). Objects that do not meet
    any of these criteria are classified as stars. This scheme is applied to our
    full working sample of 3,210,062 sources (assembled in Section~\ref{sec:sample}),
    separating it into 2,603,587 galaxies (81.1\%) and 606,475 stars (18.9\%). Once
    classified, the corresponding best-fit SED is adopted as the ``ground truth''
    spectrum for that source in the subsequent simulation.

    \begin{figure}
        \centering
        \includegraphics[width=0.48\textwidth]{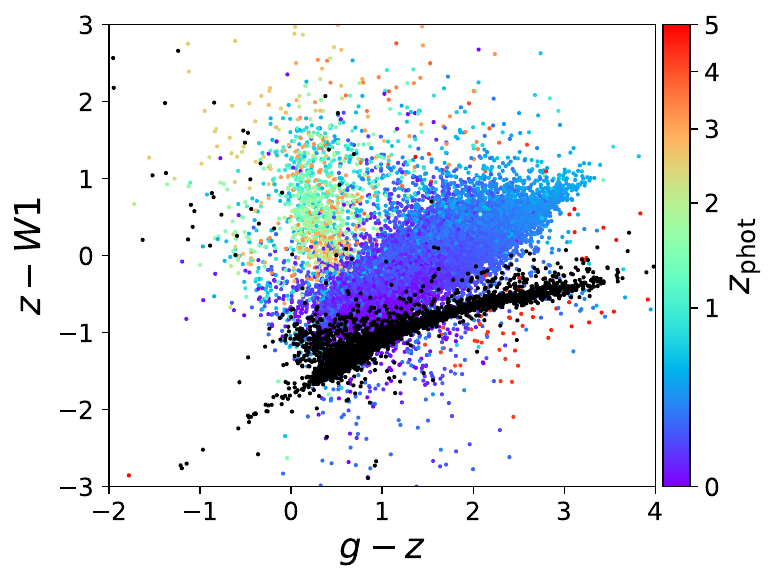}
        \caption{ Color–color diagram ($g-z$ vs.\ $z-W1$) used for star–galaxy
        separation. Sources classified as galaxies are shown with color-coding according
        to their photometric redshifts, while sources classified as stars are
        marked as black points. Only the top 1\% of objects with the highest
        signal-to-noise ratio in the Legacy Survey $r$-band flux are displayed. }
        \label{fig:sgsep}
    \end{figure}

    The effectiveness of this classification scheme is shown in Figure~\ref{fig:sgsep}.
    The color--color diagram shows a clear separation between the stellar locus,
    occupied by sources classified as stars (black points), and the broader distribution
    of galaxies (color-coded by photometric redshift). This demonstates that our
    star-galaxy separation criteria successfully distinguish the two populations.
    To quantify the classification performance at faint magnitudes where galaxies
    dominate, we measured the completeness and purity of the galaxy class as a
    function of Legacy Survey $z$-band magnitude, using spectroscopic redshifts as
    the truth label. Both metrics remain above 98\% for sources brighter than the
    SPHEREx detection limit ($m_{z,\,\rm LS}\lesssim 19$) and exceed 95\% down
    to $m_{z,\,\rm LS}\sim 22$. The discriminating power degrades only at $m_{z,\,\rm
    LS}\gtrsim 23$, confirming that the classification remains robust throughout
    the magnitude range relevant to our analysis. We further note that the $z$-band
    magnitude is a conservative proxy for SPHEREx detectability, because red and
    high-redshift galaxies are brighter in the near-infrared than in the $z$
    band; our classification nonetheless remains $\gtrsim 95\%$ complete and pure
    down to $m_{z,\,\rm LS}\sim 22$, covering the population accessible to
    SPHEREx.

    \subsubsection{Cluster Member Selection}
    \label{sec:member}

    The final step in source characterization is to identify the bona fide members
    of each galaxy cluster from the sample classified as galaxies. Using the
    photometric redshifts derived from the initial SED fitting is unreliable for
    this purpose, as the typical scatter ($\sigma_{\text{NMAD}}\approx 0.07$) would
    lead to significant contamination from foreground and background galaxies.
    Therefore, we determine cluster membership exclusively for galaxies with available
    spectroscopic redshifts.

    \begin{figure*}
        \centering
        \includegraphics[width=\textwidth]{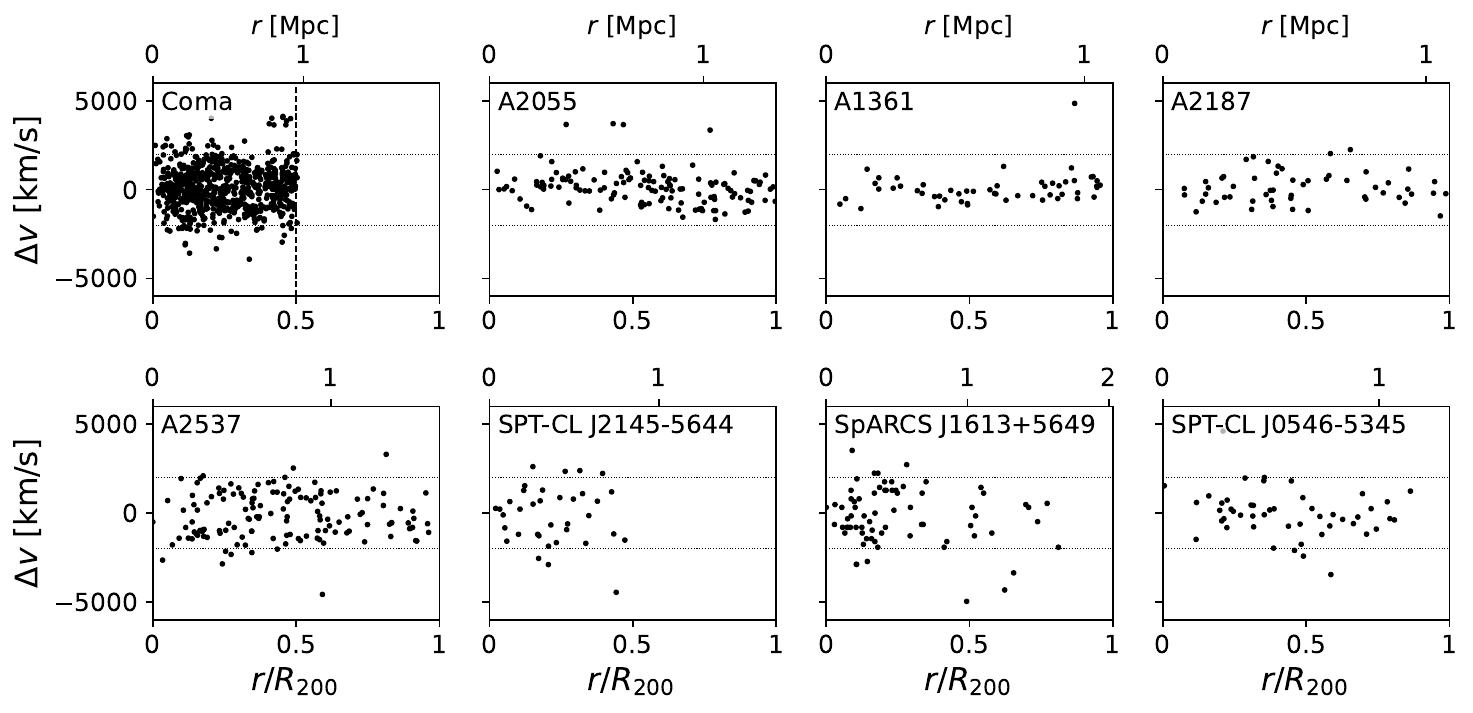}
        \caption{Phase-space diagrams ($\Delta v$ vs. projected radius) for galaxies
        with spectroscopic redshifts in the selected cluster fields. In each
        panel the bottom axis shows the projected radius normalized to the
        cluster's virial radius, $r/R_{200}$, and the corresponding physical
        scale in Mpc is shown on the top axis. Galaxies within $\pm 2000~\mathrm{km\,s^{-1}}$
        of the cluster redshift, indicated by the horizontal dotted lines, are identified
        as cluster members. For each cluster, the data extend out to $R_{200}$ (and
        to $0.5R_{200}$ for Coma), so galaxies beyond these radii are not
        included.}
        \label{fig:dvr}
    \end{figure*}

    A galaxy is identified as a cluster member if its line-of-sight velocity
    difference relative to the systemic redshift of the cluster ($z_{\rm cl}$)
    is within $\pm 2000~\mathrm{km\,s^{-1}}$. Figure~\ref{fig:dvr} displays the phase-space
    diagrams ($\Delta v$ vs. projected radius $r$) for the galaxies with
    spectroscopic redshifts in our selected cluster fields. The horizontal dotted
    lines indicate the $\pm 2000~\mathrm{km\,s^{-1}}$ velocity cut, showing a
    clear concentration of member galaxies within this range. This confirms that
    our velocity threshold is effective for selecting cluster members. A small
    number of galaxies lie just outside this boundary and may be genuine members,
    but we do not attempt to recover them. Excluding these few galaxies does not
    significantly affect our results, as our analysis requires a high-purity member
    sample rather than precise dynamical characterization of the clusters.

    \subsection{Simulation and Photometry Pipeline}
    \label{sec:pipeline}

    \begin{figure*}
        \centering
        \includegraphics[width=\textwidth]{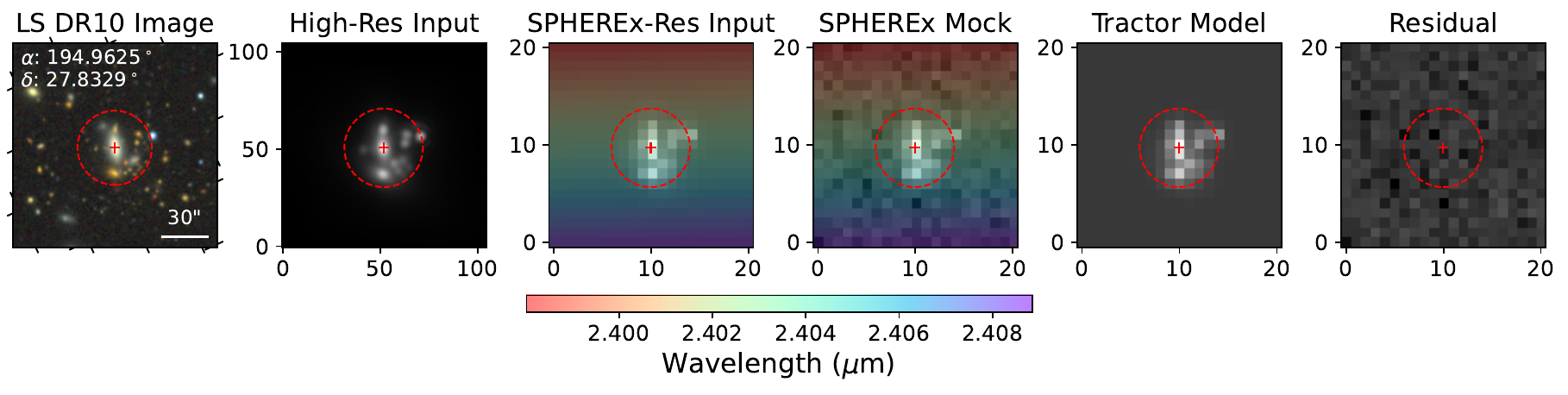}
        \includegraphics[width=\textwidth]{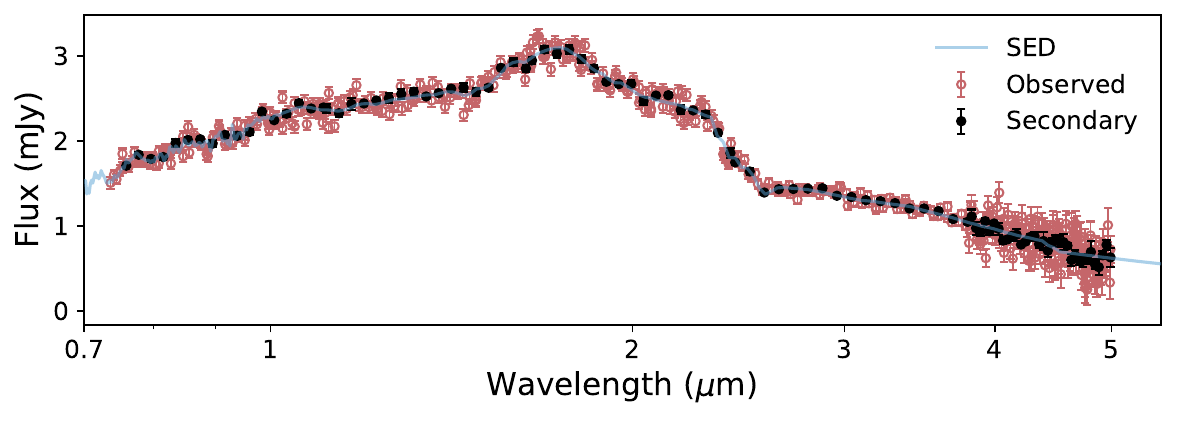}
        \caption{Illustration of the photometry pipeline for a single target
        galaxy. In the upper panels, the cross marks the target galaxy, while the
        dashed circle shows five times its half-light radius; only sources
        within this region are included in the image simulation. Starting from the
        left, the Legacy Survey image and the corresponding high-resolution
        input (five times finer than SPHEREx) are shown. This high-resolution input
        is constructed using the SED and shape parameters, then downsampled to
        the SPHEREx resolution and combined with noise to generate the mock
        SPHEREx image. Using this mock image together with the input galaxy positions
        and shapes, forced photometry is performed with \tractor{}; all sources
        within the cutout are fit simultaneously with their positions and shapes
        held fixed and the resulting residual is displayed on the right. The SPHEREx
        resolution and mock panels are color-coded by the central wavelength of the
        response curve for each pixel, illustrating that wavelength variations
        across neighboring pixels are small. The upper sequence demonstrates the
        photometry procedure for a single detector in one SPHEREx pointing. The
        lower panel shows the resulting spectrum after combining data from
        multiple detectors and pointings. Each flux measurement is associated with
        a slightly different wavelength, which is averaged into predefined 102
        channels to form the Secondary Catalog. The comparison shows that both the
        observed data points (red), the Secondary Catalog values (black), and the
        input SED (blue) are in good agreement.}
        \label{fig:photo_example}
    \end{figure*}

    Here we describe the full pipeline used to generate realistic SPHEREx mock images
    of our cluster fields and to perform photometry on them. This process is
    outlined in Figure \ref{fig:photo_example}.

    Our methodology is built upon the \simulator{}, specifically extending the
    functionality of its \texttt{QuickCatalog} module \citep[Sec. 3.3 in][]{crill25},
    which treats all sources as point sources and does not model the pixel-level
    LVF response. We introduce two major modifications to tailor the simulation
    for our goals. First, we integrate the \galsim{} software library \citep{galsim}
    to render realistic two-dimensional light profiles of extended sources based
    on their morphological parameters. Second, we implement a pixel-by-pixel
    response model to accurately simulate the chromatic effects of the LVFs on
    these extended sources.

    \subsubsection{Mock Image Generation}
    \label{sec:mockimg} The process of generating mock images begins with the
    selection of viable observations for each source in our catalog. Using the SPHEREx
    survey plan (R3.1.1), we first identify all pointings where a source would
    be observed by any of the six detectors during the mission. For each of these
    observations, we calculate the expected total flux by integrating the SED of
    sources (derived in Section~\ref{sec:source}) with the instrumental response
    curve of the central pixel on which the source lands. We then exclude any observation
    from our simulation if this flux is more than one magnitude fainter than the
    nominal 5$\sigma$ point-source sensitivity of the all-sky survey for that
    specific SPHEREx channel \citep[we use the optimistic ``current best
    estimate'', CBE value;][]{dore14}. This pre-selection ensures that our pipeline
    focuses computational resources on observations where the source is potentially
    detectable. For a passive SED at $z\approx 0.1$ ($0.3$), this pre-selection retains
    sources down to rest-frame $K_{s}\approx 21.2$ ($2 1.7$), $20.5$ ($20.7$), and
    $17.6$ ($18. 4$) for at least one, $\geq 50$, and all 102 channels,
    respectively.

    For each of the remaining valid observations, we generate a 21$\times$21 pixel
    mock cutout image, centered at a target source. The process begins by selecting
    the primary target and its neighboring sources. This selection is performed
    using an adaptive criterion that accounts for the physical extent of each
    galaxy: a neighboring source is included in the cutout if the separation is less
    than $5\,r_{e}$ of either the primary target or the neighbor itself, with a
    minimum threshold of 3 SPHEREx pixels ($\approx 18\farcs5$) to ensure that
    compact or point-like neighbors in close proximity are always accounted for.

    For each of these sources individually, we first generate a high-resolution (oversampled
    by a factor of 5) light profile using \galsim{} based on its cataloged
    morphology (point sources for those classified as stars, S\'ersic index, $r_{e}$,
    $e_{1}$, and $e_{2}$ for galaxies), which is then convolved with the appropriate
    SPHEREx PSF. We use the pre-launch, position- and wavelength-dependent PSF
    models provided by the \simulator{} \citep{crill25}. For each observation,
    we adopt the PSF at the source position and wavelength on the detector. This
    high-resolution image of a single source is then downscaled to the native SPHEREx
    pixel scale. To accurately model the chromatic effects of the LVFs, a unique
    ``response map'' is calculated for that specific source using its own SED;
    this map models the relative flux variation across the pixels and is
    multiplied by the PSF-convolved light profile of the source (see Appendix~\ref{app:ppr}
    for details). This procedure results in a set of individual, fully processed,
    native-resolution images for each source. Finally, these individual images are
    co-added onto a single 21$\times$21 pixel cutout to create the final blended
    scene, to which a realistic Zodiacal light background, evaluated at the time
    and sky position of each observation, and instrumental detector noise are
    added.

    \subsubsection{Tractor Photometry}
    We perform photometry on these mock images using \tractor{}, which employs a
    model-fitting approach. The scene is modeled as a collection of sources,
    where each source is described by a profile convolved with the local PSF. For
    galaxies, we use our S\'ersic model, a modified version of the default Mixture-of-Gaussians
    (MoG) model in \tractor{}, optimized for SPHEREx data (see Appendix~\ref{app:tractor}
    for details). \tractor{} then optimizes the model parameters to achieve a minimum
    $\chi^{2}$ between the rendered model image and the input mock data. During
    the optimization, the galaxy models are convolved with the PSF in the high-resolution
    scene before being downsampled to the SPHEREx pixel scale, ensuring accurate
    modeling of the PSF effects. The same PSF model is used for both image generation
    and fitting, so our tests do not include any error from PSF mismatch. For this
    work, we perform forced photometry by freezing the shapes and positions of
    all sources to their known input values and fitting only for their fluxes; all
    sources in a given cutout are optimized simultaneously, so that each target's
    flux is constrained jointly with those of its blended neighbors. Both priors
    are idealized: the optical morphology is held fixed across channels, neglecting
    its wavelength dependence (Section~\ref{sec:limitations}), and the positions
    are fixed to their true values, assuming perfect astrometry (Appendix~\ref{app:astrometry}).

    The result of this process is a best-fit flux and its associated uncertainty
    for every source in each observation. These uncertainties are \tractor{}'s native
    flux errors, evaluated under the Cram\'er--Rao bound and therefore computed as
    though each source were isolated; they do not propagate the source--source flux
    covariance that arises from overlapping PSFs. This covariance is
    incorporated by the \spherex{} Level~3 pipeline \citep{akeson25}, following the
    formalism of \citet{huai25}. We adopt the native uncertainties throughout this
    work and quantify the effect of this choice in Appendix~\ref{app:cov}, where
    we find the photometric-redshift performance to be insensitive to it. All individual
    measurements are then compiled into a Primary Catalog, which mimics the
    format of the official SPHEREx Level 3 data product. This catalog is subsequently
    processed to create a Secondary Catalog by binning and averaging all measurements
    for each source onto a common grid of 102 pre-defined spectral channels.
    This final binned catalog provides a single, consistently-sampled spectrum per
    object with enhanced signal-to-noise and serves as the input for the photometric
    redshift estimation with \eazypy.

    \section{Photometry Performance}
    \label{sec:phot}

    In this section, we present the results of our end-to-end pipeline,
    evaluating the performance of the photometry derived from the simulated
    SPHEREx images. The full simulation includes 49,169 unique sources,
    comprising 43,016 galaxies and 6,153 stars. Among these, 1,032 galaxies were
    identified as spectroscopically confirmed cluster members. To obtain spectrophotometry
    for these sources, a total of 12,017,922 individual mock images were
    generated and processed. We assess the performance of this extensive dataset
    in three key areas: photometric bias and error estimation, the impact of source
    blending, and the effective survey depth.

    \subsection{Bias and Error Estimation}
    \label{sec:phot-bias}

    We first evaluate the fundamental performance of our photometry pipeline by analyzing
    the residuals between the measured flux ($f$) and the known input flux ($f_{\rm
    input}$). We use two primary metrics: the fractional flux residual,
    $(f - f_{\rm input})/f_{\rm input}$, which quantifies the relative accuracy of
    the measurement across a wide range of source brightnesses; and the
    normalized flux residual, $(f - f_{\rm input})/\sigma_{f}$, which tests the
    reliability of the photometric uncertainty, $\sigma_{f}$, estimated by
    \tractor{}. An ideal error estimation would result in a standard normal
    distribution for the normalized flux residuals.

    \begin{figure*}
        \centering
        \includegraphics[width=\textwidth]{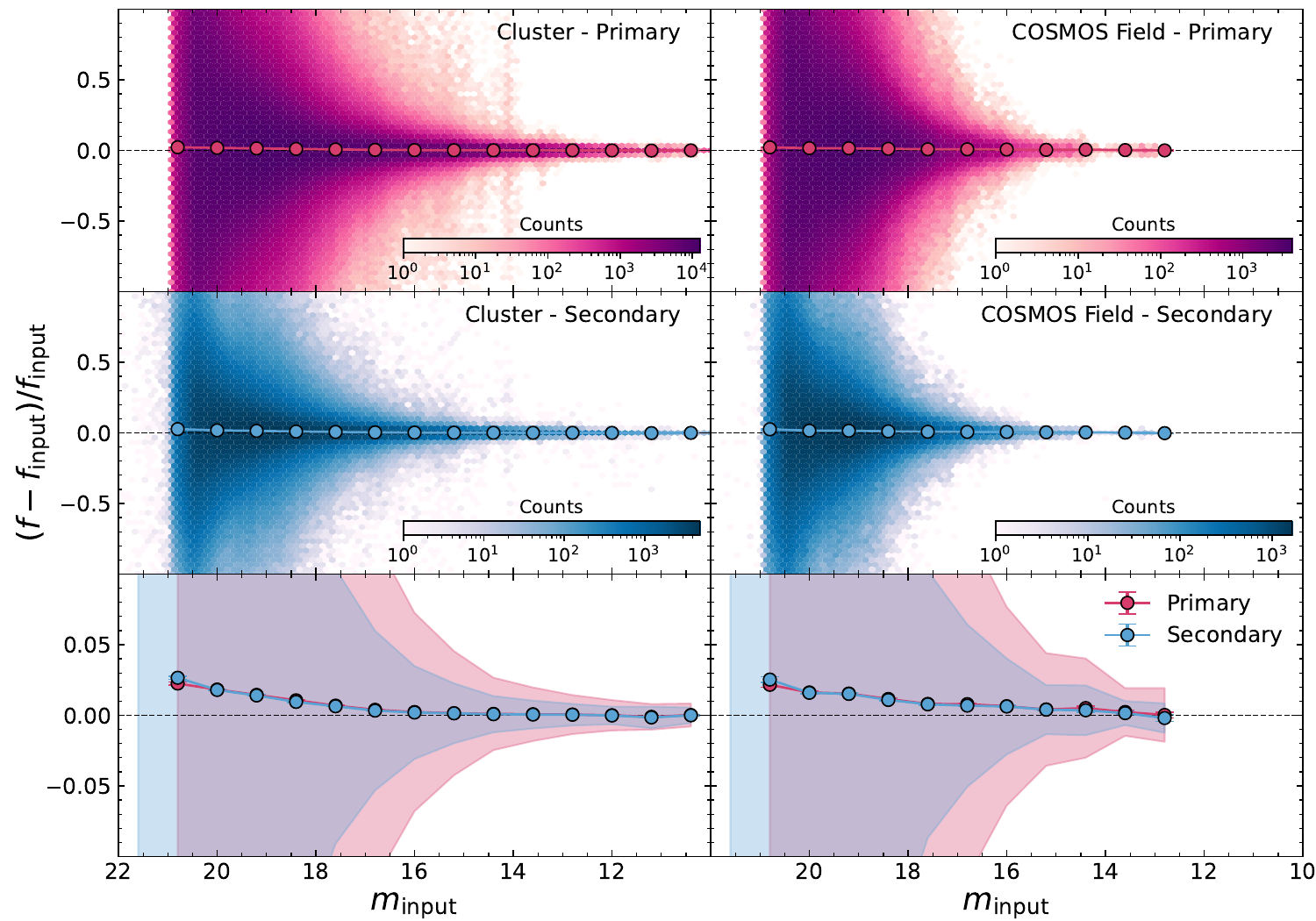}
        \caption{Fractional flux residual, $(f - f_{\rm input})/f_{\rm input}$,
        as a function of input magnitude, shown separately for the cluster and COSMOS
        field measurements. The input magnitude ($m_{\rm input}$) corresponds to
        the source brightness at the observed wavelength (per-pixel LVF bandpass
        for the Primary Catalog; per-channel bandpass for the Secondary Catalog).
        Top row: Primary Catalog (single-image photometry), for the cluster
        sample (left) and the COSMOS field sample (right). Middle row: Same for
        the Secondary Catalog (combined, channelized catalog)$^{\dagger}$. Bottom
        row: Direct Primary--Secondary comparison within each field, with the
        vertical axis zoomed in to highlight small biases. Filled circles mark the
        mean residual in each magnitude bin and shaded regions show the corresponding
        $1\sigma$ scatter. For bright sources, the bias is consistent with zero in
        both fields; at the faintest magnitudes, a small positive bias of order $\sim$2\%
        appears, but this level is generally negligible for most scientific
        applications. The cluster and COSMOS results agree closely in both bias and
        scatter, demonstrating that the pipeline performance is robust to the
        cluster environment at the field-averaged level.
        \newline
        \footnotesize{$^{\dagger}$ Sources at $m_{\rm input}\gtrsim 21$ passed the sensitivity cut via bright emission lines in the Primary data. Their input magnitudes in the Secondary Catalog are defined at fixed channel wavelengths, which often sample the fainter continuum, causing them to shift to these fainter bins.}}
        \label{fig:photbias}
    \end{figure*}

    Figure~\ref{fig:photbias} shows the fractional flux residual as a function of
    input magnitude, presented separately for the cluster and COSMOS field
    measurements. This separation directly addresses whether the photometric performance
    is sensitive to the cluster environment, which is a primary motivation of
    this work. Note that this input magnitude refers to the brightness at the specific
    wavelength of observation: for the Primary Catalog, it is derived from the input
    SED with the specific LVF response curve of the observing pixel, while for
    the Secondary Catalog, it corresponds to the magnitude within each of the 102
    standardized spectral channels. For both the Primary and Secondary catalogs,
    and for both samples, the photometry is largely unbiased for bright sources.
    A small positive bias of $\sim$2\% emerges at the faintest magnitudes. The
    binning process from the Primary to the Secondary Catalog effectively
    reduces the scatter by approximately 50\%, consistent with the averaging of
    four independent sky surveys, while the mean bias remains largely unchanged.

    This faint-end bias is primarily driven by the residual mismatch between the
    GalSim S\'ersic profiles used as input and the MoG model employed by \tractor{}
    (Appendix~\ref{app:tractor}), with a smaller additional contribution from source
    blending \citep{stickley16}. The single-source test in Section~\ref{sec:blend}
    (Figure~\ref{fig:blend2}) confirms that blending is not the dominant contributor.
    The mismatch is most severe for compact, nearly-unresolved sources ($r_{e}< 1
    ''$), which dominate the faint end (Appendix~\ref{app:psf}). More broadly,
    the simulated galaxies are themselves single-S\'ersic profiles, fitted with the
    same morphology used to generate them. The bias and scatter reported here
    are therefore best-case. The larger systematics expected for real,
    morphologically complex galaxies, and for optically derived priors that differ
    from the true NIR light profile, are discussed in Section~\ref{sec:robustness}
    and Appendix~\ref{app:tractor}.

    A direct comparison between the cluster and COSMOS columns of Figure~\ref{fig:photbias}
    shows that the field-averaged bias profiles are nearly identical, with bin-by-bin
    median differences of $|\Delta\langle\Delta f/f\rangle|\lesssim 0.5\%$ across
    the main magnitude range in both catalogs. The residual scatter is also
    closely consistent between the two samples, with the cluster sample being marginally
    tighter than COSMOS at all magnitudes; the offset is at most a few percent
    in fractional units and does not indicate a systematic bias of the
    photometry in the cluster environment. We conclude that the pipeline performance
    is robust to the cluster environment when averaged over the survey footprint
    of each sample. This field-averaged comparison, however, necessarily mixes
    high-density cluster cores with surrounding lower-density regions and does
    not yet quantify how the residuals depend on the local crowding
    configuration around each source. We turn to that analysis in \S\ref{sec:blend},
    where we examine the bias and scatter as a function of crowding metrics and also
    compare the underlying distributions of those metrics in the two samples.

    \begin{figure}
        \centering
        \includegraphics[width=0.48\textwidth]{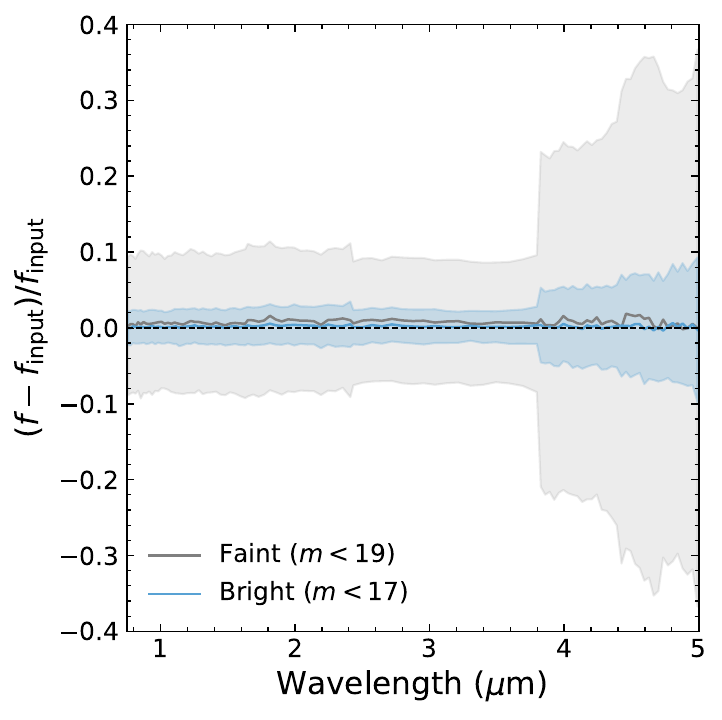}
        \caption{Fractional flux residual $(f - f_{\rm input})/f_{\rm input}$ as
        a function of wavelength for the Secondary Catalog, shown for bright ($m_{\rm
        input}< 17$; blue) and faint ($m_{\rm input}< 19$; gray) subsamples.
        Solid lines show the mean with $1\sigma$ uncertainty on the mean (error bars),
        and shaded regions show the $1\sigma$ scatter. Both subsamples exhibit a
        small positive bias that is broadly flat in wavelength, while the scatter
        increases at $\lambda \gtrsim 3.8~\mu$m due to the reduced sensitivity
        of the narrower, higher-resolution channels in Detectors~5 and~6.}
        \label{fig:wavelength_bias}
    \end{figure}

    To address whether the photometric bias and scatter exhibit a wavelength
    dependence, we examine the fractional flux residuals of the Secondary Catalog
    as a function of wavelength for two magnitude selections. Figure~\ref{fig:wavelength_bias}
    shows the mean (solid lines) and $1\sigma$ scatter (shaded regions) of
    $(f - f_{\rm input})/f_{\rm input}$ across the 102 SPHEREx channels for
    bright ($m_{\rm input}< 17$) and faint ($m_{\rm input}< 19$) subsamples. Both
    subsamples show a small positive mean bias at the level of $\sim$0.5\% for bright
    sources and $\sim$1\% for faint sources, and the bias is broadly flat in
    wavelength. The scatter, in contrast, shows a clear increase at $\lambda \gtrsim
    3.8~\mu$m, reflecting the lower sensitivity of Detectors~5 and~6. These two detectors
    are designed with higher spectral resolution ($R \approx 130$ compared to
    $R \approx 40$ for Detectors~1--4) to target ice absorption features, resulting
    in narrower bandpasses that reduce the photon collection per channel \citep{crill25}.

    \begin{figure}
        \centering
        \includegraphics[width=0.48\textwidth]{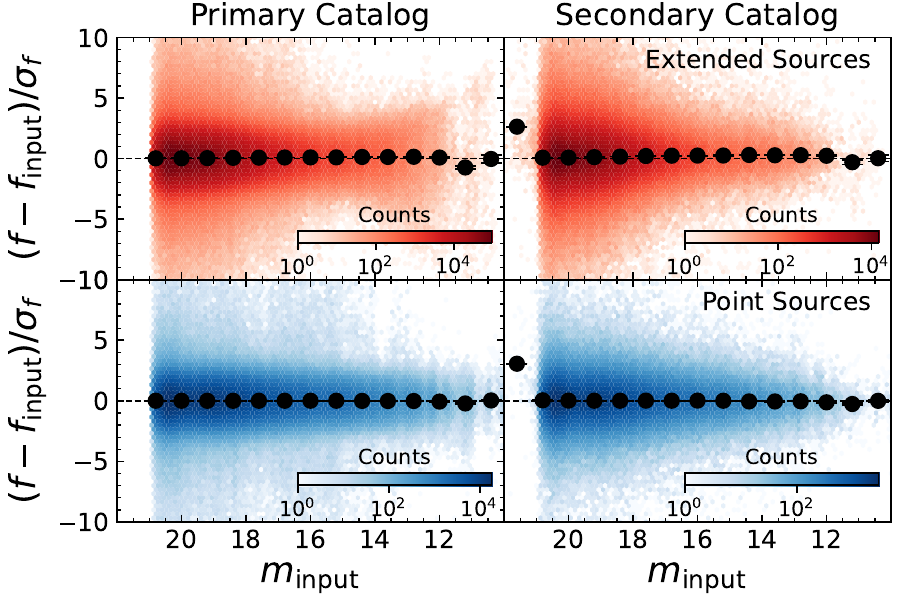}
        \caption{Bias comparison between extended and point sources in terms of
        the normalized flux residuals, $(f - f_{\mathrm{input}})/\sigma_{f}$,
        where $\sigma_{f}$ is the photometric uncertainty estimated from
        \tractor{}. The top panels show extended sources, while the bottom panels
        correspond to point sources. The left column presents results from the
        Primary Catalog, and the right column shows those from the Secondary Catalog.
        The black points mark the mean residuals in magnitude bins, indicating the
        overall unbiased performance of the photometry. }
        \label{fig:photbias_extpsf}
    \end{figure}

    In Figure~\ref{fig:photbias_extpsf}, the core of the normalized flux-residual
    distributions is consistent with a Gaussian for both extended and point
    sources\footnote{Extended sources are those assigned a resolved galaxy
    morphology in the Legacy Survey catalog, i.e., a finite half-light radius $r_{e}$
    (types REX, EXP, DEV, or SER; see Section~\ref{sec:sample}); point sources
    carry the unresolved PSF classification.}, confirming that the formal
    uncertainties from \tractor{} are broadly reliable. They are, however,
    underestimated relative to the true scatter, as expected from the Cram\'er--Rao
    bound, which provides only a theoretical lower limit on the noise estimate.
    This under-estimation is consistent with the behavior reported by \citet{huai25},
    who attribute it to the flux covariance between neighboring sources. We report
    \tractor{}'s native uncertainties directly, without rescaling; in Appendix~\ref{app:cov}
    we quantify this effect and confirm that it does not propagate into the
    photometric redshifts.

    The extended-source residuals additionally show a brightness dependence: in the
    Primary Catalog, the normalized scatter increases toward bright objects (top
    panels). This reflects the residual mismatch between the GalSim S\'ersic profiles
    used to generate the mock images and the MoG model used by \tractor{} in the
    fit (Appendix~\ref{app:tractor}). The associated fractional scatter does not
    decrease with brightness, whereas the formal errors do, so this mismatch comes
    to dominate the normalized scatter for bright, well-resolved galaxies, where
    the formal errors are smallest. Because this residual is not fully
    correlated across the repeated measurements of each source, averaging substantially
    reduces the scatter in the Secondary Catalog, while the small associated
    mean bias persists in both catalogs (Figure~\ref{fig:photbias}).

    \begin{figure}
        \centering
        \includegraphics[width=0.48\textwidth]{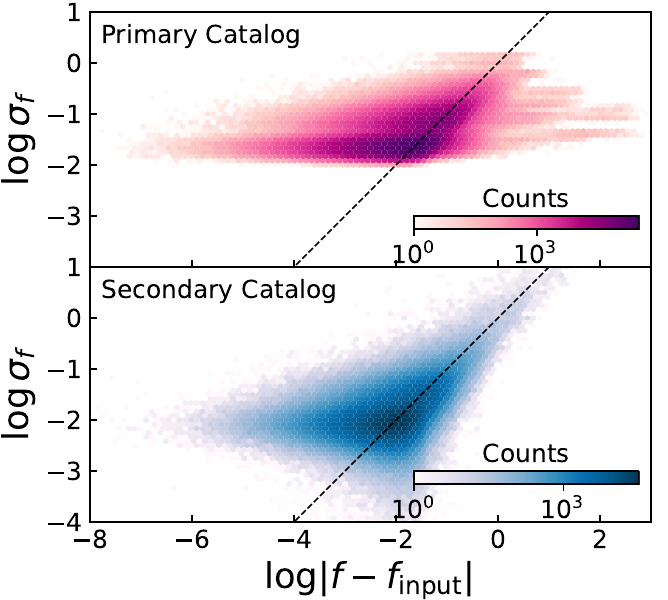}
        \caption{Comparison between flux residuals and reported measurement
        errors for the Primary (top) and Secondary (bottom) catalogs. The horizontal
        axis shows the logarithm of the absolute flux residuals (in mJy),
        $\log|f - f_{\mathrm{input}}|$, and the vertical axis shows the logarithm
        of the measurement errors, $\log \sigma_{f}$. The black dashed line indicates
        the one-to-one relation, along which the residuals would align with the
        errors in an ideal case. }
        \label{fig:photerr}
    \end{figure}

    Figure~\ref{fig:photerr} directly compares the absolute flux residuals with the
    reported measurement errors. The Secondary Catalog (bottom panel) shows a
    tighter correlation along the one-to-one line compared to the Primary
    Catalog, indicating that the binning process yields more robust error estimates
    by averaging out outliers and reducing instances of underestimated errors.

    \subsection{Blending Effects}
    \label{sec:blend}

    To specifically investigate the impact of source blending, we consider three
    metrics that characterize the crowding around a target galaxy, considering
    the neighbors selected within the $5r_{e}$ simulation boundary, subject to a
    minimum separation threshold of three \spherex{} pixels (see Section \ref{sec:mockimg}):
    (1) the flux-weighted normalized neighbor distance,
    $\langle d_{\rm neigh}/(r_{e}+r_{e, {\rm neigh}})\rangle_{f}$, where each neighbor's
    normalized distance is weighted by its flux so that bright neighbors contribute
    more, and which is small for close neighbors; (2) the total neighbor-to-target
    flux ratio, $\sum{f_{\rm neigh}/f}$; and (3) the total number of neighbors,
    $N_{\rm neigh}$. For the Secondary Catalog, these parameters are computed as
    a weighted average of the values from the contributing Primary Catalog
    measurements.

    \begin{figure*}
        \centering
        \includegraphics[width=\textwidth]{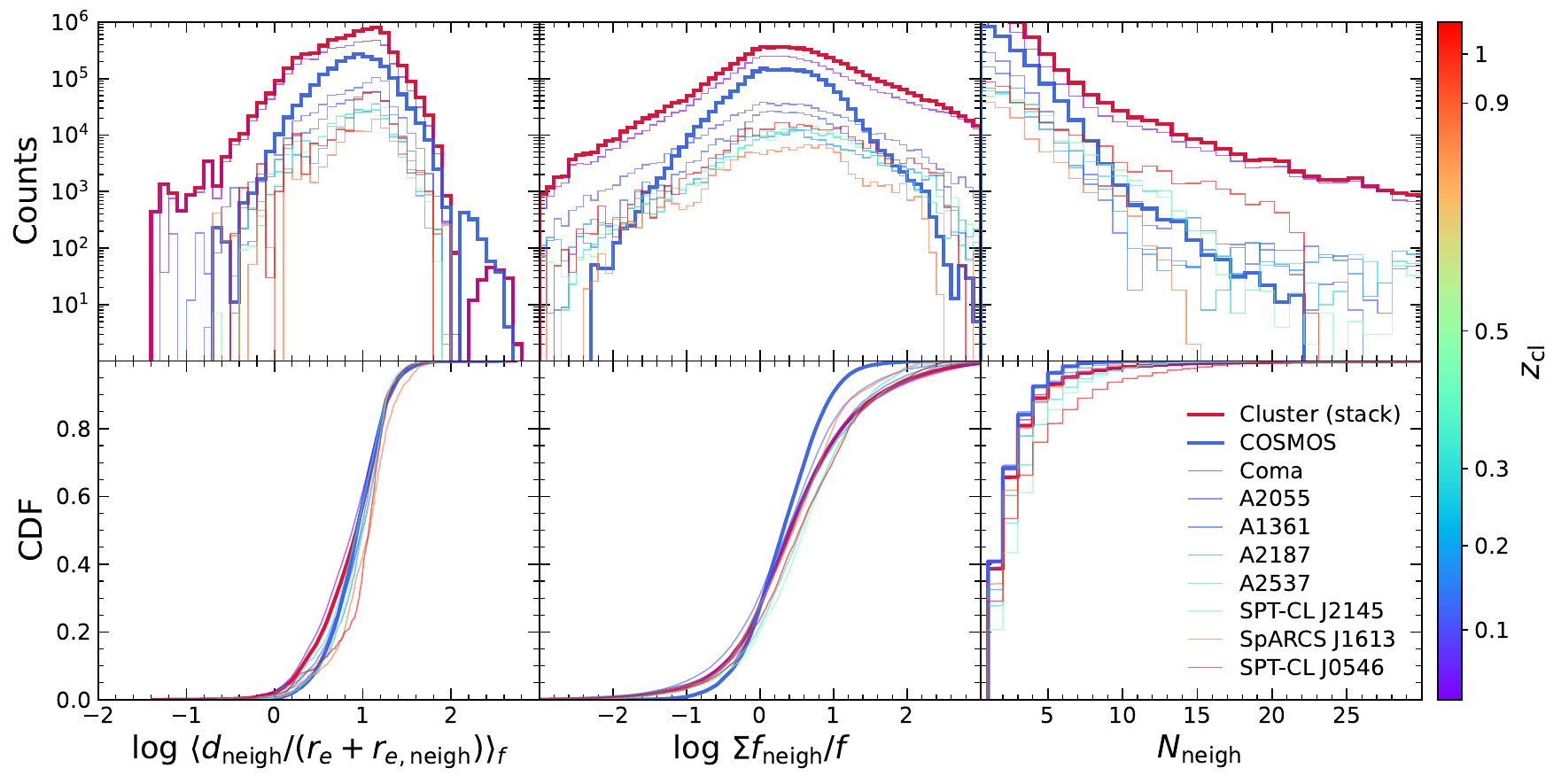}
        \caption{Distributions of the three crowding metrics for the cluster and
        COSMOS samples (Primary Catalog, sources with $N_{\rm neigh}\geq 1$).
        Columns, from left to right: flux-weighted normalized neighbor distance,
        $\log\langle d_{\rm neigh}/(r_{e}+ r_{e,\rm neigh})\rangle_{f}$; total
        neighbor-to-target flux ratio, $\log\Sigma f_{\rm neigh}/f$; number of
        neighbors, $N_{\rm neigh}$. Top row: raw histograms on a logarithmic counts
        scale. Bottom row: cumulative distribution functions. Thick red lines
        show the full cluster stack, blue lines show the COSMOS sample, and thin
        lines show individual clusters color-coded by cluster redshift (color bar).
        The bulk of all three distributions agrees between the two samples, but
        the cluster sample carries a substantially heavier high-crowding tail in
        $\Sigma f_{\rm neigh}/f$ and $N_{\rm neigh}$; the normalized neighbor-distance
        distribution is similar in the two samples, with only the lowest-redshift
        cluster (Coma) showing a modest shift toward smaller separations.}
        \label{fig:blend_dist}
    \end{figure*}

    We first compare the metrics' distributions in the cluster and COSMOS
    samples to characterize how the local crowding environment differs between
    the two samples. Figure~\ref{fig:blend_dist} shows the histograms (top row) and
    cumulative distributions (bottom row) of the three crowding metrics for sources
    with at least one neighbor in the Primary Catalog. The bulk of all three
    distributions overlaps between the two samples: the median $\log\langle d_{\rm
    neigh}/(r_{e}+ r_{e,\rm neigh})\rangle_{f}$ differs by only $\sim 0.01$~dex ($0
    .93$
    versus $0.94$), the median $\log\Sigma f_{\rm neigh}/f$ by $\sim 0.1$~dex ($0
    .41$ versus $0.32$), and the median $N_{\rm neigh}$ is identical ($2$ in
    both). The high-crowding tails, in contrast, diverge substantially: at the $9
    9$th percentile, $\Sigma f_{\rm neigh}/f$ in the cluster sample exceeds that
    in COSMOS by a factor of $\sim 20$ in linear units, and $N_{\rm neigh}$ is
    roughly twice as high ($13$ versus $7$). The normalized neighbor-distance distribution
    is similar between the two samples at all percentiles, with only Coma ($z = 0
    .02 3$) showing a modest excess at small separations.

    These differences in distributions clarify the field-averaged result of
    Section~\ref{sec:phot-bias}: the close agreement between cluster and COSMOS in
    Figure~\ref{fig:photbias} reflects the similarity of the bulk crowding configurations
    in the two samples, while the cluster-specific signature of dense
    environments is concentrated in the high-$\Sigma f_{\rm neigh}/f$ and high-$N
    _{\rm neigh}$
    tails.

    \begin{figure*}
        \centering
        \includegraphics[width=\textwidth]{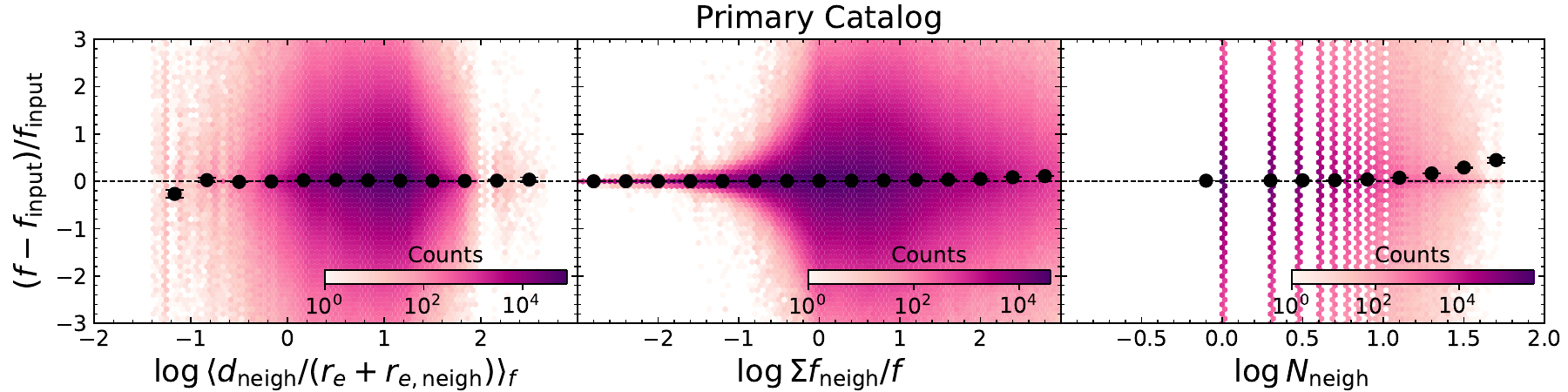}
        \includegraphics[width=\textwidth]{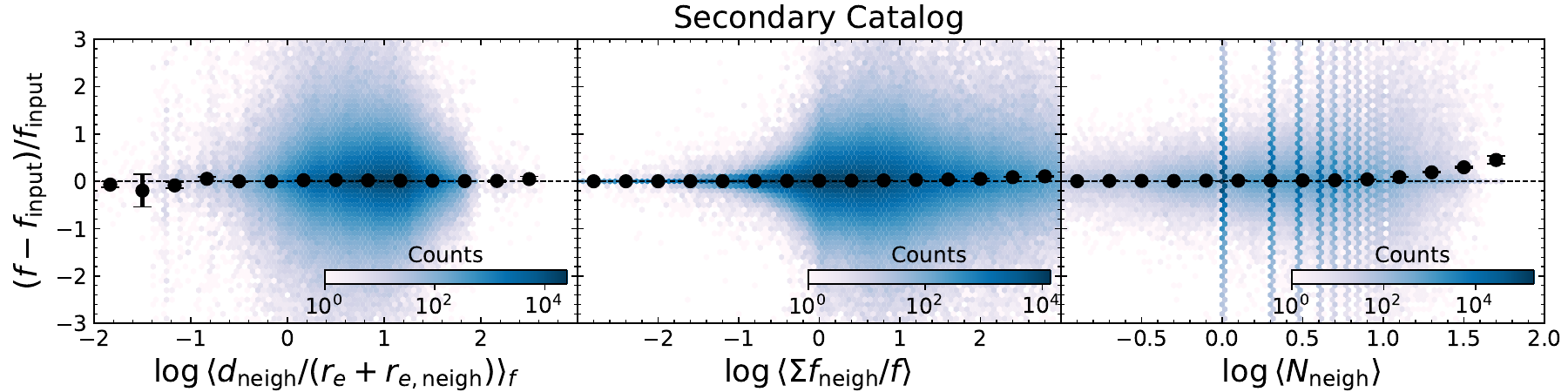}
        \caption{ The fractional flux residual as a function of three blending-related
        parameters for the Primary Catalog (top row) and Secondary Catalog (bottom
        row). The panels show the bias and scatter, with respect to the normalized
        neighbor distance (left), the total neighbor-to-target flux ratio (middle),
        and the number of neighbors (right). For the Secondary Catalog, the blending
        parameters on the horizontal axes are the weighted average of the values
        from all contributing Primary Catalog measurements. In all panels, the color
        maps show the data point density, while black points indicate the mean
        bias in each bin with error bars representing the uncertainty of the
        mean. }
        \label{fig:blend1}
    \end{figure*}

    Figure~\ref{fig:blend1} shows the fractional flux residual as a function of these
    three blending parameters. The normalized neighbor distance (left panels)
    shows only a small variation, with the bias increasing from $\sim$1\% to $\sim$3\%
    as the metric changes from $\sim30$ to $\sim3$. This suggests that our
    forced photometry pipeline is robust against close neighbors, provided they are
    not overwhelmingly bright.

    Among these three parameters, the neighbor-to-target flux ratio (middle panels)
    shows the strongest dependence. While the mean bias remains small, the
    scatter in the fractional flux residual increases dramatically when the
    combined flux of neighbors becomes comparable to the flux of the targets. The
    mean bias itself grows systematically with the flux ratio, from $\sim$1\% at
    a ratio of 1, to $\sim$2\% at a ratio of 10, and up to $\sim$10\% when
    neighbors are 100 times brighter than the target. This result implies that special
    care, such as comparison with broadband photometry, is required when analyzing
    faint sources near bright ones in cluster environments, a common scenario
    for gravitationally lensed background galaxies.

    Finally, the number of neighbors (right panels) also shows a positive bias,
    which reaches $\sim$5\% with $\sim$10 neighbors and $\sim$30\% in the most crowded
    cases with $\sim$30 neighbors. But this last trend should be interpreted with
    caution. Fewer than 300 distinct sources reach $N_{\rm neigh}\gtrsim15$, and
    only a few dozen reach $\sim$30. Each contributes many repeated observations,
    so this regime samples a small number of galaxies rather than a diverse range
    of crowding configurations. The small scatter of the binned means therefore
    understates the true uncertainty. A source's bias is set by the number and brightness
    of its neighbors together, so neighbor count alone cannot be isolated here. The
    bias is more robustly traced by the total neighbor flux (middle panels),
    with which $N_{\rm neigh}$ is correlated.

    \begin{figure}
        \centering
        \includegraphics[width=0.48\textwidth]{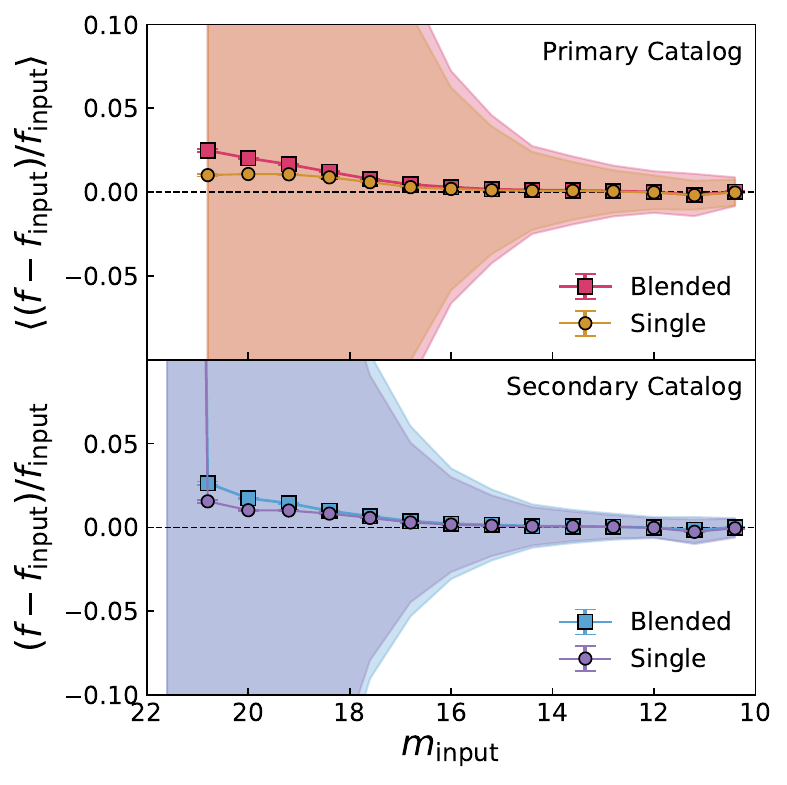}
        \caption{ Comparison of photometric bias of the fractional flux residual,
        for blended and isolated (single) sources. The blended case includes all
        neighboring galaxies within $5r_{e}$ in the image simulation, while the
        single case uses only the target galaxy, removing blending effects. Each
        point shows the mean flux bias, with error bars indicating the uncertainty
        on the mean. Shaded regions represent the $1\sigma$ scatter of
        individual measurements. The results show that blending introduces only a
        small bias for faint sources. }
        \label{fig:blend2}
    \end{figure}

    \begin{figure}
        \centering
        \includegraphics[width=0.48\textwidth]{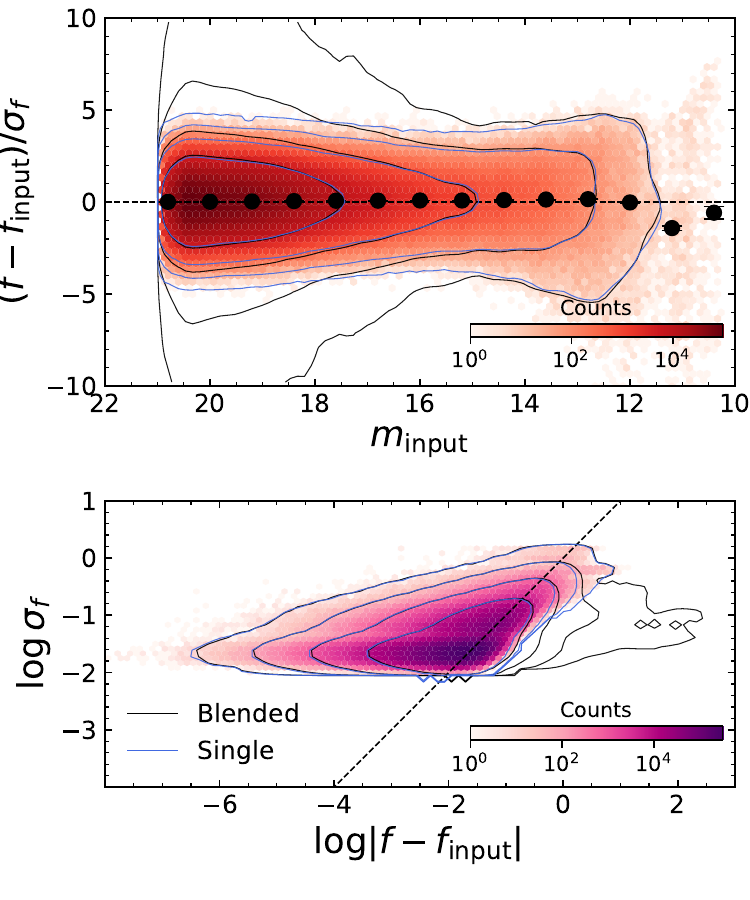}
        \caption{ Photometric performance for an idealized single-source simulation,
        isolating the pipeline performance from the effects of source blending. The
        filled color maps show the single-source run, using the same two
        diagnostics as the blended case in Figure~\ref{fig:photbias_extpsf} and
        Figure~\ref{fig:photerr}: the normalized flux residual versus input magnitude
        for extended sources (top), and the reported measurement error versus
        the absolute flux residual for all Primary sources (bottom). Black points
        in the top panel mark the mean residuals in magnitude bins. To allow a
        direct comparison, number-density contours of the blended (black) and single-source
        (blue) runs are overlaid in both panels; these correspond to the top-left
        panel of Figure~\ref{fig:photbias_extpsf} (top) and the top panel of
        Figure~\ref{fig:photerr} (bottom). }
        \label{fig:photbias_exterr_single}
    \end{figure}

    To isolate the effect of blending, we conducted a separate "single-source" simulation
    where each galaxy was rendered without any neighbors. Figure~\ref{fig:blend2}
    compares the mean bias from the blended and single-source simulations, showing
    that blending introduces only a small \textit{additional} bias (typically $\sim
    1\%$) for faint sources. A more dramatic effect is seen in the outlier population.
    In Figure~\ref{fig:photbias_exterr_single} we overlay the blended (black)
    and single-source (blue) number-density contours to compare the two runs. The
    top panel shows that removing blending suppresses the extended tails at $|(f
    - f_{\mathrm{input}})/\sigma_{f}| > 5$: the fraction of extended sources in
    these wings falls from $\sim$0.4\% in the blended run to $\lesssim$0.01\% in
    the single-source run. The bottom panel shows the same effect as a tightening
    of the distribution toward the one-to-one relation, with far fewer points
    extending to large $|f - f_{\mathrm{input}}|$ below the line, where the errors
    are underestimated. These results indicate that source blending is the
    primary driver of catastrophic photometric outliers in dense fields.

    \begin{figure}
        \centering
        \includegraphics[width=0.48\textwidth]{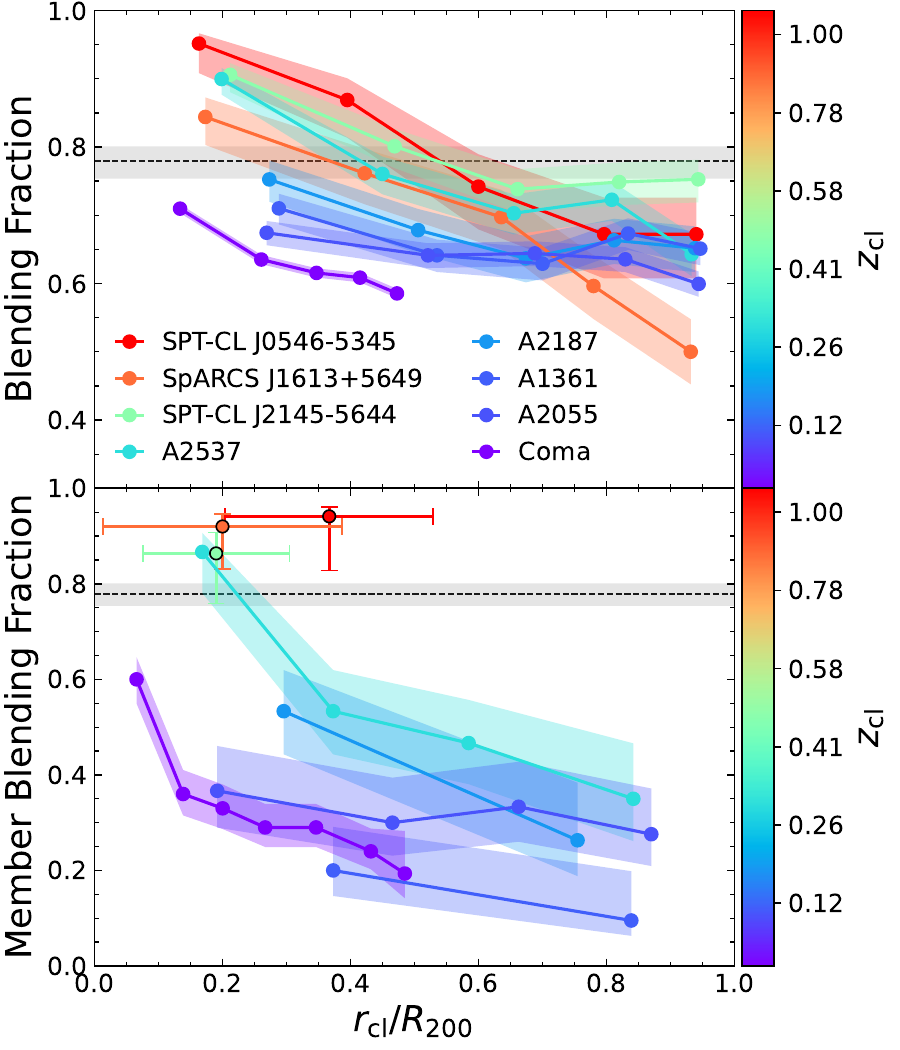}
        \caption{ Blending fraction as a function of normalized cluster-centric radius
        ($r_{\rm cl}/R_{200}$), measured from the Secondary Catalog. The blending
        fraction is defined as the proportion of sources for which the mean neighbor-to-target
        flux ratio across the 102 channels,
        $\langle \sum{f_{\rm neigh}/f}\rangle$, exceeds unity. The top panel
        shows this fraction for all sources within the cluster fields, while the
        bottom panel shows the same relation calculated exclusively for spectroscopically
        selected cluster members. For the radial profiles, data are grouped into
        bins containing an equal number of sources; this number is adjusted for
        each cluster field to ensure appropriate sampling density given the
        varying sample sizes. In the bottom panel, clusters with fewer than 30
        confirmed members are represented by a single point indicating the mean
        radius and blending fraction, with horizontal error bars showing the standard
        deviation of the radial distribution. In both panels, different colored lines
        represent different clusters, color-coded by the cluster redshift ($z_{\rm
        cl}$). The horizontal dashed line indicates the blending fraction in the
        COSMOS field for reference. }
        \label{fig:cluster_blending}
    \end{figure}

    Finally, we examine the prevalence of blending as a function of cluster-centric
    radius in Figure~\ref{fig:cluster_blending}. As expected, the blending
    fraction is highest in the cluster cores and decreases with radius for both
    the full sample (top panel) and the confirmed member galaxies (bottom panel).
    Beyond this radial trend, the blending fraction increases significantly with
    cluster redshift. Notably, for low-to-intermediate redshift clusters, the blending
    fraction in the central regions is often \textit{lower} than the field
    average.

    This counter-intuitive result is driven by the brightness contrast between
    cluster members and the background population. In low-redshift clusters,
    detected members are typically massive and bright, significantly outshining surrounding
    sources; this keeps the neighbor-to-target flux ratio low (often $<1$), resulting
    in a low blending fraction by our definition. Conversely, at higher redshifts,
    member fluxes are fainter and more comparable to the projected field
    population, driving the blending fraction up towards the field average. In
    addition, rather than fully converging to the field value (dashed line), the
    blending fractions flatten off at a level offset from it. Their systematic dependence
    on cluster-centric radius and redshift nonetheless confirms that this is a local
    environmental effect rather than a global property of the sample.
    Additionally, the fact that the member-only blending fraction (bottom panel)
    closely mirrors the full-sample trends suggests that these statistics are robust
    and not an artifact of member selection. We note, however, that these trends
    are likely modulated by the individual physical properties of the clusters (e.g.,
    mass, dynamical state) as well as the variance of the foreground and background
    distributions along the line of sight. Despite these complexities, the strong
    dependence of blending on both radius and redshift indicates that accurate spectrophotometry
    in cluster fields requires a pipeline capable of adapting to these diverse crowding
    conditions.

    \subsection{Depth Tests}
    \label{sec:depth} We define the 5$\sigma$ survey depth as the magnitude at
    which the photometric error, $\sigma_{m}$, reaches 0.217, corresponding to a
    signal-to-noise ratio of five. We compute this depth empirically by fitting
    the relationship between the input magnitude and the measured photometric
    error from our simulations.

    \begin{figure*}
        \centering
        \includegraphics[width=\textwidth]{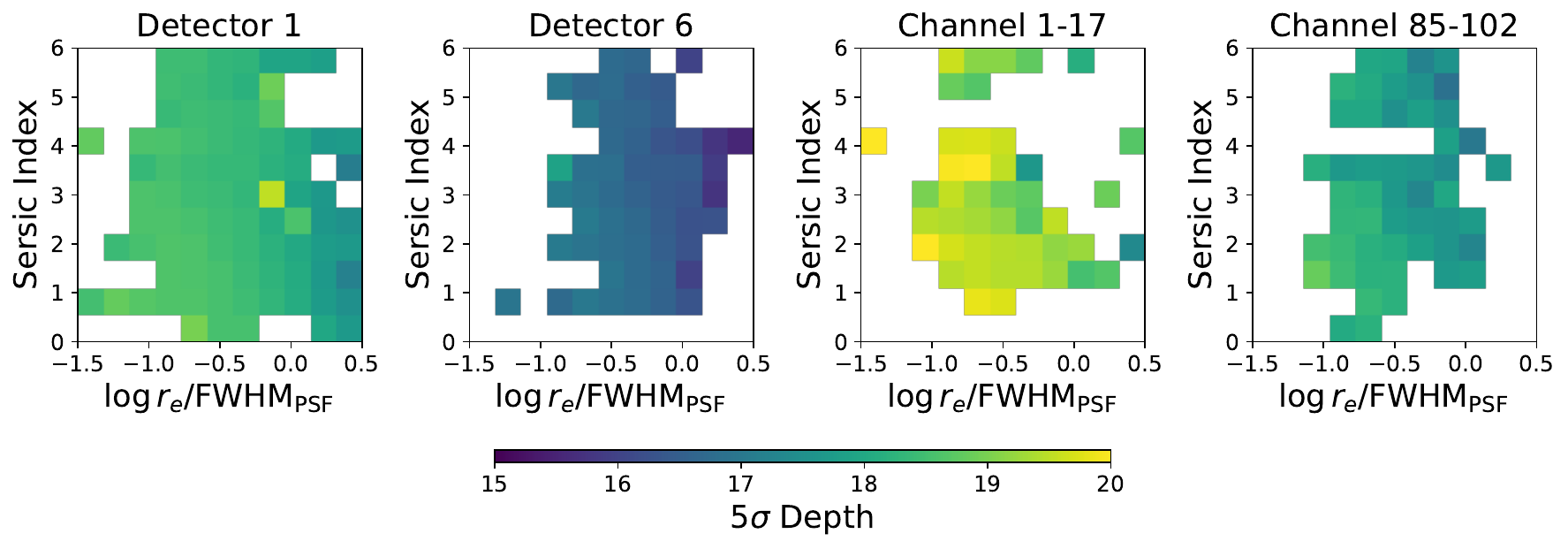}
        \caption{$5\sigma$ depth as a function of source size, normalized by the
        PSF FWHM, and S\'ersic index. Each panel shows results for different detector
        or channel subsets: (left to right) Detector~1 (Primary Catalog; $0.75$--$1
        .09\,\mu$m), Detector~6 (Primary Catalog; $4.42$--$5.00\,\mu$m),
        Secondary Catalog channels~1--17 ($0.75$--$1.12\,\mu$m), and channels~85--102
        ($4.37$--$5.01\,\mu$m). Depth is largely insensitive to S\'ersic index, but
        becomes shallower for larger sources. White regions indicate bins with
        insufficient statistics, where fewer than 10 sources satisfy $0.15 < \sigma
        _{m}< 0.25$. }
        \label{fig:depthmap}
    \end{figure*}

    Figure~\ref{fig:depthmap} shows this 5$\sigma$ depth as a 2D map in the
    plane of galaxy morphology, defined by the S\'ersic index and the effective radius
    normalized by the PSF Full Width at Half Maximum (FWHM; $r_{e}/\text{FWHM}_{\text{PSF}}$).
    The depth is largely insensitive to the S\'ersic index but becomes significantly
    shallower for larger, more resolved sources (higher $r_{e}/\text{FWHM}_{\text{PSF}}$),
    as their flux is spread over a larger number of pixels. This trend holds for
    both the Primary Catalog and the binned Secondary Catalog.

    \begin{figure}
        \centering
        \includegraphics[width=0.48\textwidth]{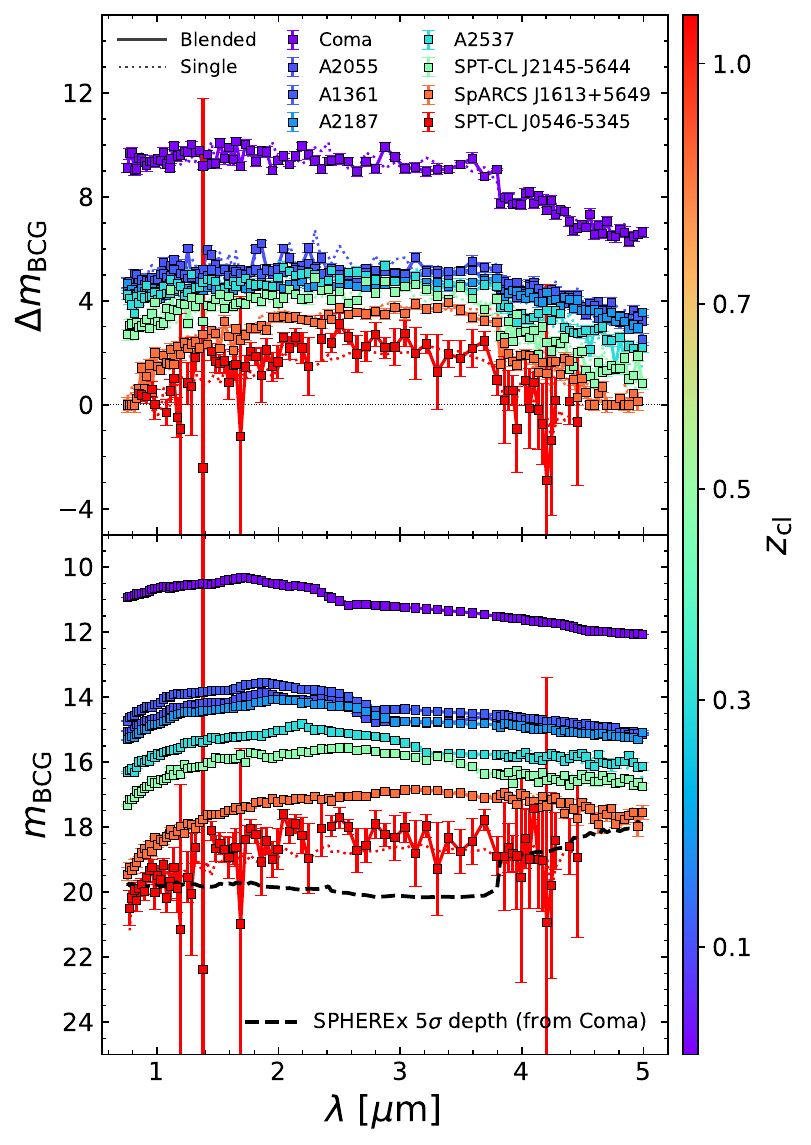}
        \caption{ Detection limit of member galaxies in the Secondary Catalog as
        a function of wavelength, across the 102 channels. Upper panel: the
        magnitude difference $\Delta m_{\rm BCG}$ between the BCG and the
        faintest member galaxy detected with $\mathrm{S/N}>5$ in each channel; solid
        lines with square markers show the main blended-field simulation and dotted
        lines the single-source counterpart (Section~\ref{sec:photoz_blending}).
        Lower panel: the observed BCG magnitude, with the dashed line showing
        the 5$\sigma$ point-source depth as a function of wavelength, measured
        in the Coma-field simulation ($K_{s}\approx 20$). The color of each point
        indicates the cluster redshift. }
        \label{fig:depth_dm}
    \end{figure}

    To assess the depth in a scientifically relevant context, we measure the
    magnitude difference ($\Delta m_{\rm BCG}$) between the BCG and the faintest
    spectroscopically confirmed member detected with S/N$>$5 in each channel. The
    upper panel of Figure~\ref{fig:depth_dm} shows this limit as a function of wavelength
    for each cluster, while the lower panel shows the observed BCG magnitude
    with the per-channel 5$\sigma$ depth (dashed; $K_{s}\approx 20$ at $2.2\,\mu$m).
    Because the faintest detectable member sits at this near-constant depth, $\Delta
    m_{\rm BCG}$ is mostly set by the BCG brightness. For the nearby Coma
    cluster, where the BCG is bright ($K_{s}\approx 1 0.6$), SPHEREx detects
    members 7--9 magnitudes fainter than the BCG; at $z \sim 1$ the BCG has
    faded to $K_{s}\approx 18$ (SPT-CL J0546$-$5345), leaving only 1--2
    magnitudes. Source blending is sub-dominant, changing $\Delta m_{\rm BCG}$
    by less than 0.5~mag in every cluster (single-source run, dotted curves).
    The contraction is also not a spectroscopic-incompleteness artifact: at $z>0.
    7$, the faintest detected member already reaches the depth. The few negative
    $\Delta m_{\rm BCG}$ values, confined to SPT-CL J0546$-$5345, correspond to confirmed
    members rather than interlopers, arising in low-S/N channels where noise
    drives the floor-level BCG flux below another member's.

    It is useful to compare this detection depth with the confusion limit set by
    source crowding. From our injected source counts and the simulated per-channel
    PSF, a notional confusion limit \citep{condon1974} is $m\approx20$--$2 1.5$
    over $1$--$5\,\mu$m for a typical high-latitude field, brightening toward longer
    wavelengths as the beam widens. It is fainter than the $5\sigma$ depth at
    every wavelength, so a typical field is sensitivity- rather than confusion-limited,
    consistent with the zodiacal-background-limited sensitivity reported by \citet{crill25}.
    Within the densest cluster cores, however, the local source density rises by
    a factor of a few and brightens the confusion onset to a level comparable to
    the detection depth; this is the regime in which the high-resolution
    positional priors adopted here (Section~\ref{sec:mockimg}) are most
    important for reliable deblending.

    \section{Photometric Redshift Performance}
    \label{sec:photoz}

    The final step in our pipeline is to derive photometric redshifts for each source
    using the binned spectrophotometry from the Secondary Catalog. We again
    employ the \eazypy{} software, following a procedure similar to the initial
    SED modeling described in Section~\ref{sec:sed} but with parameters
    optimized for redshift estimation from SPHEREx data. We fit the 102-channel SPHEREx
    spectrophotometry of each source using the same combined library of galaxy templates
    (we only perform photometric redshift estimation for galaxies). Channels
    without valid measurements due to the sensitivity pre-selection (Section~\ref{sec:mockimg})
    are omitted from the fit and do not contribute to the $\chi^{2}$. For this final
    fit, the systematic error floor was reduced to 1\%, and the redshift grid
    was refined to span from $z=0.001$ to $3$ with a finer step size of
    $\delta z = 0.001$. The resulting redshift probability distribution, $P(z)$,
    for each source is then used to determine its final photometric redshift and
    associated uncertainty, where we adopt the maximum likelihood redshift as
    the best estimate. We define 3$\sigma$ catastrophic outliers ($\eta_{3\hat{\sigma}}$)
    as sources where the input redshift, $z_{\rm input}$, falls outside the 99.7\%
    confidence interval of the $P(z)$ derived by \eazypy.

    \subsection{Overall Performance and Sample Selection}

    For the full sample of 43,016 galaxies, the photometric redshift performance
    is characterized by a bias of 0.016, a scatter of
    $\sigma_{\text{NMAD}}= 0.46 1$, and high outlier fractions of $\eta_{0.15}= 5
    2.8\%$ and $\eta_{3\hat{\sigma}}= 36.1\%$. However, these statistics are dominated
    by the large number of faint sources for which the spectrophotometry has low
    signal-to-noise (S/N). To assess the performance for a more scientifically
    useful sample, we apply selections based on brightness and data quality,
    which significantly improves accuracy and reduces the rate of catastrophic
    outliers.

    First, we select sources based on their synthetic $K_{s}$-band magnitude,
    derived from the Secondary Catalog spectrophotometry \citep{crill25}. For a
    bright sample with $K_{s}< 19$ (6,905 sources, 16\% of the total), the performance
    improves dramatically to a bias of $-0.001$, $\sigma_{\text{NMAD}}= 0.016$, $\eta
    _{0.15}= 4.6\%$ and $\eta_{3\hat{\sigma}}= 27.5\%$. A stricter cut of
    $K_{s}< 17.5$ (1,618 sources, 4\%) further improves the scatter to
    $\sigma_{\text{NMAD}}= 0.004$ with a bias of $-0.004$.

    \begin{figure*}
        \centering
        \includegraphics[width=\textwidth]{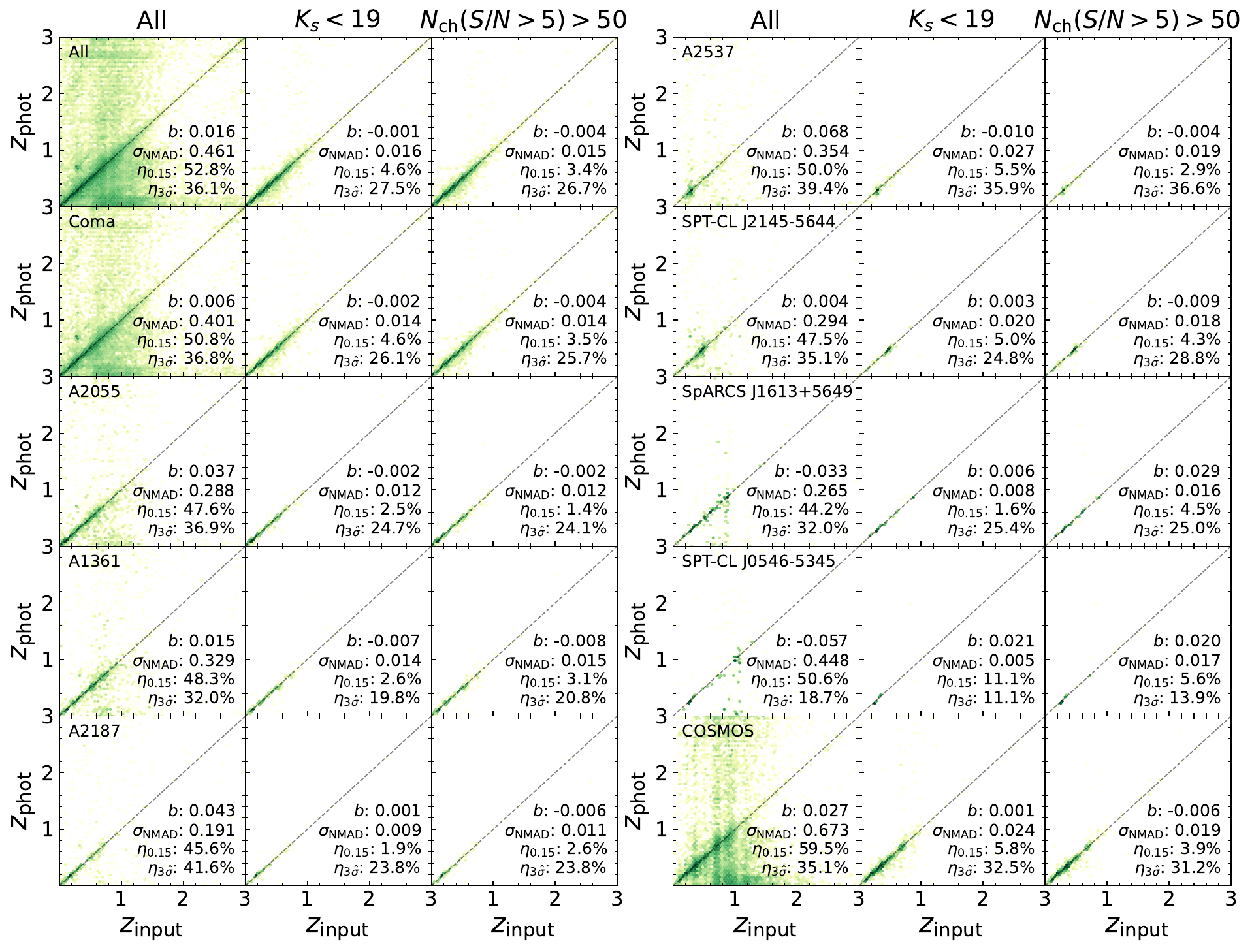}
        \caption{ Photometric redshift performance for individual cluster fields
        and the COSMOS field. Each block displays five fields, with the top left
        row showing results for all fields combined. Subsequent rows corresponds
        to individual cluster fields and the COSMOS field (right bottom). Within
        each block, the three panels compare $z_{\mathrm{input}}$ and
        $z_{\mathrm{phot}}$ for different galaxy selections: all galaxies (left),
        galaxies with synthetic $K_{s}< 19$, and galaxies detected in more than 50
        channels with $S/N > 5$ (right). For each panel, the bias $b$, scatter
        $\sigma_{\mathrm{NMAD}}$, and outlier fractions $\eta_{0.15}$ and $\eta_{3\hat{\sigma}}$
        are reported.}
        \label{fig:photoz_all}
    \end{figure*}

    Second, we select galaxies based on the number of spectral channels in which
    they are detected with S/N$>$5. A cut requiring more than 50 such channels
    yields a sample with performance ($\sigma_{\text{NMAD}}= 0.015$) comparable
    to the $K_{s}< 19$ cut. For the highest quality sample, with S/N$>$5 in all
    102 channels (372 sources, 1\%), the scatter is as low as $\sigma_{\text{NMAD}}
    = 0.002$. These tests show that applying simple cuts in either magnitude or
    data quality can effectively isolate a sample with high-fidelity photometric
    redshifts. Table~\ref{tab:photoz_perf_compactN} and Figure~\ref{fig:photoz_all}
    show the performance after applying these cuts for each of our individual cluster
    fields, confirming that these trends are robust and not dependent on a
    single field.

    \begin{table*}
        \centering
        \caption{Photometric redshift performance per field. Rows are split into
        $N$, $b$, $\sigma_{\rm NMAD}$, and $\eta$ (0.15 / $3\hat{\sigma}$). Columns
        list selections including the full sample.}
        \begin{tabular}{l lccccc}
            \toprule Field                                               & Metric                           & All galaxies    & $K_{s}<19$      & $N_{\rm ch}(S/N>5)>50$ & $K_{s}<17.5$   & $N_{\rm ch}(S/N>5)= 102$ \\
            \multirow{4}{*}{All}                                         & $N$                              & 43016           & 6905            & 6605                   & 1618           & 372                      \\
                                                                         & $b$                              & 0.016           & -0.001          & -0.004                 & -0.004         & -0.001                   \\
                                                                         & $\sigma_{\rm NMAD}$              & 0.4606          & 0.0164          & 0.0149                 & 0.0042         & 0.0016                   \\
                                                                         & $\eta\ (0.15\ /\ 3\hat{\sigma})$ & 52.8\% / 36.1\% & 4.6\% / 27.5\%  & 3.4\% / 26.7\%         & 1.1\% / 14.3\% & 0.3\% / 7.8\%            \\
            \midrule \multirow{4}{*}{Coma}                               & $N$                              & 22702           & 3777            & 3803                   & 1024           & 286                      \\
                                                                         & $b$                              & 0.006           & -0.002          & -0.004                 & -0.005         & -0.001                   \\
                                                                         & $\sigma_{\rm NMAD}$              & 0.4008          & 0.0137          & 0.0136                 & 0.0037         & 0.0015                   \\
                                                                         & $\eta\ (0.15\ /\ 3\hat{\sigma})$ & 50.8\% / 36.8\% & 4.6\% / 26.1\%  & 3.5\% / 25.7\%         & 1.7\% / 13.7\% & 0.3\% / 8.0\%            \\
            \midrule \multirow{4}{*}{Abell 2055}                         & $N$                              & 2553            & 511             & 498                    & 153            & 34                       \\
                                                                         & $b$                              & 0.037           & -0.002          & -0.002                 & -0.005         & -0.001                   \\
                                                                         & $\sigma_{\rm NMAD}$              & 0.2877          & 0.0123          & 0.0120                 & 0.0037         & 0.0015                   \\
                                                                         & $\eta\ (0.15\ /\ 3\hat{\sigma})$ & 47.6\% / 36.9\% & 2.5\% / 24.7\%  & 1.4\% / 24.1\%         & 0.7\% / 14.4\% & 0.0\% / 5.9\%            \\
            \midrule \multirow{4}{*}{Abell 1361}                         & $N$                              & 1831            & 303             & 351                    & 61             & 13                       \\
                                                                         & $b$                              & 0.015           & -0.007          & -0.008                 & -0.002         & -0.001                   \\
                                                                         & $\sigma_{\rm NMAD}$              & 0.3286          & 0.0138          & 0.0149                 & 0.0027         & 0.0008                   \\
                                                                         & $\eta\ (0.15\ /\ 3\hat{\sigma})$ & 48.3\% / 32.0\% & 2.6\% / 19.8\%  & 3.1\% / 20.8\%         & 0.0\% / 11.5\% & 0.0\% / 7.7\%            \\
            \midrule \multirow{4}{*}{Abell 2187}                         & $N$                              & 776             & 160             & 189                    & 47             & 9                        \\
                                                                         & $b$                              & 0.043           & 0.001           & -0.006                 & 0.000          & 0.000                    \\
                                                                         & $\sigma_{\rm NMAD}$              & 0.1913          & 0.0094          & 0.0107                 & 0.0036         & 0.0017                   \\
                                                                         & $\eta\ (0.15\ /\ 3\hat{\sigma})$ & 45.6\% / 41.6\% & 1.9\% / 23.8\%  & 2.6\% / 23.8\%         & 0.0\% / 10.6\% & 0.0\% / 0.0\%            \\
            \midrule \multirow{4}{*}{Abell 2537}                         & $N$                              & 1045            & 256             & 175                    & 43             & 3                        \\
                                                                         & $b$                              & 0.068           & -0.010          & -0.004                 & -0.001         & 0.002                    \\
                                                                         & $\sigma_{\rm NMAD}$              & 0.3537          & 0.0271          & 0.0191                 & 0.0083         & 0.0038                   \\
                                                                         & $\eta\ (0.15\ /\ 3\hat{\sigma})$ & 50.0\% / 39.4\% & 5.5\% / 35.9\%  & 2.9\% / 36.6\%         & 0.0\% / 18.6\% & 0.0\% / 0.0\%            \\
            \midrule \multirow{4}{*}{\shortstack[l]{SPT-CL\\J2145-5644}} & $N$                              & 795             & 141             & 139                    & 20             & 5                        \\
                                                                         & $b$                              & 0.004           & 0.003           & -0.009                 & -0.000         & 0.002                    \\
                                                                         & $\sigma_{\rm NMAD}$              & 0.2938          & 0.0200          & 0.0184                 & 0.0060         & 0.0047                   \\
                                                                         & $\eta\ (0.15\ /\ 3\hat{\sigma})$ & 47.5\% / 35.1\% & 5.0\% / 24.8\%  & 4.3\% / 28.8\%         & 0.0\% / 15.0\% & 0.0\% / 0.0\%            \\
            \midrule \multirow{4}{*}{\shortstack[l]{SpARCS\\J1613+5649}} & $N$                              & 437             & 63              & 88                     & 16             & 6                        \\
                                                                         & $b$                              & -0.033          & 0.006           & 0.029                  & -0.001         & -0.002                   \\
                                                                         & $\sigma_{\rm NMAD}$              & 0.2652          & 0.0083          & 0.0156                 & 0.0038         & 0.0017                   \\
                                                                         & $\eta\ (0.15\ /\ 3\hat{\sigma})$ & 44.2\% / 32.0\% & 1.6\% / 25.4\%  & 4.5\% / 25.0\%         & 0.0\% / 18.8\% & 0.0\% / 16.7\%           \\
            \midrule \multirow{4}{*}{\shortstack[l]{SPT-CL\\J0546-5345}} & $N$                              & 235             & 18              & 36                     & 3              & 1                        \\
                                                                         & $b$                              & -0.057          & 0.021           & 0.020                  & -0.002         & -0.004                   \\
                                                                         & $\sigma_{\rm NMAD}$              & 0.4484          & 0.0054          & 0.0169                 & 0.0012         & 0.0000                   \\
                                                                         & $\eta\ (0.15\ /\ 3\hat{\sigma})$ & 50.6\% / 18.7\% & 11.1\% / 11.1\% & 5.6\% / 13.9\%         & 0.0\% / 33.3\% & 0.0\% / 100.0\%          \\
            \midrule \multirow{4}{*}{COSMOS}                             & $N$                              & 12642           & 1676            & 1326                   & 251            & 15                       \\
                                                                         & $b$                              & 0.027           & 0.001           & -0.006                 & -0.001         & 0.000                    \\
                                                                         & $\sigma_{\rm NMAD}$              & 0.6728          & 0.0244          & 0.0192                 & 0.0065         & 0.0022                   \\
                                                                         & $\eta\ (0.15\ /\ 3\hat{\sigma})$ & 59.5\% / 35.1\% & 5.8\% / 32.5\%  & 3.9\% / 31.2\%         & 0.0\% / 16.7\% & 0.0\% / 6.7\%            \\
            \bottomrule
        \end{tabular}
        \label{tab:photoz_perf_compactN}
    \end{table*}

    \subsection{Performance for Cluster Members}
    \label{sec:photoz_member}

    \begin{table*}
        \centering
        \caption{Photometric redshift performance for spectroscopically
        confirmed cluster members. Same format as Table~\ref{tab:photoz_perf_compactN},
        but restricted to member galaxies only.}
        \begin{tabular}{l lccccc}
            \toprule Field                                               & Metric                           & All members     & $K_{s}<19$     & $N_{\rm ch}(S/N>5)>50$ & $K_{s}<17.5$   & $N_{\rm ch}(S/N>5)= 102$ \\
            \multirow{4}{*}{All}                                         & $N$                              & 1032            & 901            & 881                    & 519            & 208                      \\
                                                                         & $b$                              & 0.0169          & 0.0028         & 0.0011                 & -0.0003        & -0.0004                  \\
                                                                         & $\sigma_{\rm NMAD}$              & 0.0064          & 0.0051         & 0.0048                 & 0.0024         & 0.0011                   \\
                                                                         & $\eta\ (0.15\ /\ 3\hat{\sigma})$ & 5.2\% / 19.7\%  & 1.6\% / 17.1\% & 1.0\% / 16.2\%         & 0.2\% / 7.5\%  & 0.0\% / 4.8\%            \\
            \midrule \multirow{4}{*}{Coma}                               & $N$                              & 632             & 577            & 576                    & 365            & 174                      \\
                                                                         & $b$                              & 0.0269          & 0.0062         & 0.0026                 & -0.0001        & -0.0004                  \\
                                                                         & $\sigma_{\rm NMAD}$              & 0.0043          & 0.0036         & 0.0036                 & 0.0021         & 0.0011                   \\
                                                                         & $\eta\ (0.15\ /\ 3\hat{\sigma})$ & 3.6\% / 17.1\%  & 1.4\% / 14.0\% & 0.7\% / 13.2\%         & 0.3\% / 5.2\%  & 0.0\% / 4.0\%            \\
            \midrule \multirow{4}{*}{Abell 2055}                         & $N$                              & 120             & 115            & 116                    & 77             & 22                       \\
                                                                         & $b$                              & 0.0212          & -0.0003        & -0.0003                & -0.0001        & -0.0006                  \\
                                                                         & $\sigma_{\rm NMAD}$              & 0.0053          & 0.0052         & 0.0052                 & 0.0029         & 0.0017                   \\
                                                                         & $\eta\ (0.15\ /\ 3\hat{\sigma})$ & 1.7\% / 15.8\%  & 0.0\% / 16.5\% & 0.0\% / 16.4\%         & 0.0\% / 13.0\% & 0.0\% / 9.1\%            \\
            \midrule \multirow{4}{*}{Abell 1361}                         & $N$                              & 52              & 52             & 52                     & 29             & 5                        \\
                                                                         & $b$                              & 0.0056          & 0.0056         & 0.0056                 & -0.0034        & -0.0010                  \\
                                                                         & $\sigma_{\rm NMAD}$              & 0.0070          & 0.0070         & 0.0070                 & 0.0023         & 0.0001                   \\
                                                                         & $\eta\ (0.15\ /\ 3\hat{\sigma})$ & 1.9\% / 13.5\%  & 1.9\% / 13.5\% & 1.9\% / 13.5\%         & 0.0\% / 13.8\% & 0.0\% / 20.0\%           \\
            \midrule \multirow{4}{*}{Abell 2187}                         & $N$                              & 50              & 50             & 50                     & 26             & 4                        \\
                                                                         & $b$                              & 0.0007          & 0.0007         & 0.0007                 & 0.0006         & -0.0001                  \\
                                                                         & $\sigma_{\rm NMAD}$              & 0.0065          & 0.0065         & 0.0065                 & 0.0034         & 0.0003                   \\
                                                                         & $\eta\ (0.15\ /\ 3\hat{\sigma})$ & 0.0\% / 18.0\%  & 0.0\% / 18.0\% & 0.0\% / 18.0\%         & 0.0\% / 11.5\% & 0.0\% / 0.0\%            \\
            \midrule \multirow{4}{*}{Abell 2537}                         & $N$                              & 111             & 86             & 69                     & 19             & 2                        \\
                                                                         & $b$                              & -0.0040         & -0.0120        & -0.0081                & -0.0014        & 0.0031                   \\
                                                                         & $\sigma_{\rm NMAD}$              & 0.0343          & 0.0290         & 0.0215                 & 0.0073         & 0.0019                   \\
                                                                         & $\eta\ (0.15\ /\ 3\hat{\sigma})$ & 9.0\% / 39.6\%  & 4.7\% / 40.7\% & 4.3\% / 42.0\%         & 0.0\% / 15.8\% & 0.0\% / 0.0\%            \\
            \midrule \multirow{4}{*}{\shortstack[l]{SPT-CL\\J2145-5644}} & $N$                              & 23              & 16             & 14                     & 2              & 1                        \\
                                                                         & $b$                              & -0.0096         & -0.0155        & -0.0189                & 0.0010         & 0.0026                   \\
                                                                         & $\sigma_{\rm NMAD}$              & 0.0217          & 0.0158         & 0.0146                 & 0.0036         & 0.0000                   \\
                                                                         & $\eta\ (0.15\ /\ 3\hat{\sigma})$ & 4.3\% / 17.4\%  & 6.2\% / 12.5\% & 7.1\% / 14.3\%         & 0.0\% / 0.0\%  & 0.0\% / 0.0\%            \\
            \midrule \multirow{4}{*}{\shortstack[l]{SpARCS\\J1613+5649}} & $N$                              & 26              & 4              & 4                      & 1              & 0                        \\
                                                                         & $b$                              & -0.0549         & -0.0021        & -0.0021                & -0.0007        & --                       \\
                                                                         & $\sigma_{\rm NMAD}$              & 0.3141          & 0.0164         & 0.0164                 & 0.0000         & --                       \\
                                                                         & $\eta\ (0.15\ /\ 3\hat{\sigma})$ & 34.6\% / 42.3\% & 0.0\% / 25.0\% & 0.0\% / 25.0\%         & 0.0\% / 0.0\%  & --                       \\
            \midrule \multirow{4}{*}{\shortstack[l]{SPT-CL\\J0546-5345}} & $N$                              & 18              & 1              & 0                      & 0              & 0                        \\
                                                                         & $b$                              & -0.0177         & -0.0533        & --                     & --             & --                       \\
                                                                         & $\sigma_{\rm NMAD}$              & 0.2778          & 0.0000         & --                     & --             & --                       \\
                                                                         & $\eta\ (0.15\ /\ 3\hat{\sigma})$ & 44.4\% / 5.6\%  & 0.0\% / 0.0\%  & --                     & --             & --                       \\
            \bottomrule
        \end{tabular}
        \label{tab:photoz_perf_members}
    \end{table*}

    \begin{figure*}
        \centering
        \includegraphics[width=\textwidth]{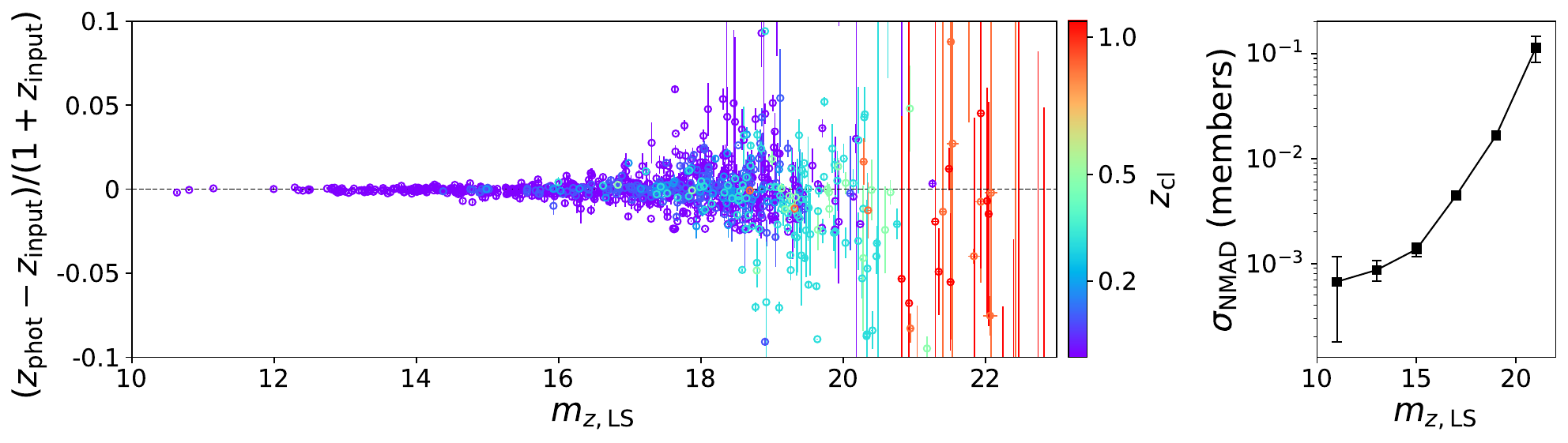}
        \caption{Scatter in member-galaxy photometric redshift with respect to
        Legacy Survey $z$-band magnitude, with cluster redshift color-coded (left)
        and $\sigma_{\rm NMAD}$ shown for binned members (right). }
        \label{fig:photoz_mem}
    \end{figure*}

    To specifically evaluate the photometric redshift performance for cluster
    galaxies, we analyze the sample of member galaxies with spectroscopic redshifts.
    Table~\ref{tab:photoz_perf_members} summarizes the performance statistics
    for each cluster, following the same format and selection criteria as Table~\ref{tab:photoz_perf_compactN}.
    For the combined member sample with $K_{s}< 19$ ($N_{\rm ch}(\mathrm{S/N}>5)
    > 50$), we find $\sigma_{\rm NMAD}\approx 0.005$ ($0.005$) with $\eta_{0.15}\approx
    1 .6\%$ ($1.0\%$), significantly better than the full field-galaxy sample
    due to the brighter magnitudes of confirmed members. Figure~\ref{fig:photoz_mem}
    shows the photometric redshift scatter for these members as a function of their
    Legacy Survey $z$-band magnitude. The performance is strongly dependent on
    galaxy magnitude rather than cluster redshift. For the brightest members,
    such as those in the Coma cluster with $m_{z, \text{LS}}\approx 11$, the scatter
    is extremely low, $\sigma_{\text{NMAD}}\approx 0.0007$. The scatter remains at
    $\sigma_{\text{NMAD}}\le 0.003$ for members as faint as $m_{z, \text{LS}}= 15$,
    before increasing to 0.01 at 19th magnitude and 0.1 at 21st magnitude.

    However, this strong magnitude dependence imposes an effective redshift limit
    on reliable cluster member characterization, considering the survey depth
    analysis in Section~\ref{sec:depth}. As the cluster redshift increases, the
    apparent magnitude of the member population fades, shifting the bulk of galaxies
    into fainter bins where photometric uncertainties rise. Consequently, for clusters
    at $z \sim 1$, photometric redshifts of $\sigma_{\text{NMAD}}\sim 0. 01$ are
    attainable only for the brightest members comparable to the BCGs. For the general
    member population in these distant systems, the utility of \spherex{} photometric
    redshifts will be limited by the fainter magnitudes, unlike in local
    clusters where a broad range of the luminosity function is accessible with
    high precision.

    \begin{figure*}
        \centering
        \includegraphics[width=\textwidth]{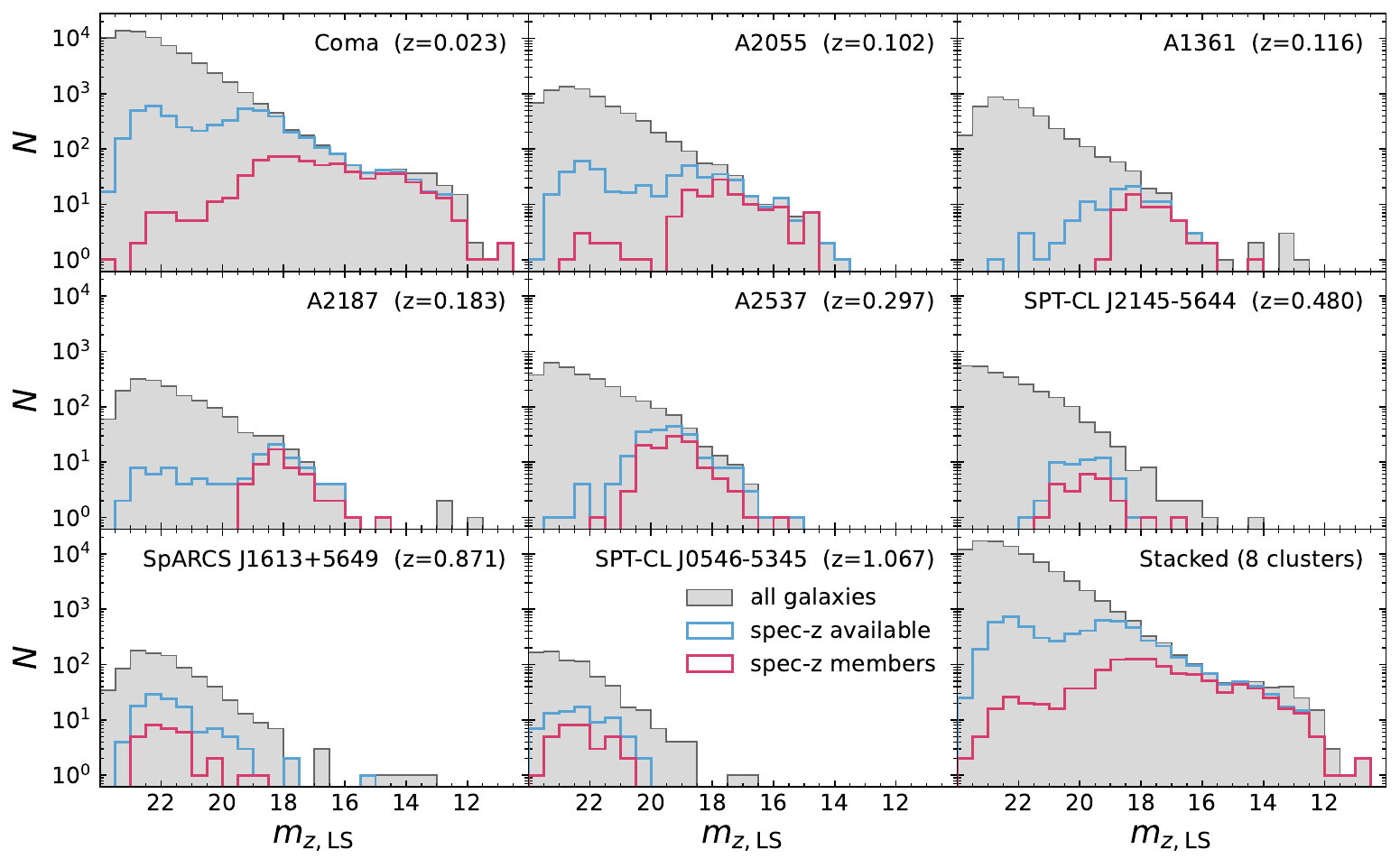}
        \caption{Apparent-magnitude coverage of the spectroscopic member sample.
        For each cluster field, and stacked over all eight clusters (bottom right),
        the gray filled histogram shows the LS DR10 $z$-band magnitude distribution
        of all galaxies in the field. The blue and red histograms show the
        subsets with an available spectroscopic redshift and with confirmed
        cluster membership ($|c\,\Delta z| < 2000\ \mathrm{km\,s^{-1}}$), respectively.
        Confirmed members occupy the bright end of the distribution, illustrating
        the brightness bias of the sample used for the photometric-redshift assessment
        in Section~\ref{sec:photoz_member}.}
        \label{fig:specz_coverage}
    \end{figure*}

    Because this assessment rests on spectroscopically confirmed members (Section~\ref{sec:member}),
    its results apply most directly to the bright end of the cluster population.
    Figure~\ref{fig:specz_coverage} shows that the confirmed members occupy the bright
    tail of the field galaxy distribution, particularly at low redshift, and grow
    incomplete toward faint magnitudes. The photometric-redshift performance, however,
    is set by apparent brightness rather than membership status: Figure~\ref{fig:photoz_mem}
    shows a smooth scatter--magnitude relation with no residual dependence on
    cluster redshift. Photometrically selected members of comparable brightness should
    therefore follow the same relation, so our results extend to the broader
    member population at fixed magnitude. A full validation on such members requires
    a non-spectroscopic membership method, which we leave to future work (Section~\ref{sec:implications}).

    \subsection{Cluster Redshift Estimation}
    \label{sec:cluster_z}

\begin{figure*}
        \centering
        \includegraphics[width=\textwidth]{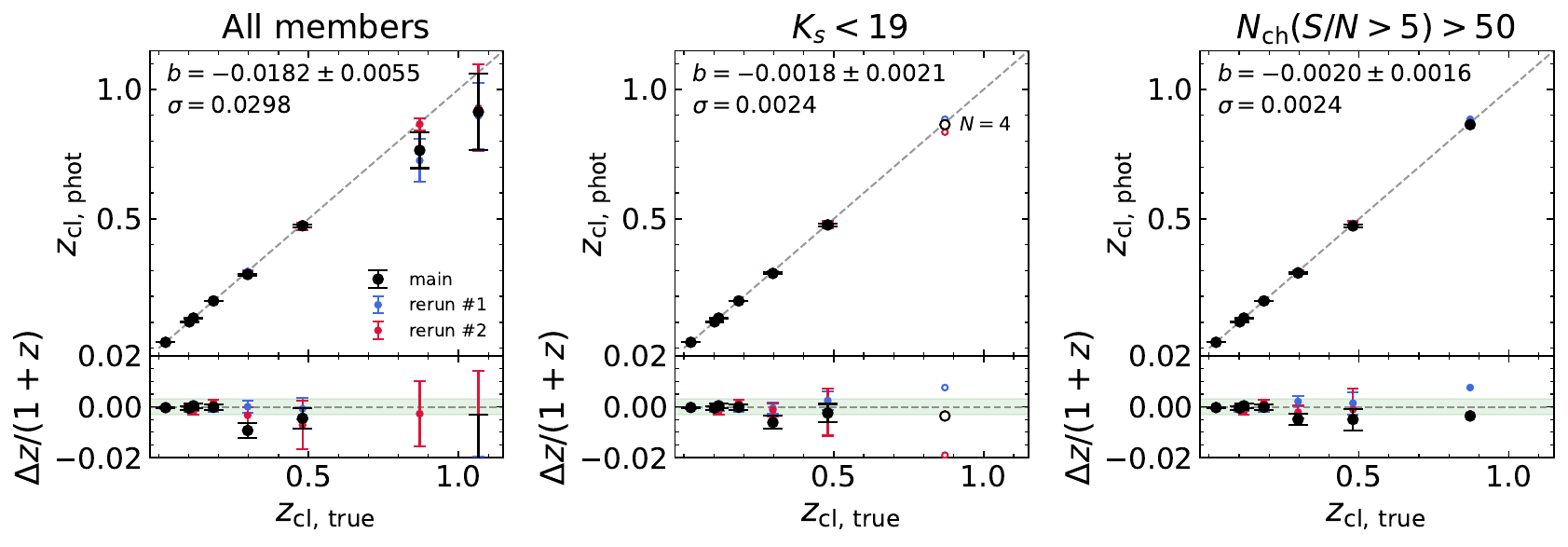}
        \caption{Cluster redshift estimation from the combined biweight location
        of the member photometric redshift estimates. Upper panels: recovered
        cluster redshift $z_{\rm cl,phot}$ versus true redshift
        $z_{\rm cl,true}$. Lower panels: normalized residual $\Delta z/(1+z)$, where
        the green shaded band indicates the $|\Delta z|/(1+z) < 0.003$ bias floor
        required for cluster cosmology \citep{lima07}; the requirement spans $0.0
        01$--$0.005$ depending on survey design \citep{huterer04}. From left to
        right: all spectroscopic members, members with $K_{s}< 19$, and members with
        $N_{\rm ch}(S/N > 5) > 50$. The three points at each cluster show three
        independent end-to-end realizations (image simulation, photometry, and photometric-redshift
        fitting) under identical settings; the black point is our main realization
        and the blue and red points are the two repeats. Error bars are the
        $1{,}000$-sample member bootstrap uncertainty on the biweight location
        for each realization. The open symbol in the $K_{s}<19$ panel marks SpARCS~J1613+5649,
        for which only four members survive the cut (no bootstrap error is shown);
        SPT-CL~J0546$-$5345 has no members passing the $K_{s}<19$ or $N_{\rm ch}>
        50$ cuts and is omitted from those panels. Annotations give the bias
        averaged over the cluster sample for the main realization, $b=\langle\Delta
        z/(1+z)\rangle$, with its run-to-run uncertainty (the standard deviation
        across the three realizations), and the cluster-to-cluster standard deviation
        $\sigma$ of the main residuals. With quality-selected members, the cluster
        redshifts are recovered with a bias consistent with zero and a scatter ($\sigma
        \approx 0.002$) well below the corresponding $\sim 0.03$ scatter requirement,
        meeting the cosmological precision requirements for clusters at
        $z \lesssim 0.5$.}
        \label{fig:cluster_z}
    \end{figure*}

    A key application of member galaxy photometric redshifts is the estimation of
    cluster redshifts by combining the individual measurements. We compute the
    cluster photometric redshift, $z_{\rm cl,phot}$, as the biweight location
    estimator \citep{beers90} of the member photometric redshift values, which
    provides a robust estimate of the central tendency in the presence of
    outliers. To assess the stability of the recovered cluster redshifts against
    realization noise, we repeat the full end-to-end analysis (image simulation,
    photometry, and photometric-redshift fitting) three times under identical
    settings. Figure~\ref{fig:cluster_z} reports the values from our main realization,
    and we use the run-to-run spread to quantify the uncertainty on the bias.

    Figure~\ref{fig:cluster_z} shows the recovered cluster redshifts as a function
    of the true redshift for three member selections. When all spectroscopic
    members are included, the two highest-redshift clusters, SpARCS J1613+5649 ($z
    = 0.87$) and SPT-CL J0546$-$5345 ($z = 1.07$), show significant deviations from
    the true redshifts, driven by the large photometric redshift scatter of
    their faint member populations. The resulting mean bias of $b = -0.018$ (averaged
    over the clusters) exceeds the cosmological bias floor of $\lesssim 0.003$--$0
    .005$
    (Section~\ref{sec:intro}), and the dispersion of $\sigma = 0.03$ is comparable
    to the corresponding scatter requirement ($\sim 0.03$).

    Applying a brightness cut of $K_{s}< 19$ or a data quality cut of $N_{\rm ch}
    (S/N > 5) > 50$ dramatically improves the precision. With either selection, the
    mean residual across the cluster sample is consistent with zero,
    $\langle\Delta z/(1+z)\rangle = -0.002 \pm 0.002$, where the average is taken
    over the clusters and the uncertainty is the run-to-run spread across three independent
    realizations. This bias is smaller than the cluster-to-cluster dispersion ($\sigma
    = 0.002$, itself stable to within $0.001$ across realizations). Both the
    bias and the dispersion lie within the cosmological requirements on bias ($\lesssim
    0.003$--$0.005$) and scatter ($\sim 0.03$; Section~\ref{sec:intro}).

    The largest single-cluster deviation in our main realization is Abell~2537 ($z
    = 0.30$), with $|\Delta z|/(1+z) = 0.006$. As Figure~\ref{fig:cluster_z}
    shows, this is driven by realization noise rather than a systematic trend with
    redshift: the two repeat realizations of the same cluster fall close to zero,
    and the member bootstrap error bar, which holds the member sample fixed, does
    not capture this run-to-run variation. Such a deviation in a single cluster
    therefore does not affect the sample-averaged bias, which remains well within
    the requirement. At $z \gtrsim 0.8$, the number of members passing the quality
    cuts drops to only a few, which limits the precision of the combined
    estimate; SPT-CL J0546$-$5345 in particular has no members meeting the $N_{\rm
    ch}> 50$ criterion and is omitted from that panel.

    These results demonstrate that \spherex{} can deliver cluster redshifts of
    sufficient precision for cosmological applications at $z \lesssim 0.5$, which
    spans the redshift range of current all-sky cluster-abundance cosmology samples
    \citep[$0.1 < z < 0.8$;][]{erass, ghirardini24}, where a substantial number of
    bright members are available. The individual member redshifts already reach
    $\sigma_{\rm NMAD}\sim 0.003$--$0.01$ for quality-selected members, and
    combining members tightens the cluster redshift further as
    $\sigma_{\rm cl}\approx \sigma_{\rm NMAD}/\sqrt{N_{\rm member}}$.

    Two practical considerations bear on how many members are needed in practice.
    First, our test uses spectroscopically confirmed members, whereas a cosmological
    analysis must identify members through photometric selection, whether from
    the \spherex{} photometry itself or from complementary optical data (for
    example, by red-sequence selection); the accuracy of that membership selection,
    including interloper contamination and missed members, adds an uncertainty not
    captured here. We nonetheless expect a similar precision in practice, provided
    that the membership selection prioritizes sources with reliable photometric redshifts,
    because our quality-selected samples already consist mostly of such sources.
    Second, the $\sim 0.03$ requirement derives from forecasts targeting a
    $\sim 10\%$ measurement of the dark energy equation of state \citep{huterer04, lima07},
    and more ambitious analyses may demand tighter control. For both reasons,
    securing a large number of member redshifts per cluster remains valuable: it
    improves the cluster-redshift precision and helps absorb the additional uncertainties
    of a realistic analysis. At higher redshifts, the utility of \spherex{}
    cluster redshifts will depend on the number of members above the photometric
    quality threshold, and may benefit from combination with complementary optical--infrared
    photometry.

    \subsection{The Impact of Source Blending}
    \label{sec:photoz_blending}

    To isolate the impact of source blending on photometric redshift performance,
    we compare the results from our main simulation (\textit{Blended}) with those
    from the \textit{Single} source simulation. The global performance statistics
    after applying magnitude or S/N cuts are largely consistent between the two cases.
    For instance, for the $K_{s}< 19$ sample in the single-source run, we find a
    bias of $-0.005$, $\sigma_{\text{NMAD}}= 0.014$, and
    $\eta_{3\hat{\sigma}}= 26.5\%$, all of which are very similar to the blended
    case.

    \begin{figure}
        \centering
        \includegraphics[width=0.48\textwidth]{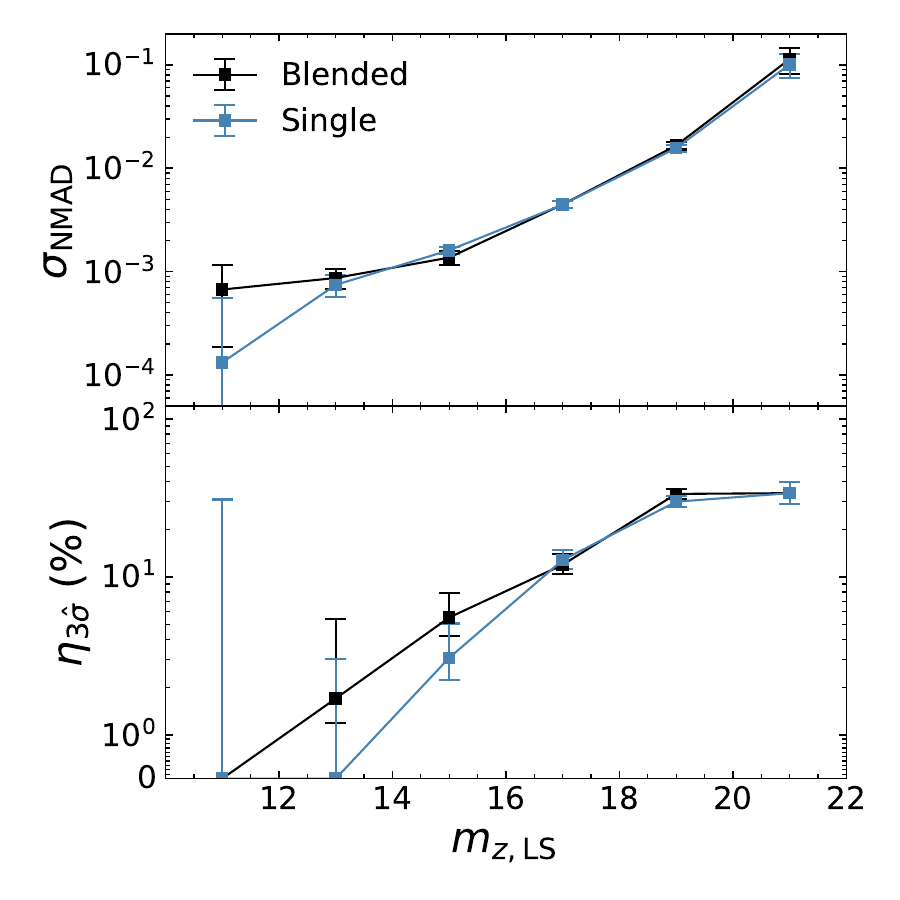}
        \caption{ Comparison of photometric redshift performance for member galaxies
        as a function of Legacy Survey $z$-band magnitude. The top panel shows
        the photometric redshift scatter, quantified by the normalized median absolute
        deviation ($\sigma_{\mathrm{NMAD}}$), while the bottom panel shows the
        $3 \hat{\sigma}$ outlier fraction ($\eta_{3\hat{\sigma}}$). Results are shown
        for the \textit{Blended} (black) and \textit{Single} (blue) simulations.
        The photometric redshift scatter and the outlier fraction are consistent
        between the two scenarios. }
        \label{fig:photoz_blending}
    \end{figure}

    Figure~\ref{fig:photoz_blending} provides a more direct comparison for the
    member galaxy sample. The photometric redshift scatter ($\sigma_{\mathrm{NMAD}}$,
    top panel) is consistent between the blended and single-source cases within
    the $1\sigma$ bootstrap errors. The outlier fraction ($\eta_{3\hat{\sigma}}$,
    bottom panel), however, shows a tendency to be lower in the single-source
    case for bright galaxies, though the difference is not highly statistically significant
    given the small number of sources in these bins. This suggests that source
    blending may primarily increase the rate of catastrophic redshift failures
    rather than the typical scatter for the bulk of the population. We further investigate
    this behavior by examining individual redshift probability density functions
    (PDFs) in Appendix~\ref{app:photoz_pdf}. As shown in Figure~\ref{fig:photoz_pdf},
    blending typically introduces additional noise that broadens and shifts the PDF,
    rather than creating distinct secondary peaks at the neighbor's redshift. This
    degradation of spectra leads to less precise photometric redshift estimates,
    particularly for faint sources where such distortions can easily lead to
    catastrophic errors.

    The same robustness extends to the treatment of the flux uncertainty itself.
    As shown in Appendix~\ref{app:cov}, replacing the native \tractor{} errors with
    the full source--source covariance leaves the redshift scatter and outlier
    rates unchanged at every crowding level. After channelization, the Secondary-Catalog
    error budget is dominated not by the source--source covariance that the native
    errors omit, but by a deterministic bandpass-sampling mismatch. This mismatch
    is intrinsic to any template fit to the channelized catalog, and is
    therefore present for real data as well. As a result, the redshift point
    estimates that drive our cluster-member results are insensitive to this
    choice. Quantities defined directly on the per-source redshift uncertainty would
    be more affected, such as the precision-bin selection used in cosmology
    forecasts \citep{huai25}.

    \section{Discussion}
    \label{sec:discussion}

    Our end-to-end simulations provide a quantitative forecast for the
    performance of \spherex{} forced photometry in galaxy cluster environments.
    The comprehensive tests on flux measurements, blending effects, survey depth,
    and photometric redshifts indicate that \spherex{} will be an effective tool
    for cluster science. Our results suggest that for nearby clusters with bright
    member galaxies, such as Coma, the recovery of spectrophotometric information
    is robust and the photometric redshifts are recovered with high precision.
    However, for high-redshift clusters whose members are fainter, the
    combination of source blending and shallower effective depth requires more
    careful consideration, particularly when analyzing sources near the detection
    limit. In the following subsections, we examine the robustness of our pipeline
    against systematic effects not captured by the baseline simulation (Section~\ref{sec:robustness}),
    evaluate the performance specifically for the quiescent galaxy populations
    that dominate cluster environments (Section~\ref{sec:cluster_populations}), place
    our results in context with previous SPHEREx performance predictions (Section~\ref{sec:comparison}),
    address the practical considerations that arise when applying the pipeline
    to real survey data (Section~\ref{sec:limitations}), and discuss the implications
    for cluster cosmology and future survey strategies (Section~\ref{sec:implications}).

    \subsection{Robustness of the Photometry Pipeline}
    \label{sec:robustness}

    Our simulation pipeline excludes observations where the expected source flux
    falls more than one magnitude below the $5\sigma$ point-source sensitivity
    of the corresponding SPHEREx channel (Section~\ref{sec:mockimg}). While this
    pre-selection is introduced to reduce the computational cost, it omits the
    flux contribution of sub-threshold sources that would be present in real
    SPHEREx observations. To quantify the impact of this omission, we perform
    two additional sets of simulations for all cluster fields except Coma, whose
    large angular extent makes the full simulation computationally impractical.
    In the \textit{Confusion-only} simulation, all cataloged sources contribute
    flux to the mock images, but only the threshold-selected sources are modeled
    by \tractor{} during forced photometry. In the \textit{Full} simulation, all
    sources are both rendered in the images and simultaneously fitted. We refer
    to the original simulation described in Section~\ref{sec:pipeline} as the
    \textit{Baseline}. This design isolates two distinct effects: direct flux contamination
    from unmodeled neighbors (Effect~A), measured by comparing the Confusion-only
    simulation to the Baseline, and fitting degeneracy arising from the simultaneous
    modeling of many faint sources (Effect~B), measured by comparing the Full
    run to the Confusion-only run.

\begin{figure}
        \centering
        \includegraphics[width=0.48\textwidth]{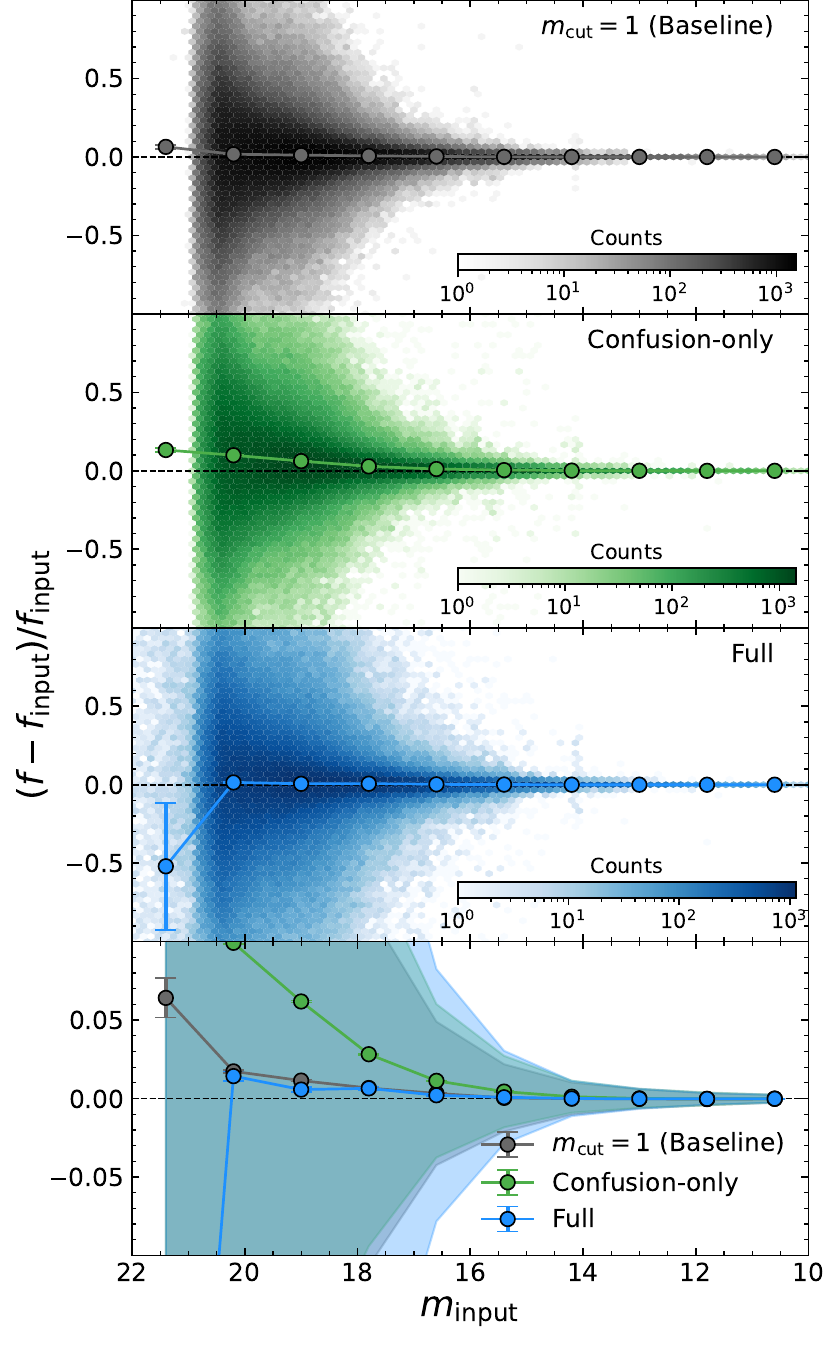}
        \caption{Fractional flux residual versus input magnitude for the
        Baseline (gray), Confusion-only (green), and Full (blue) simulations, combining
        all cluster fields except Coma. The bottom panel compares the mean bias
        (points) and $1\sigma$ scatter (shaded regions). The Confusion-only case
        shows a positive bias at faint magnitudes with scatter comparable to the
        Baseline, while the Full case shows dramatically increased scatter
        despite a similar mean bias.}
        \label{fig:confusion_phot}
    \end{figure}

    Figure~\ref{fig:confusion_phot} compares the fractional flux residuals
    across the three scenarios. The Confusion-only simulation reveals a systematic
    positive bias for faint sources ($m_{\rm input}\gtrsim 18$), reaching $\sim$5--13\%
    at the faintest magnitudes. This bias arises because unmodeled sub-threshold
    sources contribute flux to the target pixels, which \tractor{} attributes to
    the target. The effect is more pronounced for extended sources, whose
    broader light profiles overlap more with neighboring unmodeled sources: the mean
    normalized residual ($(f-f_{\rm input})/\sigma_{f}$) for extended sources reaches
    $\sim$0.7 at $m_{\rm input}\sim 16$, compared to $\sim$0.2 for point sources
    at the same magnitude. This is particularly relevant for cluster fields, as
    the member galaxies of interest are predominantly extended sources in dense
    environments where source confusion is most severe. Despite this bias, the
    photometric scatter in the Confusion-only case remains comparable to the
    Baseline, indicating that the confusion noise adds a systematic offset without
    significantly increasing random errors.

    The Full simulation presents a qualitatively different picture. While the
    mean bias remains similar to the Baseline (as positive and negative flux errors
    from degeneracy tend to cancel on average) the scatter increases
    dramatically. This is the expected behavior of an underconstrained fit: when
    many faint sources with negligible signal are simultaneously modeled, their flux
    amplitudes become degenerate. The effect is amplified by the large pixel
    scale of \spherex{}, which causes multiple sources to share fewer
    independent pixels, further reducing the constraints available to disentangle
    individual flux contributions. This excess scatter reflects the joint fit of
    all overlapping sources rather than the mere presence of faint neighbors: the
    Confusion-only run, which renders the same sub-threshold sources without
    fitting them, shows no comparable increase in scatter. Propagating the full
    source--source covariance recalibrates the flux uncertainties but does not reduce
    this scatter in the point estimates (Appendix~\ref{app:cov}); the broadening
    is intrinsic to the degenerate fit itself.

\begin{figure}
        \centering
        \includegraphics[width=0.48\textwidth]{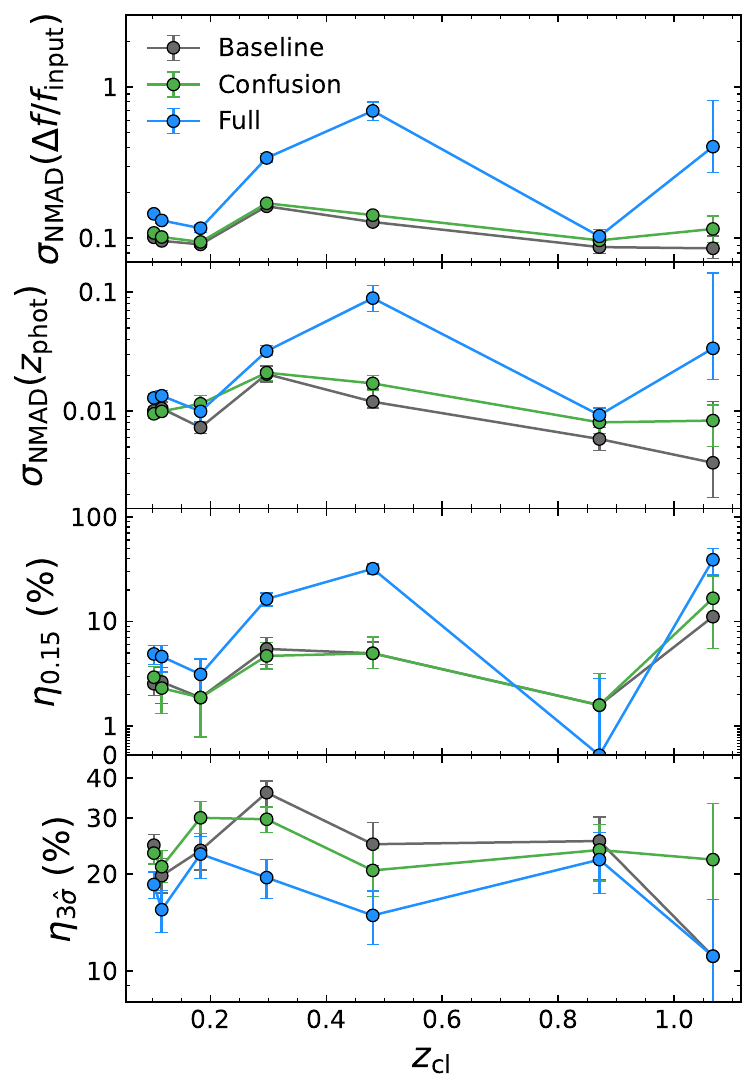}
        \caption{Flux and photometric redshift performance as a function of
        cluster redshift for the Baseline (gray), Confusion-only (green), and
        Full (blue) simulations. All metrics are computed for galaxies with $K_{s}
        < 19$. From top to bottom: the flux-residual scatter $\sigma_{\rm NMAD}(\Delta
        f/f_{\rm input})$, the photometric redshift scatter
        $\sigma_{\rm NMAD}(z_{\rm phot})$, and the outlier fractions $\eta_{0.15}$
        and $\eta_{3\hat{\sigma}}$. The Confusion-only results closely track the
        Baseline, while the Full simulation shows elevated scatter and $\eta_{0.15}$
        at $z_{\rm cl}\gtrsim 0.3$. }
        \label{fig:confusion_zcl}
    \end{figure}

    Figure~\ref{fig:confusion_zcl} shows these effects as a function of cluster
    redshift, through the flux-residual scatter $\sigma_{\rm NMAD}(\Delta f/f_{\rm
    input})$, the photometric-redshift scatter $\sigma_{\rm NMAD}(z_{\rm phot})$,
    and the outlier fractions $\eta_{0.15}$ and $\eta_{3\hat{\sigma}}$. The
    Confusion-only results closely track the Baseline in all four metrics, confirming
    that unmodeled flux contamination alone does not significantly degrade photometric
    redshift performance. The Full simulation shows elevated scatter ($\sigma_{\rm
    NMAD}$ in both flux and photometric redshift) and outlier fraction
    $\eta_{0.15}$ at $z_{\rm cl}\gtrsim 0.3$, most pronounced in the crowded,
    faint, high-$z$ fields. The $3\hat{\sigma}$ fraction $\eta_{3\hat{\sigma}}$
    is instead slightly lower, reflecting the wider photometric redshift uncertainties
    that loosen the $3\hat{\sigma}$ threshold in the degenerate regime. Per-cluster
    values for the sparsely sampled high-$z$ systems (e.g.\ $N \sim18$ for SPT-CL
    J0546$-$5345) carry limited statistical weight.

    These results validate the source pre-selection strategy adopted in our Baseline
    pipeline. By excluding sources well below the detection threshold, the pre-selection
    effectively prevents fitting degeneracy (Effect~B) while accepting a modest positive
    bias from residual confusion noise (Effect~A). For dense environments such
    as galaxy clusters, where the high surface density of sources amplifies the degeneracy,
    applying a similar pre-selection based on broadband photometry from the
    Reference Catalog may be particularly beneficial. Although the expected flux
    estimates derived from broadband magnitudes will not be exact, sources that
    fall several magnitudes below the SPHEREx detection threshold can be
    reliably identified and excluded regardless of the assumed SED shape.

    We also note that our pre-selection threshold is referenced to the
    optimistic CBE sensitivity \citep{dore14}; its relation to the as-launched
    depths is discussed in Section~\ref{sec:limitations}. Adopting the on-orbit sensitivities
    in place of the CBE would shift only a small number of faint neighbors across
    the modeling boundary. Because our Confusion-only and Full runs bracket the
    extremes of fully unmodeled and fully modeled neighbors, any such shift is already
    encompassed by the range these two simulations span.

    The residual confusion bias (Effect~A), which is most significant for faint
    extended sources, does not propagate into a comparable degradation in photometric
    redshift performance, as demonstrated by the consistency between the
    Confusion-only and Baseline results (Figure~\ref{fig:confusion_zcl}). This suggests
    that the channel-dependent bias largely preserves the overall spectral shape,
    and can be further mitigated by anchoring the \spherex{} spectrophotometry to
    deep broadband photometry, which provides an independent flux scale
    unaffected by \spherex{}-specific confusion.

    As an alternative to explicit pre-selection, the fitting degeneracy (Effect~B)
    can also be suppressed by regularized forced photometry, most directly through
    $L_{1}$-penalized regression \citep[Lasso;][]{lasso}. In this approach, a
    penalty proportional to the sum of the absolute source fluxes (their $L_{1}$
    norm) is added to the least-squares photometric objective. Because an $L_{1}$
    penalty drives weakly constrained amplitudes exactly to zero, it can
    suppress the fluxes of undetected sources without requiring a prior flux
    estimate for each source. The trade-off is that the penalty also imposes a shrinkage
    bias on the surviving fluxes, which must be corrected with a debiasing refit,
    and its strength must be calibrated to the per-channel noise properties of \spherex{}.
    A full evaluation of regularized photometry, which may be more valuable in
    the absence of reliable prior flux estimates, is left for future work.

    \subsection{Performance for Cluster Galaxy Populations}
    \label{sec:cluster_populations}

    Galaxy clusters are dominated by quiescent, early-type galaxies,
    particularly at low redshift. Given that the analysis presented in Sections~\ref{sec:phot}
    and~\ref{sec:photoz} treats all galaxies uniformly, we examine here whether
    the performance differs for the passive population that is most relevant to
    cluster science.

    We classify galaxies as passive or non-passive based on their input SED
    template type: galaxies best fit by elliptical, S0, or Sa templates from the
    \citet{brown14} and \citet{ilbert09} libraries are designated as passive. Of
    the 160 galaxy templates used in our fitting, 30 are classified as passive.
    Among the 1,032 spectroscopically confirmed cluster members, 72\% are
    classified as passive, consistent with the expected dominance of quiescent populations
    in massive clusters. This fraction ranges from $\sim$50\% to $\sim$79\%
    across individual clusters.

    \begin{figure}
        \centering
        \includegraphics[width=0.48\textwidth]{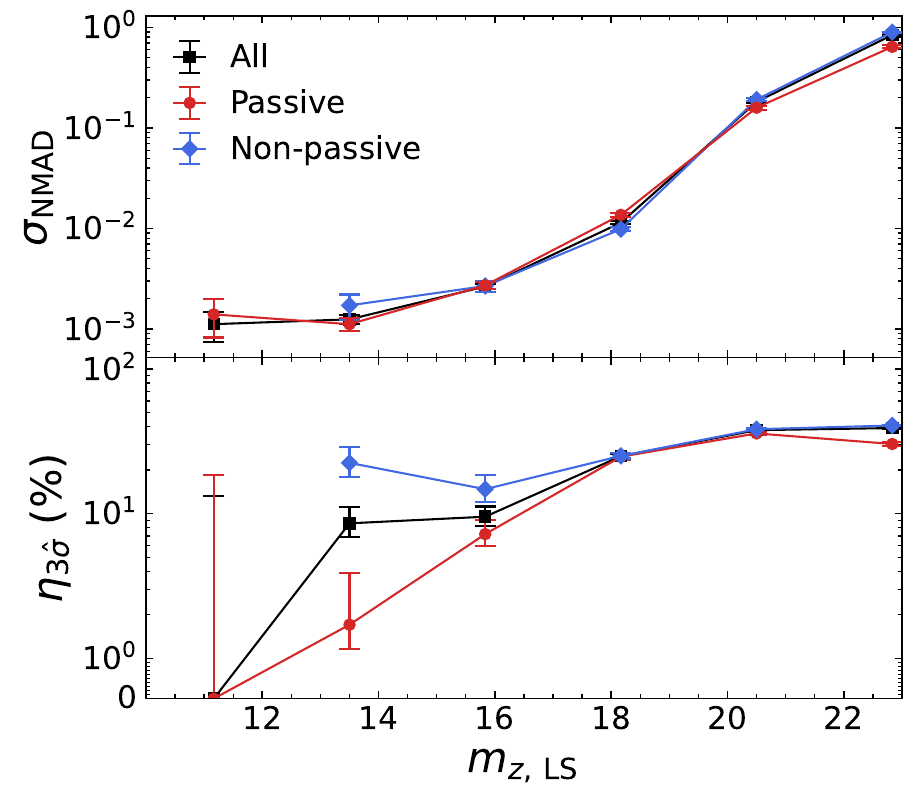}
        \caption{Photometric redshift performance as a function of Legacy Survey
        $z$-band magnitude for galaxies in the cluster fields, separated by input
        SED type: all galaxies (black), passive (red; E, S0, and Sa templates), and
        non-passive (blue). Upper panel: photometric redshift scatter $\sigma_{\rm
        NMAD}$. Lower panel: $3\hat{\sigma}$ catastrophic outlier fraction $\eta_{3\hat{\sigma}}$.
        Passive galaxies show substantially lower outlier rates at bright
        magnitudes, with the two populations converging at intermediate magnitudes
        before diverging again at faint magnitudes.}
        \label{fig:passive_photoz}
    \end{figure}

    Figure~\ref{fig:passive_photoz} compares the photometric redshift performance
    as a function of magnitude for passive and non-passive galaxies in the cluster
    fields. At bright magnitudes, passive galaxies are more reliable in both metrics:
    the scatter ($\sigma_{\rm NMAD}$, upper panel) is $\sim$0.001 versus $\sim$0.002
    for non-passive galaxies at $m_{z,\,\rm LS}\sim 13$, and the catastrophic
    outlier fraction ($\eta_{3\hat{\sigma}}$, lower panel) is only $\sim$2\% versus
    $\sim$22\% at $m_{z,\,\rm LS}\lesssim 14$. This gap narrows through
    intermediate magnitudes, with the outlier fractions converging at $m_{z,\,\rm
    LS}\sim 18$ where both reach $\sim$25\%. At the faintest magnitudes ($m_{z,\,\rm
    LS}\gtrsim 21$), the scatter of the two populations remains comparable, but
    the outlier fraction diverges again, with passive galaxies at $\sim$30\% while
    non-passive galaxies reach $\sim$41\%. This persistent outlier-rate
    advantage at faint magnitudes indicates that the smooth continua of
    quiescent galaxies, and in particular the 1.6~$\mu$m stellar bump, constrain
    the redshift well even in the low signal-to-noise regime, whereas the
    continua of star-forming galaxies, shaped by a broader mix of stellar populations
    and dust, constrain the redshift more weakly.

    \begin{figure}
        \centering
        \includegraphics[width=0.48\textwidth]{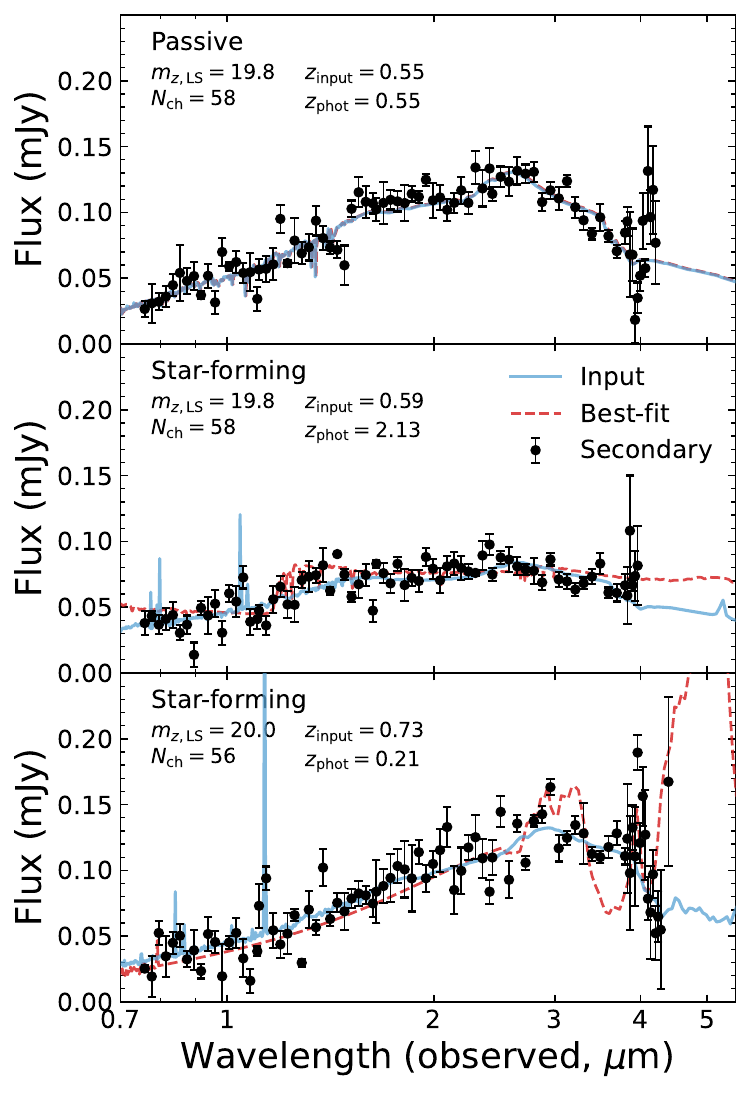}
        \caption{Input and best-fit SEDs for a passive galaxy and two star-forming
        galaxies of comparable brightness and spectral coverage ($m_{z,\,\rm LS}\approx
        19.8$--$20.0$, $N_{\rm ch}=56$--$58$ channels at S/N $>5$). Each panel
        gives the input SED (blue), the Secondary Catalog spectrophotometry (black
        points), and the best-fit template at the photometric redshift (red dashed).
        The passive galaxy (top) is recovered in both template and redshift ($z_{\rm
        input}=0.55$, $z_{\rm phot}=0.55$), whereas the two star-forming galaxies
        become catastrophic outliers at the same depth, driven to high redshift
        ($z_{\rm input}=0.59$, $z_{\rm phot}=2.13$) and to low redshift ($z_{\rm
        input}=0.73$, $z_{\rm phot}=0.21$).}
        \label{fig:sf_passive_sed}
    \end{figure}

    To illustrate why this type dependence arises at fixed signal-to-noise,
    Figure~\ref{fig:sf_passive_sed} shows three galaxies of comparable Legacy
    Survey magnitude ($m_{z,\,\rm LS}\approx20$) and comparable spectral
    coverage ($N_{\rm ch}\approx57$ channels at S/N $>5$). The passive galaxy (top
    panel) has a smooth continuum with a well-defined 1.6~$\mu$m stellar bump, and
    both its template and redshift are recovered. The star-forming galaxies (middle
    and bottom panels) instead have continua that lack a single strong
    localizing feature. At the SPHEREx spectral resolution such continua allow several
    template-redshift solutions of similar likelihood, and the fit can be driven
    to a catastrophically incorrect redshift in either direction. Nebular emission
    lines provide little additional leverage at this resolution. Because
    magnitude and channel count are matched across the three panels, this behaviour
    reflects the intrinsic diversity of star-forming SEDs rather than signal-to-noise
    alone.

    These results confirm that the quiescent population dominating cluster membership
    benefits from a consistent advantage in photometric redshift reliability.
    Considering that this population also defines the cluster cosmology
    observables of richness and red-sequence calibration, this advantage is particularly
    valuable for such applications.

    \subsection{Comparison with Previous SPHEREx Predictions}
    \label{sec:comparison}

    Our photometric redshift performance is broadly consistent with previous
    SPHEREx forecasts, though direct comparison requires care due to differences
    in sample selection, ancillary data, and treatment of source blending. For
    bright galaxies detected in all 102 SPHEREx channels, we achieve $\sigma_{\rm
    NMAD}= 0.002$, in good agreement with the \citet{feder24} prediction of $\sigma
    _{z}< 0.003(1+z)$ for $\sim$19 million galaxies. At brighter magnitudes ($K_{s}
    < 17.5$), our $\sigma_{\rm NMAD}= 0.007$ is comparable to the SPHEREx-only
    results in \citet{bae2025}, where $\sigma_{\rm NMAD}\approx 0.003$--$0.005$ is
    obtained for GAMA-like galaxies ($i < 18$) using multiple photometric redshift
    codes including \eazypy. The somewhat larger scatter in our results likely
    reflects differences in sample selection (NIR vs.\ optical magnitude cuts) and
    our inclusion of realistic galaxy--galaxy blending in the simulation. For the
    full $K_{s}< 19$ sample, our $\sigma_{\rm NMAD}= 0.024$ is larger than the
    $\sigma_{\rm NMAD}\approx 0.009$ reported by \citet{feder24} at comparable depths
    ($18.5<W1<19.0$). Beyond the different sample selections, several factors likely
    contribute to this difference, including the incorporation of ancillary
    DECaLS $grz$ and WISE photometry in that analysis, which provides additional
    constraints at optical wavelengths, as well as the absence of source
    blending in their simulation. The blending fractions at this depth are $\sim$0.7\%
    for the all-sky survey \citep{dachan24}, and \citet{huai25} find that the
    additional photometric uncertainty from blending of targeted neighbors
    removes $\sim$9\% of galaxies from the cosmology sample by degrading their
    redshift precision below the cosmology threshold.

    Both our work and \citet{feder24} employ the same 160-template Brown+COSMOS library
    for both mock generation and template fitting. This template self-consistency
    makes the quoted $\sigma_{\rm NMAD}$ values somewhat optimistic relative to
    real SPHEREx data, where true galaxy SEDs will exhibit greater diversity than
    the template library can capture. As investigated by \citet{bae2025}, the
    performance of \eazypy{} is comparable to the official in-house SPHEREx codes,
    though further improvements may be expected from pipelines optimized for
    SPHEREx data, for instance through the application of internal extinction correction,
    which \citet{bae2025} identify as a significant factor. Moreover, the development
    of specialized template libraries, such as those for active galactic nuclei,
    will be crucial for extending reliable photometric redshift estimation to galaxy
    populations not fully represented by the current templates.

    \subsection{Practical Considerations for Application to Survey Data}
    \label{sec:limitations}

    Our results are based on controlled simulations; applying the pipeline to real
    \spherex{} data raises several practical considerations, spanning the assumed
    survey sensitivity, the cost of the photometric modeling, and the morphological
    and astrometric assumptions, which we make explicit here.

    A first consideration concerns the survey sensitivity adopted in our
    simulations. The photometric noise in our mock images derives from the \simulator{}
    Zodiacal-light backgrounds \citep{crill25}, based on the pre-launch
    instrument characterization, and the source pre-selection (Section~\ref{sec:pipeline})
    is referenced to the CBE point-source sensitivity \citep{dore14}. Both are broadly
    consistent with the as-launched all-sky performance reported by \citet{spherex},
    so we do not expect the choice of sensitivity to affect our results. Two
    localized exceptions are the degraded sensitivity near $1.08\,\mu$m, from
    photon noise associated with terrestrial \ion{He}{1} airglow, and near $2.4\,
    \mu$m, from the dichroic beam splitter separating the two detector focal planes
    \citep{spherex}; the reduced depth in these channels would preferentially affect
    the faintest cluster members detected in only a few channels.

    A key methodological choice in our pipeline is the use of full morphological
    modeling for extended sources. While this approach is essential for accuracy,
    the additional convolution step makes it computationally more expensive than
    simple PSF photometry. For an all-sky survey like \spherex{}, this
    computational cost is a significant consideration, especially if shape parameters
    are also allowed to vary during the fit (in our tests, freeing the S\'ersic shape
    parameters increases the fitting wall time by a factor of $\sim$4 on average).
    A practical compromise for future pipeline development could be to apply PSF
    photometry to sources that are very compact relative to the \spherex{} PSF size.
    As detailed in Appendix~\ref{app:psf}, our tests on sources with $r_{e}< 1\arcsec$
    show that while this introduces a systematic flux bias, it may be an
    acceptable trade-off depending on the specific science goal. As the \spherex{}
    Science Data Center provides tools for users to perform their own forced
    photometry\footnote{\dataset[https://irsa.ipac.caltech.edu/applications/spherex]{https://irsa.ipac.caltech.edu/applications/spherex}},
    one can choose the optimal method for their science case.

    A second methodological choice concerns the galaxy light profile itself. Our
    pipeline applies a correction for the flux deficit intrinsic to \tractor{}'s
    MoG approximation of a S\'ersic profile, which is required for recovering unbiased
    fluxes in our simulations but whose application to real \spherex{} data
    depends on the downstream use case; we refer the reader to Appendix~\ref{app:tractor}
    for details.

    A separate idealization concerns the wavelength dependence of source morphology.
    As described in Section~\ref{sec:pipeline}, each source is modeled with a single,
    wavelength-independent morphology derived from its optical (DESI Legacy)
    profile and held fixed across all SPHEREx channels. Holding optically derived
    positions and morphologies fixed while fitting only the fluxes is the standard
    forced-photometry approach for extracting longer-wavelength fluxes from high-resolution
    optical priors \citep[e.g.,][]{lang16}, and it is also the procedure adopted
    by the official \spherex{} Level~3 pipeline, which performs forced
    photometry at the positions and shapes of \spherex{} Reference Catalog sources
    \citep{akeson25}. It nonetheless represents a best-case scenario, since the
    morphological parameters used in the fit are identical to those used to
    generate the mock sources. In practice, only the optical morphology is
    typically available, whereas galaxy sizes are known to decrease toward the NIR
    by $\sim$30\% from $g$ to $K$ for early-type galaxies \citep{labarbera10}, with
    comparable trends across galaxy types \citep{kelvin12, vulcani14}.

    Adopting a fixed optical morphology across all channels could therefore
    introduce a bias not captured by our simulations. An alternative is to leave
    the S\'ersic shape parameters free during the fit. We tested this by
    repeating the photometry with $r_{e}$, $b/a$, and position angle all free, with
    the fit initialized at the true morphology. Even in this best case, releasing
    the shape prior inflates the per-channel flux scatter by a factor of $\sim$1.8
    and degrades the bright-sample ($K_{s}<17.1$) photometric redshifts by a
    comparable factor ($\sigma_{\rm NMAD}\approx0.006\rightarrow0.010$). The recovered
    shapes also frequently drift into unphysical regions (e.g., $r_{e}<0$ or
    $b/a>1$). The morphology prior thus acts as a useful regularizer, and a prior-based
    fit is preferable to a free fit whenever a reasonable morphology estimate is
    available. Assessing the bias from genuine optical-NIR morphology mismatch
    requires a test in which the assumed morphology differs from the truth, which
    we leave to future work.

    The fixed-position assumption proves similarly benign. \spherex{} astrometric
    calibration is accurate to $0\farcs1$--$0\farcs4$ \citep{akeson25}, and
    repeating the analysis with position errors of this size (Appendix~\ref{app:astrometry})
    leaves the photometric redshifts of the science samples essentially
    unchanged. A coherent mission-wide offset can shift the mean flux, but it
    shifts all channels by a similar amount and preserves the spectral shape
    that carries the redshift information. The photometric-redshift bias of the $K
    _{s}<19$ and $K_{s}<17.5$ samples therefore stays below $10^{-3}$ in $\Delta
    z/(1+z)$, with the scatter and outlier fraction unchanged within the
    uncertainties. A clear degradation appears only for position errors above $\sim
    0\farcs5$, beyond the expected level.

    \subsection{Implications for Cluster Cosmology and Future Work}
    \label{sec:implications}

    Our simulations confirm that \spherex{} can deliver photometric redshifts of
    the quality required for cluster-cosmology applications (Section~\ref{sec:cluster_z}).
    While specific forecasts for constraining cosmological models with \spherex{}
    cluster samples have not yet been made, the achieved precision indicates that
    the data will be able to make meaningful contributions. Furthermore, the rich
    NIR spectrophotometry from \spherex{}, when combined with ancillary data from
    optical-infrared \citep[e.g., 7DS, DESI, LSST, Euclid;][]{7dt, desi, lsst, euclid},
    X-ray, and SZ surveys, has the potential to reduce the scatter in the mass-observable
    scaling relations whose calibration is the leading systematic in current
    cluster-abundance cosmology \citep{bocquet24} and that are crucial for self-calibration
    techniques in cluster cosmology \citep{rozo14, huterer18}.

    Looking forward, the development of dedicated cluster analysis methods for \spherex{}
    data will be essential for producing Level 4 legacy science products, such as
    a \spherex{} Galaxy Cluster Catalog. This will require new cluster detection
    algorithms (e.g., matched filters) tailored to \spherex{} data, careful source
    pre-selection strategies for forced photometry in dense environments (Section~\ref{sec:robustness}),
    as well as methods to assign membership probabilities using the full
    spectrophotometric information. For instance, the detailed NIR spectra from
    \spherex{} could enable more robust red-sequence selection than is possible with
    broadband photometry alone, building on the favorable photometric redshift
    performance for passive galaxies demonstrated in Section~\ref{sec:cluster_populations}.

    In production, such a catalog would build directly on the official \spherex{}
    photometry and redshift pipeline. The Level~3 photometry follows the same forced-photometry
    approach we validate here \citep{akeson25}, while additionally incorporating
    artifact rejection and masking on real data, together with the flux-covariance
    treatment of \citet{huai25}. The pipeline presented in this work thus serves
    as a prototype for this production path.

    The photometric systematics we characterize here, namely residual flux biases,
    blending-induced catastrophic redshift outliers, and their radial dependence
    toward cluster cores (Sections~\ref{sec:blend}, \ref{sec:photoz_blending}), are
    inputs to the completeness and purity of any such cluster finder:
    photometric-redshift errors propagate into cluster-finder failures such as
    projection, fragmentation, and over-merging. For optical and X-ray samples,
    neglecting this selection function has been shown to bias dark energy
    parameters by more than $2\sigma$ \citep{aguena18}, and current surveys
    construct it explicitly from end-to-end simulations \citep[e.g.,][]{clerc24}.
    These demonstrations involve different selection observables than \spherex{},
    but they establish that characterizing such photometric systematics is a prerequisite
    for the selection-function modeling that a \spherex{}-based cluster
    cosmology analysis would require.

    As demonstrated in Section~\ref{sec:cluster_z}, combining photometric redshifts
    of quality-selected individual members yields cluster redshifts with a scatter
    of $\sigma \approx 0.002$ at $z \lesssim 0.5$, comfortably within the
    precision required by current cluster-abundance and clustering analyses. For
    number-count analyses the binding redshift systematic is the cluster-redshift
    bias rather than its scatter \citep{huterer04, lima07}, which our quality-selected
    estimates keep to $|\Delta z|/(1+z) < 0.002$; for clustering analyses, where
    the cluster-redshift scatter instead dilutes the signal, forecasts adopt
    $\sigma_{z0}\approx 0.005$--$0.01$ \citep{fumagalli25}, well above the scatter
    we achieve.

    The $z \lesssim 0.5$ range over which we demonstrate cosmology-grade cluster
    redshifts is a conservative baseline, not a hard limit. Because the precision
    is set by apparent magnitude rather than cluster redshift (Section~\ref{sec:photoz_member}),
    the limit at higher redshift is the number of members above our quality
    threshold, not their individual quality. The brightest members, such as the BCG
    and the most luminous early-type galaxies, remain bright enough for
    $\sigma_{\rm NMAD}\sim0.01$ to $z\sim1$ and can anchor a cluster redshift as
    the population fades. The framework also extends to clusters detected by wide
    optical surveys such as Rubin/LSST. These surveys identify the cluster and
    its candidate members from red-sequence colors or photometric-redshift over-densities,
    and \spherex{} then measures precise redshifts for those members \citep{dore16, dore18}.
    In that case, membership would be photometric, from the optical selection or
    from the \spherex{}-based methods noted above. Both can supply more members
    per cluster than our spectroscopic census and tighten the cluster redshift. Improvements
    to the redshift estimation, such as machine-learning methods \citep{feder26}
    or refined template libraries, could extend the usable range further.

    Finally, the ability to accurately recover SEDs, as demonstrated in this
    work, opens the door to detailed studies of galaxy evolution within cluster environments.
    Applying techniques similar to those used for resolved sources \citep{lee25}
    to our well-defined cluster member samples will allow for robust constraints
    on their stellar masses, ages, and star formation histories, providing new
    insights into the environmental processes that shape galaxies.

    \section{Conclusion}
    \label{sec:conclusion}

    In this work, we have performed an end-to-end simulation to quantitatively
    assess the performance of the \spherex{} forced photometry pipeline for galaxy
    cluster science. By generating realistic mock observations for eight galaxy
    clusters spanning a wide redshift range ($z \approx 0.02$--$1$) and processing
    them through our pipeline, we have tested the key aspects of photometric
    accuracy, source blending, survey depth, and photometric redshift accuracy in
    dense environments. Our main findings are summarized as follows:

    \begin{enumerate}
        \item Photometry Performance: Our pipeline robustly recovers the fluxes
            of both point-like and extended sources with minimal bias (typically
            $<2\%$) for most magnitude ranges. The binning of individual measurements
            from the Primary Catalog into a Secondary Catalog effectively reduces
            the photometric scatter by $\sim$50\%, consistent with expectations
            from four all-sky surveys, and yields more reliable error estimates.
            A small positive bias at faint magnitudes is primarily driven by
            compact ($r_{e}< 1\arcsec$) extended sources.

        \item Source Blending: Source blending is a primary driver of
            catastrophic outliers in the photometry. The photometric bias and scatter
            are most strongly correlated with the total neighbor-to-target flux ratio,
            with the bias reaching up to $\sim$10\% when neighbors are 100 times
            brighter than the target. While blending increases the rate of outliers,
            it introduces only a small additional mean bias for the general
            population.

        \item Survey Depth: The 5$\sigma$ point-source depth is a function of galaxy
            morphology, becoming shallower for sources with larger effective
            radii relative to the PSF size. For cluster members, our simulations
            show that \spherex{} can detect galaxies 7--9 magnitudes fainter
            than the BCG in nearby clusters like Coma, whereas this limit is reduced
            to 1--2 magnitudes for clusters at $z \sim 1$, primarily reflecting the
            BCG fading toward the survey depth with redshift.

        \item Photometric Redshift Accuracy: With appropriate sample selections
            based on brightness or data quality, \spherex{} can achieve a
            photometric redshift precision of $\sigma_{\rm NMAD}\approx 0.004$
            for bright cluster galaxies ($K_{s}< 17.5$), reaching $\sigma_{\rm
            NMAD}\approx 0.002$ for the highest-quality sample (S/N $>$ 5 in all
            102 channels). We find that source blending does not significantly
            degrade the photometric redshift scatter for the bulk of the
            population but may increase the outlier fraction.

        \item Cluster Redshift Estimation: Combining the photometric redshifts
            of quality-selected members recovers cluster redshifts to a bias of
            $| \Delta z|/(1+z) < 0.002$ and a scatter of $\sigma \approx 0.002$
            at $z \lesssim 0.5$, within the precision required by current cluster-abundance
            and clustering analyses (Section~\ref{sec:implications}).
    \end{enumerate}

    Our results demonstrate that, despite its large pixel scale, \spherex{} can
    deliver reliable photometry and photometric redshifts for cluster galaxies. For
    bright members, the achieved precision is well within the level required by
    cluster-abundance analyses, making \spherex{} a promising source of large, well-characterized
    cluster samples; fully realizing their cosmological potential will
    additionally require the cluster detection, membership selection, and mass
    calibration developed in future work (Section~\ref{sec:implications}). Furthermore,
    the rich NIR spectrophotometry, when combined with data from other surveys,
    will enhance self-calibration techniques by helping to break degeneracies in
    mass-observable scaling relations.

    This work serves as a foundational study for the development of the \spherex{}
    Level 4 legacy science products, particularly the forthcoming SPHEREx Galaxy
    Cluster Catalog. Future work should focus on developing cluster detection and
    membership selection algorithms specifically tailored to \spherex{} data, such
    as robust red-sequence finders that leverage the 102-channel spectrophotometry.
    Such methods, combined with the high-fidelity photometric redshifts demonstrated
    here, will enable precise cluster redshift measurements for the bulk of the
    cluster population. Ultimately, by accurately recovering the SEDs of member
    galaxies, the methodology validated in this paper paves the way for detailed
    studies of galaxy evolution in the most extreme environments across cosmic time.

    \begin{acknowledgments}
        The work of HB was supported by Basic Science Research Program through the
        National Research Foundation of Korea (NRF) funded by the Ministry of Education
        (RS-2025-25403440). HSH acknowledges support from the National Research
        Foundation of Korea (NRF) funded by the Korea government (MSIT; RS-2026-25482692)
        and the Global-LAMP Program funded by the Ministry of Education (RS-2023-00301976).
        Bomee Lee is supported by the National Research Foundation of Korea (NRF)
        NRF grant funded by the Korea government(MSIT), 2022R1C1C1008695. J.H.L.
        acknowledges support from the Basic Science Research Program through the
        National Research Foundation of Korea (NRF) funded by the Ministry of
        Education (No. RS-2024-00452816). Part of the research described in this
        paper was carried out at the Jet Propulsion Laboratory, California Institute
        of Technology, under a contract with the National Aeronautics and Space
        Administration (80NM0018D0004).
    \end{acknowledgments}

    \facilities{\spherex}

    \software{\tractor, \eazypy, \simulator, \galsim}

    \appendix
    \restartappendixnumbering

    \section{Impact of Pixel-by-pixel Response}
    \label{app:ppr}

    A key component of the \spherex{} instrument is the LVFs, which causes the
    central wavelength and shape of the spectral bandpass to vary continuously
    with position on the detector. For a point source, the photometry can be
    well-approximated using the response curve of the central pixel upon which
    the source lands. For an extended source such as a galaxy, however, different
    parts of the object are observed through different bandpasses, leading to a
    spatially varying flux response across the image. To accurately model this effect
    in our simulation, we introduce a Pixel Response Map, $R(x, y)$.

    The response map quantifies the relative flux of a given source as seen through
    the bandpass of any pixel $(x, y)$ in a cutout, normalized to the flux
    observed through the bandpass of the source center $(x_{c}, y_{c})$. The observed
    flux, $F_{\text{obs}}(x, y)$, for a source with an intrinsic SED $S(\lambda)$
    at pixel $(x, y)$ is given by the integral over the local bandpass transmission
    function, $T(x, y; \lambda )$:
    \begin{equation}
        F_{\text{obs}}(x, y) = \frac{\int S(\lambda) T(x, y; \lambda) d\lambda}{\int
        T(x, y; \lambda) d\lambda}. \\
    \end{equation}
    The response map $R(x, y)$ is then defined as the ratio:
    \begin{equation}
        R(x, y) = \frac{F_{\text{obs}}(x, y)}{F_{\text{obs}}(x_{c}, y_{c})}.
    \end{equation}
    In our simulation pipeline, this response map is multiplied by the intrinsic,
    PSF-convolved 2D light profile of the galaxy, $I_{\text{intrinsic}}(x, y)$,
    to produce the final flux distribution, $I_{\text{final}}(x, y) = I_{\text{intrinsic}}
    (x, y) \times R(x, y)$, before the addition of background and noise. This process
    ensures a physically accurate representation of how light from each part of an
    extended source is modulated by the LVFs. It is particularly important for
    galaxies with strong color gradients or prominent emission line features, and
    is a step in generating realistic mock images to validate the photometry pipeline.

    A key motivation for this test is to validate the simplifying assumptions made
    during the photometry stage. While our simulation pipeline carefully models
    the pixel-by-pixel response variations across the cutout, the model fitting performed
    by \tractor{}, by contrast, assumes a single, uniform response for all pixels
    within that same cutout. If this uniform-response assumption were a
    significant source of error, we would expect the photometry performed on the
    more realistic ``Pixel-by-Pixel Response'' simulation to exhibit larger scatter
    or bias compared to the ``Uniform Response'' case, due to the mismatch
    between the complex data and the simple model.

\begin{figure}
        \centering
        \includegraphics[width=0.48\textwidth]{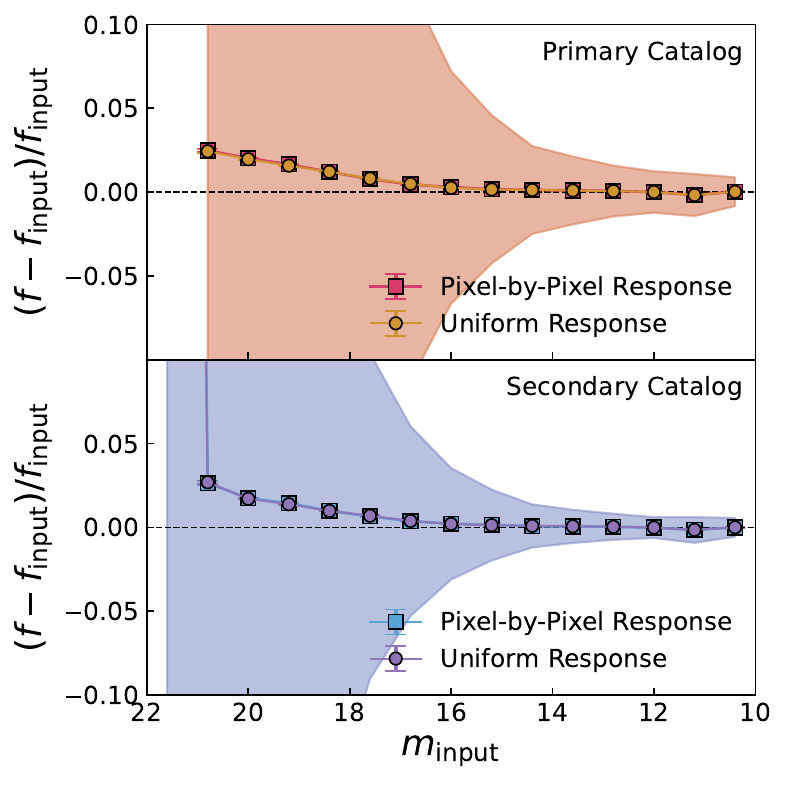}
        \caption{ Impact of pixel-response variations on photometry. The panels
        show fractional flux residuals as a function of input magnitude for the
        primary (top) and secondary (bottom) catalogs. The ``Pixel-by-Pixel Response''
        case adopts the actual detector response map in the image simulation,
        while the ``Uniform Response'' case applies the response curve of the
        central pixel to all cutout pixels as an approximation. The comparison illustrates
        how spatial variations in pixel response can affect measured fluxes. }
        \label{fig:app-pixel}
    \end{figure}

    However, as shown in Figure~\ref{fig:app-pixel}, the results from both
    simulations are remarkably similar. The lack of significant degradation in performance
    indicates that the assumption of a uniform response is a valid approximation
    for \spherex{} data. This suggests that the wavelength variation across a small
    cutout is not substantial enough to significantly alter the broadband morphological
    profile of the sources, and therefore does not introduce a major systematic
    error into the forced photometry.

    \section{Tractor S\'ersic Modeling Details}
    \label{app:tractor}

    The photometry in this work is performed using \tractor{}, which employs a
    forward model-fitting approach. To model the smooth light profiles of galaxies,
    \tractor{} approximates a S\'ersic profile as a MoG \citep{hogg13}, using a pre-computed
    look-up table of MoG parameters as a function of the S\'ersic index ($n$).
    Because the individual Gaussian components in the mixture fall off more
    rapidly than the target S\'ersic profile, the MoG approximation systematically
    underrepresents both the central cusp and the extended wings of high-$n$ profiles.
    The public version of \tractor{} mitigates the central deficit with a tabulated
    ``core'' flux correction during the fit, but the wings receive no such
    correction by default.

\begin{figure}
        \centering
        \includegraphics[width=0.48\textwidth]{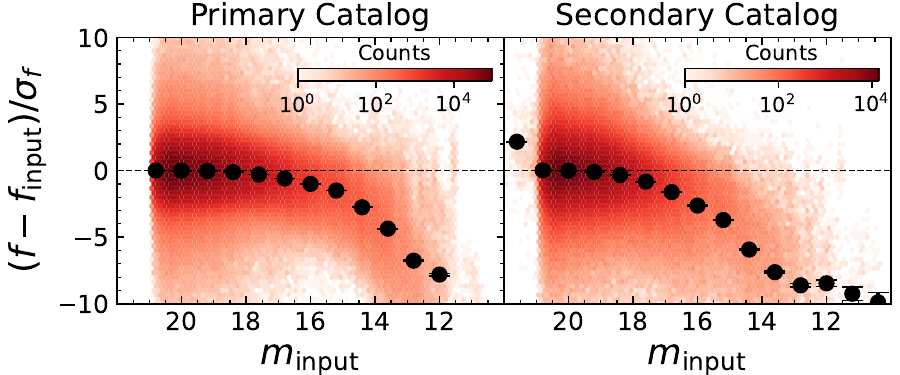}
        \caption{ Same format as Figure~\ref{fig:photbias_extpsf} (extended
        sources), but using the default S\'ersic MoG model in \tractor{}. Unlike
        our modified model, the default implementation shows a clear bias for bright
        galaxies, systematically underestimating their fluxes. This highlights the
        necessity of our improved Sérsic model for accurate \spherex{}
        photometry.}
        \label{fig:app-tractor}
    \end{figure}

\begin{figure}
        \centering
        \includegraphics[width=0.48\textwidth]{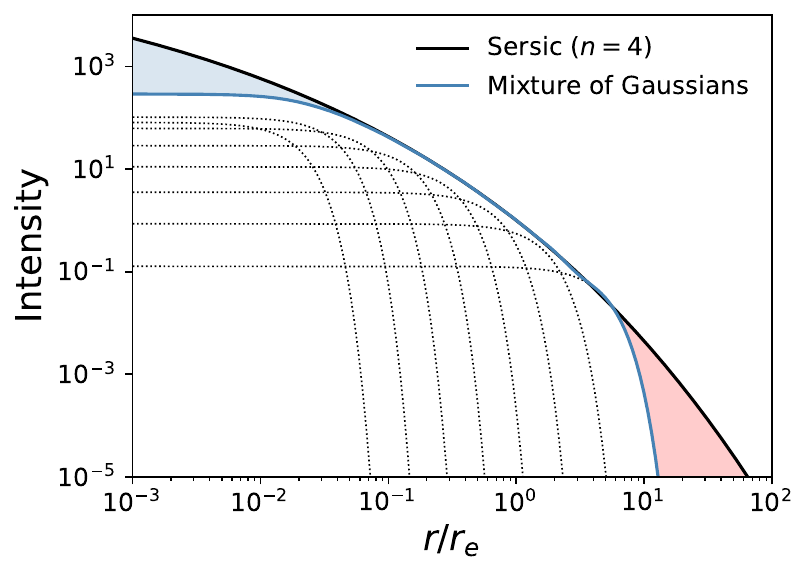}
        \caption{ Comparison of a true S\'ersic profile ($n=4$; black) with MoG
        approximation of \tractor{} (blue). The MoG model systematically
        underestimates the flux in two regions: the central cusp (blue shading) and
        the extended outer wings (red shading). For this $n=4$ example, these deficits
        correspond to approximately 1\% and 5.8\% of the total flux,
        respectively. While the standard \tractor{} implementation corrects only
        for the core deficit, our pipeline adds a precomputed correction for the
        outer-wing deficit. This additional correction is crucial for \spherex{}
        data, where the large pixel scale would otherwise lead to flux biases
        for bright, high-$n$ galaxies. }
        \label{fig:app-mog}
    \end{figure}

    For our simulated \spherex{} images, galaxies are drawn from S\'ersic
    profiles by \galsim{}. The MoG approximation therefore integrates to a
    smaller total flux than the input profile, with a deficit that grows with $n$.
    For $n=4$, the core and wing deficits correspond to approximately 1\% and
    5.8\% of the total flux, respectively (Figure~\ref{fig:app-mog}). Fitting our
    simulated images with the default \tractor{} S\'ersic model consequently
    produces a clear underestimation of fluxes for bright, high-$n$ galaxies (Figure~\ref{fig:app-tractor}).

    To ensure that our pipeline recovers unbiased input fluxes in the simulation,
    we implement a modified S\'ersic model that supplements the standard core
    correction with an analogous correction for the outer wings. We precompute the
    wing-flux deficit as a function of $n$ by numerically integrating both the S\'ersic
    profile and its MoG approximation, and apply the tabulated correction as an
    additional flux contribution during optimization. Figure~\ref{fig:app-tractor}
    demonstrates that this correction removes the model-induced bias for bright,
    high-$n$ sources. The correction is best understood as a profile-consistency
    adjustment: it ensures that the MoG model used during the fit integrates to
    the same total flux as the S\'ersic profile adopted as input by \galsim{},
    thereby isolating \spherex{}-specific systematics (blending, pixelization,
    PSF matching) from bias intrinsic to the MoG approximation itself.

    The relevance of this correction to real \spherex{} data, however, depends
    on the intended downstream use. \tractor-based reference catalogs such as Legacy
    Surveys and COSMOS report fluxes from the native (uncorrected) MoG model,
    and therefore carry the same few-percent deficit internally. When \spherex{}
    photometry is cross-matched or combined with such catalogs (for example, for
    SED fitting or photometric redshift estimation) omitting the correction generally
    yields better cross-survey consistency. We find empirically that uncorrected
    \spherex{} fluxes agree more closely with Legacy Survey broad-band photometry
    for the same sources than corrected fluxes do. We therefore recommend that
    users combining \spherex{} forced photometry with Legacy Surveys, COSMOS, or
    similar \tractor-based catalogs use the native \tractor{} output without the
    S\'ersic correction described above.

    For applications aiming to estimate intrinsic total fluxes rather than
    maintain cross-catalog consistency, the MoG approximation is not the dominant
    morphological systematic. Real galaxies are not single-S\'ersic profiles: bulge--disk
    structure, spiral arms, and the extended envelopes of cD and BCG-like
    galaxies introduce morphological deviations whose photometric impact
    typically exceeds the few-percent MoG--S\'ersic mismatch \citep[e.g.,][]{simard11, bernardi14, meert15, sonnenfeld22, kluge20}.
    In such regimes the single-S\'ersic assumption itself is the dominant bias, and
    multi-component modeling or carefully chosen aperture measurements would in principle
    be preferable.

    Finally, while our work focused on improvements to the galaxy model, pipeline
    efficiency could be further enhanced by using a pre-computed Fast Fourier Transform
    (FFT) of the pixel-based PSF, which is likely more efficient than
    approximating the complex \spherex{} PSF with a large MoG model \citep{lang20}.

    \section{Point-source Photometry within $r<1\arcsec$}
    \label{app:psf}

    While our main pipeline models the full morphological profile of each galaxy,
    this process is computationally intensive. A common simplification,
    particularly for sources that are nearly unresolved, is to approximate them as
    point sources and perform pure PSF-based photometry. In this appendix, we quantify
    the photometric bias introduced by such an approximation for compact extended
    sources within the \spherex{} survey.

\begin{figure}
        \centering
        \includegraphics[width=0.48\textwidth]{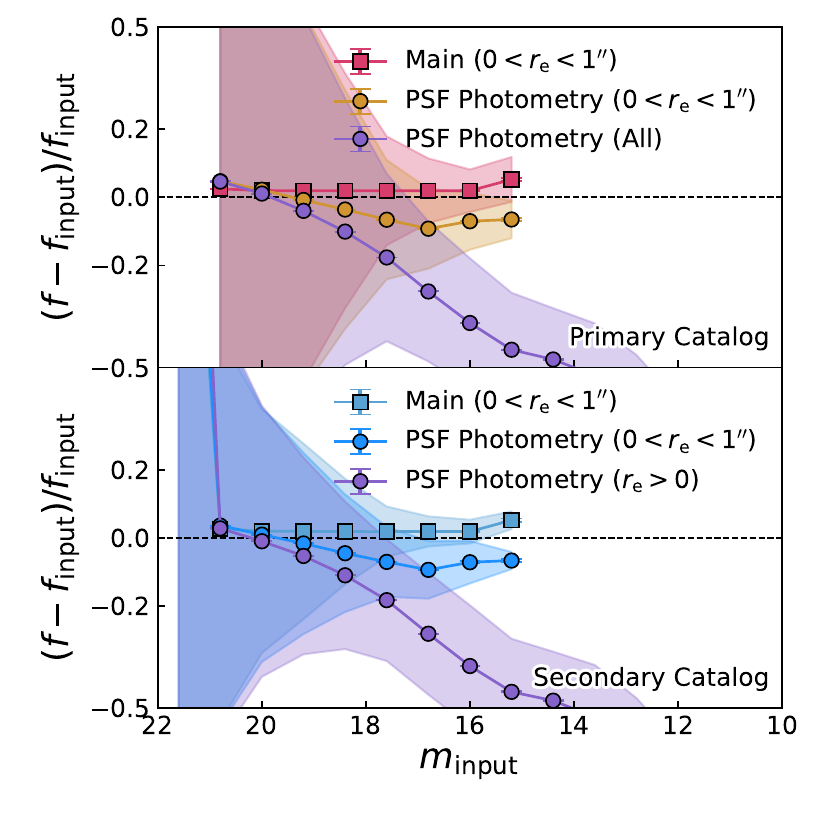}
        \caption{ Photometric bias test for compact sources with half-light radii
        $r_{e}< 1''$. Since many users might want to treat such small sources as
        point sources to reduce computational cost, we compare the resulting flux
        estimates. The ``Main'' photometry (squares) uses the full shape-convolved
        model, while the ``PSF Photometry'' applies pure PSF fitting. For
        sources with $r_{e}< 1''$ (circles), PSF-based measurements underestimate
        the fluxes by $\sim 10$--$20\%$, while including larger sources ($r_{e}>
        0$; purple) leads to even stronger underestimates. Even the Main photometry
        shows a small $\sim 2\%$ bias, suggesting that these very compact
        extended sources are primarily responsible for the faint-end bias seen
        in Figure~\ref{fig:photbias}. }
        \label{fig:app-ps}
    \end{figure}

    Figure~\ref{fig:app-ps} shows the results of this test for sources with half-light
    radii $r_{e}< 1\arcsec$. We compare our standard pipeline results (``Main''
    photometry), which use the full shape-convolved model, against ``PSF Photometry,''
    which treats these sources as unresolved points. The comparison reveals that
    approximating these compact sources with a PSF model leads to a significant and
    systematic underestimation of their flux, typically as much as 20\% (typically
    $\sim$5--10\%). As expected, this bias becomes even more severe when PSF
    photometry is applied to all extended sources ($r_{e}> 0$). This demonstrates
    that systematic bias arises because even compact galaxies ($r_{e}< 1''$) possess
    extended wings that are not captured by the PSF model. In undersampled data
    like \spherex, this subtle morphological mismatch leads to significant flux losses
    when a point-source profile is enforced \citep[e.g.,][]{lauer99}. Together
    with the results of Appendix~\ref{app:tractor}, this illustrates a broader pattern:
    adopting a profile model that is simpler than the true light distribution
    systematically underestimates the recovered flux. Even for sources appearing
    nearly unresolved in the coarse \spherex{} pixels, full morphological modeling
    is necessary for unbiased photometry.

    Furthermore, even our main photometry retains a small residual bias of
    $\sim$2\% for this compact population, comparable to the faint-end bias seen
    in our main results (Figure~\ref{fig:photbias}). Because these nearly-unresolved
    sources dominate the faint end, this residual ties the main-result bias to the
    residual mismatch between the input S\'ersic profile and the \tractor{} MoG
    model (Appendix~\ref{app:tractor}).

    \section{Impact of Astrometric Uncertainty}
    \label{app:astrometry}

    Throughout this work our photometry freezes every source at its input
    position (Section~\ref{sec:pipeline}), which assumes that the \spherex{}
    astrometric solution is exact. In practice the solution is expected to be accurate
    only to about $0\farcs1$--$0\farcs4$ \citep{akeson25}, a small fraction of
    the $6\farcs15$ pixel but not zero, so each source model is placed on a slightly
    mis-centered pixel relative to the data. To assess whether this assumption
    affects our results, here we relax this idealization, injecting realistic
    position offsets into the simulation and tracing their effect on the
    recovered fluxes and photometric redshifts. We find the redshifts essentially
    unaffected at the expected \spherex{} accuracy, with any degradation becoming
    significant only for offsets well beyond it.

    We reproduce an astrometric error by shifting only the assumed source positions,
    displacing them by a Gaussian of width $\sigma_{\rm astro}$ per axis from
    their true values while leaving the simulated pixels themselves unchanged. \tractor{}
    then fits each source at its shifted position, so its model is mis-centered relative
    to the data by $\sigma_{\rm astro}$, exactly as an astrometric error would
    do. We take $\sigma_{\rm astro}=0\farcs3$ as our representative value and scan
    from $0$ to $1\arcsec$. To bracket how the offset correlates across the
    mission, we consider two limiting cases. In the \emph{uncorrelated} case each
    observation receives an independent offset; in the \emph{coherent} case a single
    offset is shared by all observations. The uncorrelated offsets partly
    average down over the 102 channels of the Secondary Catalog, whereas the coherent
    offset does not.

    The size of this effect depends on how bright a source is and on how much flux
    its neighbors contribute, so a random subset of cluster members would barely
    sample the crowded, bright-neighbor cases where astrometric errors matter most.
    We therefore bin Abell 2055 members on a $3\times3$ grid of input magnitude
    ($15<K_{s}<17$, $17$--$19$, $19$--$21$) and neighbor-to-target flux ratio (isolated
    $\sum f_{\rm neigh}/f<1$, moderate $1$--$10$, crowded $\ge10$) and aim for 30
    targets per cell. The bright, crowded cells fall short of this because such sources
    are rare: the brightest bin yields only a single crowded target and 17
    moderately blended ones, against the 30 drawn elsewhere, for 228 targets in
    total. The grid also keeps the sample small enough to repeat many times: for
    each value of $\sigma_{\rm astro}$ and each correlation case we run 15
    realizations, each an independent simulation with a fresh random offset.

\begin{figure}
        \centering
        \includegraphics[width=0.48\textwidth]{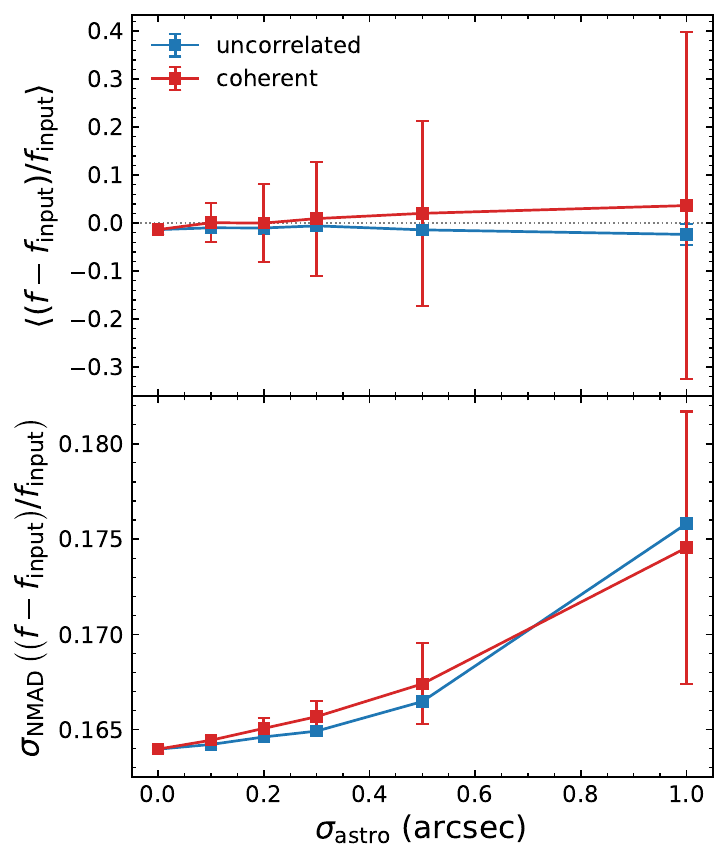}
        \caption{Flux-residual bias (top) and scatter $\sigma_{\rm NMAD}$ (bottom)
        of $(f-f_{\rm input})/f_{\rm input}$ versus injected astrometric error
        $\sigma_{\rm astro}$, for the stratified Abell 2055 subsample. Blue: uncorrelated
        (independent offset per observation) case; red: coherent (single mission-wide
        offset) case. Error bars show the $1\sigma$ realization-to-realization
        spread, that is, the range a single mission-wide offset can produce, not
        the uncertainty on the mean.}
        \label{fig:astrometry_scatter}
    \end{figure}

    Figure~\ref{fig:astrometry_scatter} shows the flux-residual bias and scatter
    as a function of $\sigma_{\rm astro}$. In both panels the central value
    behaves similarly for the two cases, the scatter growing by only about 1\%
    at $\sigma_{\rm astro}=0\farcs3$ and rising steeply only beyond $0\farcs5$. The
    cases differ instead in their realization-to-realization spread (the error
    bars), the range a single survey realization can produce. For the uncorrelated
    case the spread is small, because each source receives many independent
    offsets across its observations that average down. For the coherent case it does
    not average down: a single mission-wide offset shifts every channel of every
    source the same way, so a real survey carries a net bias set by the offset direction.
    This spread is largest in the flux bias (upper panel): at $\sigma_{\rm astro}
    =0 \farcs3$ it reaches about $\pm12\%$ ($1\sigma$) around a small central value
    that is slightly positive for this blending-balanced sample, widening to
    roughly $\pm36\%$ at $1\arcsec$, while in the scatter (lower panel) it stays
    within $\sim1\%$ at $0\farcs3$.

\begin{figure*}
        \centering
        \includegraphics[width=\textwidth]{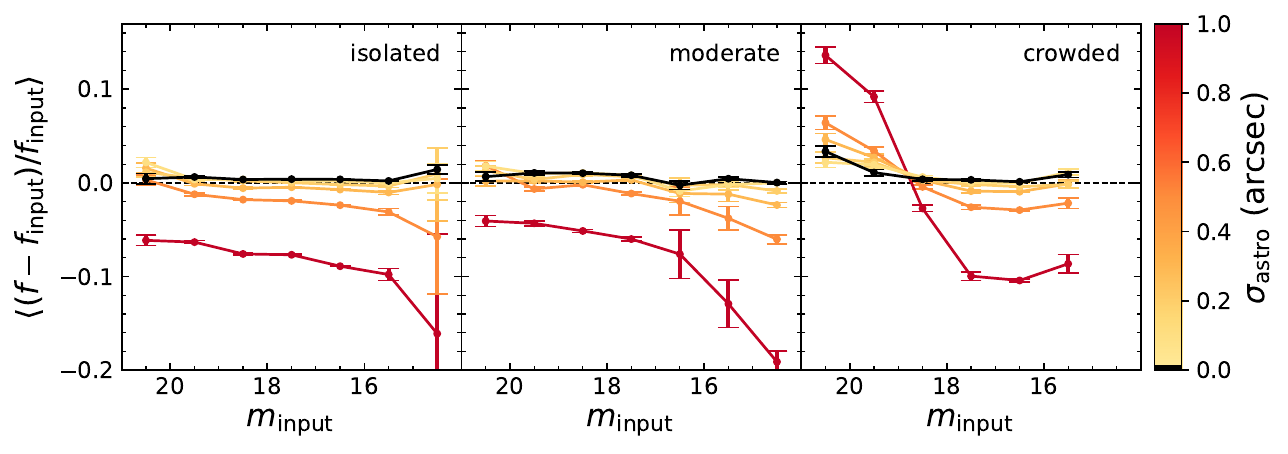}
        \caption{Mean fractional flux residual
        $\langle(f-f_{\rm input})/f_{\rm input}\rangle$ versus input magnitude
        for the stratified Abell 2055 subsample, split by neighbor-to-target flux
        ratio into isolated ($\sum f_{\rm neigh}/f<1$), moderate ($1\le\sum f_{\rm
        neigh}/f<10$), and crowded ($\ge10$) bins; color encodes
        $\sigma_{\rm astro}$ (black is the $\sigma_{\rm astro}=0$ baseline). For
        isolated and moderate sources the bias turns negative with increasing $\sigma
        _{\rm astro}$, most strongly at faint magnitudes. For crowded sources the
        faint end instead turns positive, as a mis-centered faint model absorbs
        flux from a bright neighbor.}
        \label{fig:astrometry_blendbias}
    \end{figure*}

    The sign of the bias depends on the local environment, as shown in Figure~\ref{fig:astrometry_blendbias}.
    For isolated and moderately blended sources the bias is negative and grows
    toward faint magnitudes, as a mis-centered model recovers less of the source's
    own flux. For crowded sources the faint end instead develops a positive bias,
    because a mis-centered faint model absorbs a fraction of a bright neighbor's
    flux, which is large relative to the target's own flux. Both behaviors are small
    at the expected $\sigma_{\rm astro}\approx0\farcs2$--$0\farcs3$ and become
    pronounced only for $\sigma_{\rm astro}\gtrsim0\farcs5$.

    Despite the wider flux-bias range of the coherent case, the impact on the
    photometric redshifts is small, because a coherent offset displaces all channels
    similarly and largely preserves the spectral shape that constrains the redshift.
    At the expected $\sigma_{\rm astro}=0\farcs3$, relative to the frozen-position
    baseline, the bias of the $K_{s}<19$ and $K_{s}<17.5$ samples changes by
    less than $10^{-3}$ in $\Delta z/(1+z)$, their redshift scatter $\sigma_{\rm
    NMAD}(z)$ by at most $\sim5\%$, and the catastrophic-outlier fraction
    $\eta_{0.15}$ by under $0.2$ percentage points; the uncorrelated and coherent
    cases agree within these bounds. We therefore treat the frozen-position
    configuration as a best-case assumption whose relaxation does not alter our
    conclusions at the expected \spherex{} astrometric accuracy.

    \section{Source--Source Covariance in the Flux Uncertainties}
    \label{app:cov}

    As stated in Sections~\ref{sec:pipeline} and~\ref{sec:phot-bias}, the uncertainties
    we report are \tractor{}'s native flux errors, evaluated under the Cram\'er--Rao
    bound and therefore computed as if each source were isolated; the normalized
    residuals are correspondingly broader than a unit Gaussian. \citet{huai25} attribute
    this excess to the flux covariance between sources with overlapping PSFs. To
    verify this interpretation and to justify our use of the native errors, we re-ran
    the entire pipeline (photometry, channelization, and photometric-redshift
    estimation) over all nine fields with the source--source covariance
    propagated through the inverse Fisher information matrix, following \citet{huai25}.
    This run serves only as a cross-check; our main results retain the native
    uncertainties. To characterize the error model in the crowded, fully-blended
    regime, we additionally repeat this native-versus-covariance comparison for the
    Confusion-only and Full sub-threshold scenarios of Section~\ref{sec:robustness},
    on the same seven fields used there (all clusters except the wide-field Coma).

    \begin{figure*}
        \centering
        \includegraphics[width=\textwidth]{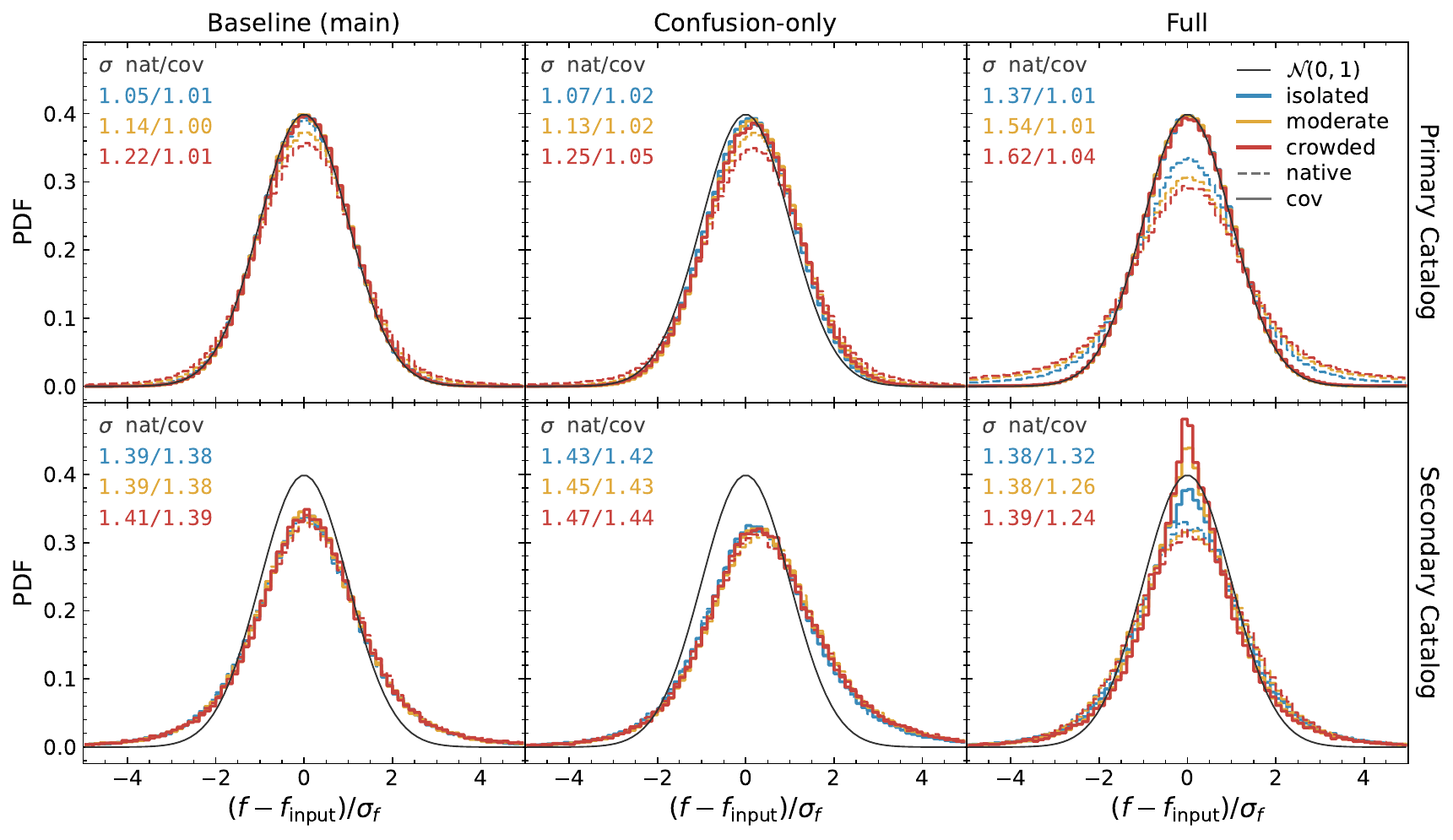}
        \caption{Normalized flux residual $(f-f_{\rm input})/\sigma_{f}$ for the
        Primary (top row) and Secondary (bottom row) catalogs, comparing the native
        \tractor{} errors (dashed) with the Fisher-covariance run (solid), split
        by the neighbor-to-target flux ratio into isolated ($\sum f_{\rm neigh}/f
        <1$), moderate ($1\le\sum f_{\rm neigh}/f<10$), and crowded ($\ge 10$) bins.
        Columns show the Baseline, Confusion-only, and Full scenarios of Section~\ref{sec:robustness},
        on the seven cluster fields analyzed there (Coma excluded). The native
        Primary width grows with crowding within each scenario and more strongly
        between scenarios (crowded bin: $1.22$, $1 .25$, $1.62$ for Baseline, Confusion-only,
        and Full), while the covariance restores near-unit width in every cell ($\le
        1.05$). The Secondary width is set by the bandpass-sampling floor ($\sim$$1
        .4$) and is largely insensitive to the covariance, except for a modest reduction
        in the crowded Full bin ($1.39\rightarrow1.24$). Quoted $\sigma$ are
        clipped to the plotted $[-5,5]$ window; the heavy non-Gaussian tails of the
        native Full run are discussed in the text.}
        \label{fig:app-cov-pull}
    \end{figure*}

    When the covariance is included, the normalized-residual width of the
    Primary Catalog converges to near unity in every crowding regime and every
    scenario (top row of Figure~\ref{fig:app-cov-pull}). In the native run this width
    grows both with the neighbor-to-target flux ratio and, more strongly, with the
    scenario. For the Baseline it rises from $1.05$ (isolated) to $1.14$ (moderate)
    to $1.22$ (crowded), consistent with the full nine-field sample ($1.05$, $1.1
    6$, $1.28$); the Confusion-only run is nearly identical ($1.07$, $1.13$, $1.2
    5$), whereas the Full run is markedly broader ($1.37$, $1.54$, $1.62$). The
    close match between Baseline and Confusion-only shows that the per-observation
    error mis-calibration is a property of the joint fit of all overlapping
    sources, not of the mere presence of faint neighbors in the scene. With the
    covariance these widths collapse to $1.00$--$1.05$ across all bins and all three
    scenarios.

    The quoted core widths are clipped to the plotted window and understate the failure
    in the densest regions: the native Full crowded bin develops a heavy non-Gaussian
    tail, with $\sim$16\% of observations beyond $5\sigma$ (and $\sim$38\% for
    SPT-CL J0546$-$5345 alone, $z=1.067$, where the predominantly faint, low-S/N
    sources drive the largest native-error underestimation). The covariance suppresses
    these tails by one-to-two orders of magnitude (crowded fraction beyond
    $5\sigma$: $16.4\%\rightarrow2.7\%$), repairing the core and the tail together.
    This reproduces, with our own crowding metric, the $z$-score recalibration
    of \citet{huai25}.

    The bias and the scatter of the residuals separate cleanly between the two effects
    of Section~\ref{sec:robustness}. The pull centroid (the mean of $(f-f_{\rm
    input})/\sigma_{f}$) is essentially identical in the native and covariance
    runs---the covariance rescales the error, not the flux---and tracks Effect~A:
    it is small in the Baseline ($\sim$$+0.02$--$0.04\,\sigma$ for the Primary Catalog),
    large in the Confusion-only run, where flux from un-modeled neighbors is
    misattributed to the fitted sources ($\sim$$+0.12$--$0.18\,\sigma$ Primary,
    $\sim$$+0.3$--$0.4\,\sigma$ Secondary), and collapses back to the Baseline
    level in the Full run ($\sim$$+0.02$--$0.06\,\sigma$) once those neighbors
    are themselves fit and deblended. The residual width and outlier fraction instead
    track Effect~B, the joint-fit degeneracy, and are the quantities the covariance
    repairs. The confusion bias is therefore an error-model-independent flux
    systematic, while the broadening of the normalized residuals is an error-model
    artifact that the operational covariance treatment removes.

    After channelization, the Secondary-Catalog residual width is $\sim$$1 .4$ almost
    independently of crowding bin, and the covariance leaves it nearly unchanged
    for the Baseline and Confusion-only runs (bottom row of Figure~\ref{fig:app-cov-pull}).
    In the Full run the covariance reduces it modestly, and more so in the
    crowded bin ($1.39\rightarrow1.24$), reflecting the larger residual source--source
    covariance when all overlapping sources are fit.

    This behavior follows from how the Secondary flux is constructed: it is an inverse-variance-weighted
    average of the per-observation measurements within each channel, with
    weights $\propto 1/\sigma_{f}^{2}$. The covariance, which corrects the under-estimated
    $\sigma_{f}$ of blended observations, therefore propagates into the Secondary
    point estimate through these weights, down-weighting the most contaminated
    measurements and slightly reducing the Secondary flux scatter in the crowded
    regime; this is the one path by which the covariance affects a point estimate
    rather than only an error bar. The pull width itself, however, is largely
    unmoved: it is set by the deterministic bandpass-sampling mismatch. The
    channel ``truth'' is integrated over a single fiducial bandpass, whereas the
    binned flux samples the slightly offset LVF wavelengths of the contributing measurements.
    This SED-slope-driven systematic is not part of the formal flux uncertainty
    and dominates the Secondary residual budget over the source--source covariance
    that the native errors omit (consistent with the discussion of Figure~\ref{fig:photbias}).
    It is not specific to our simulation: the same systematic enters any template-based
    redshift fit to a binned \spherex{} spectrum, and the self-consistency between
    our mock-generation and fitting templates makes our estimate of its magnitude
    a likely-conservative one.

    \begin{figure*}
        \centering
        \includegraphics[width=\textwidth]{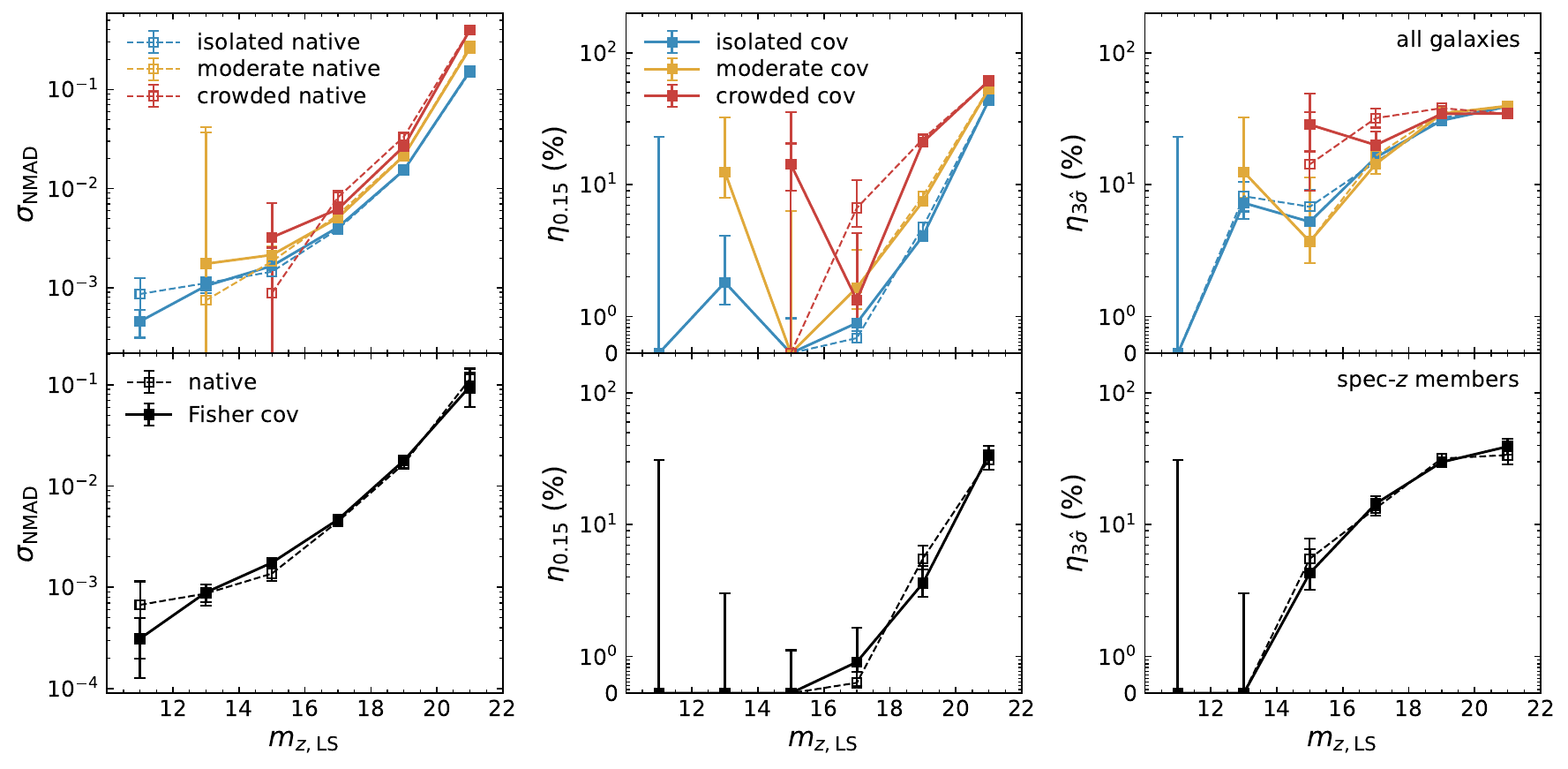}
        \caption{Photometric-redshift performance with the native \tractor{} errors
        (dashed, open symbols) versus the Fisher-covariance run (solid, filled),
        as a function of Legacy Survey $z$-band magnitude. Columns show, from
        left to right, $\sigma_{\rm NMAD}$, the $|\Delta z|/(1+z)>0.15$ outlier
        fraction $\eta_{0.15}$, and the $3\hat{\sigma}$ outlier fraction $\eta_{3\hat{\sigma}}$.
        \textit{Top row:} all galaxies split into isolated ($\sum f_{\rm neigh}/f
        <1$), moderate ($1\le\sum f_{\rm neigh}/f<10$), and crowded ($\ge10$) bins
        (colors). \textit{Bottom row:} the spectroscopically confirmed cluster members.
        In both views the native and covariance curves agree within the
        estimated uncertainties at every magnitude and crowding level, while the
        scatter is set primarily by source brightness; this confirms that the
        redshift estimates (and in particular the performance of our primary member
        sample) are insensitive to the flux-uncertainty treatment.}
        \label{fig:app-cov-photoz}
    \end{figure*}

    Because the photometric redshifts inherit this channelized error budget, they
    are correspondingly insensitive to the covariance treatment (Figure~\ref{fig:app-cov-photoz}).
    The global redshift scatter changes only marginally ($\sigma_{\rm NMAD}=0.269
    \rightarrow0.261$), the mean bias is unchanged ($0.016\rightarrow0.015$),
    and the outlier fraction is stable ($\eta_{0.15}=52.8\%\rightarrow52.6\%$). Within
    each crowding bin the native and covariance results agree to within the estimated
    uncertainty (e.g.\ in the crowded bin $\sigma_{\rm NMAD}=0.448$ vs.\ $0.442$),
    the spectroscopically confirmed cluster members show the same agreement
    across the full magnitude range (Figure~\ref{fig:app-cov-photoz}, bottom row),
    and the improvement from a bright-source or high-S/N selection (Section~\ref{sec:photoz})
    is recovered identically in the two runs.

    The same insensitivity holds in the Full run of Section~\ref{sec:robustness}:
    even where the per-observation Primary pulls develop the heavy tails described
    above, the channelized Secondary fluxes that feed the redshift fit remain
    averaging- and bandpass-limited, so the native and covariance photometric
    redshift metrics remain consistent within their estimated uncertainties. More
    generally, because we adopt the maximum-likelihood redshift, a a change to the
    flux-error treatment leaves the point estimates (and hence
    $\sigma_{\rm NMAD}$, the bias, and $\eta_{0.15}$) nearly unchanged, altering
    chiefly the width of the $P(z)$ and the $3\hat\sigma$ outlier rate; our redshift
    accuracy is therefore largely insensitive to the precise error normalization.
    Finally, the covariance calculation adds negligible computational overhead,
    with per-field photometry wall-time ratios of $0.95$--$1 .06$. We therefore
    report the native uncertainties directly, noting that the \spherex{} Level~3
    pipeline applies the full covariance formalism by default.

    \section{Impact of Blending on Redshift PDFs}
    \label{app:photoz_pdf}

    In Section \ref{sec:photoz_blending}, we observed that source blending tends
    to increase the rate of catastrophic photometric redshift failures rather
    than merely inflating the scatter. To investigate the physical mechanism behind
    these failures, we examine the full redshift PDFs ($p(z)$) for a subset of
    sources. We specifically target cases with high neighbor-to-target flux ratios
    to highlight the effects of severe contamination. We compare the standard
    pipeline results against a control simulation where the same sources were
    injected and extracted in isolation (see Section \ref{sec:blend}).

\begin{figure*}
        \centering
        \includegraphics[width=0.9\textwidth]{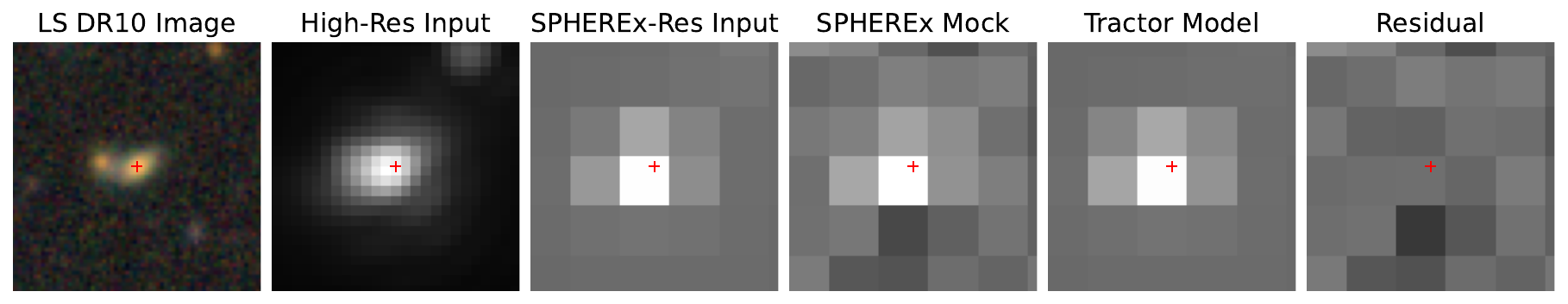}
        \includegraphics[width=0.9\textwidth]{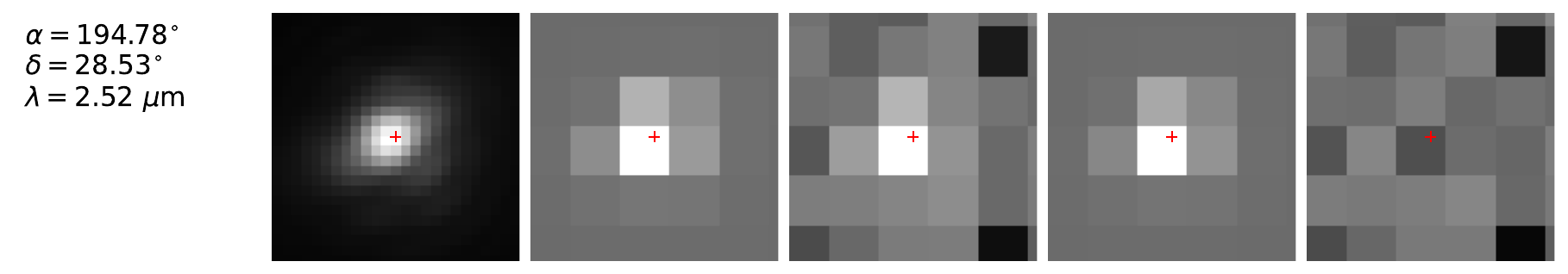}
        \includegraphics[width=0.9\textwidth]{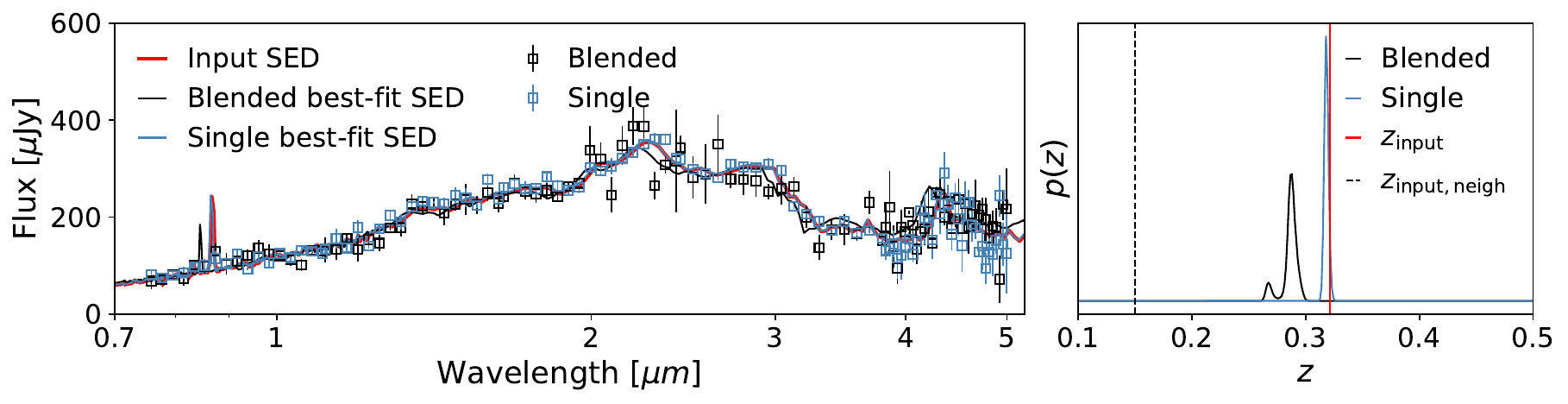}
        \includegraphics[width=0.9\textwidth]{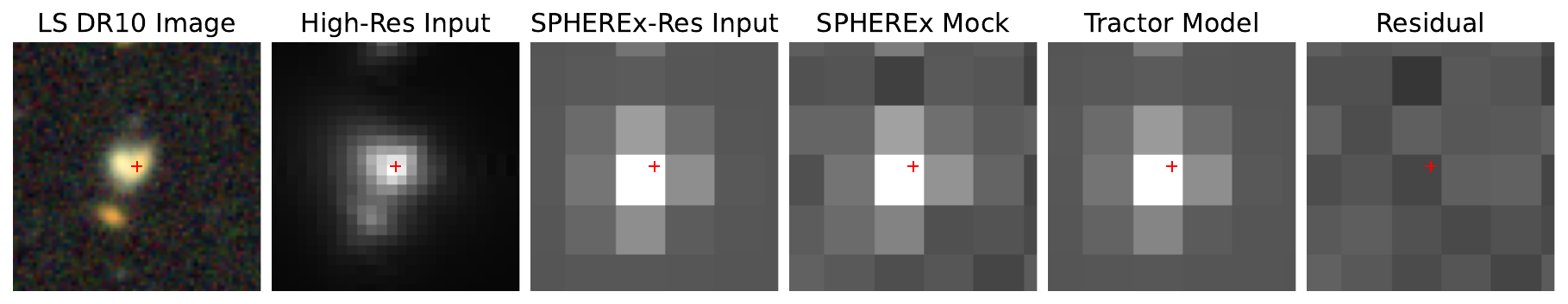}
        \includegraphics[width=0.9\textwidth]{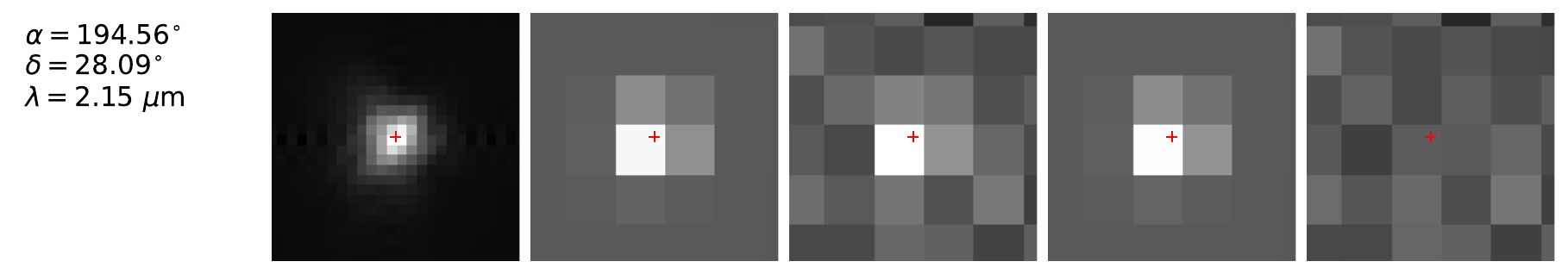}
        \includegraphics[width=0.9\textwidth]{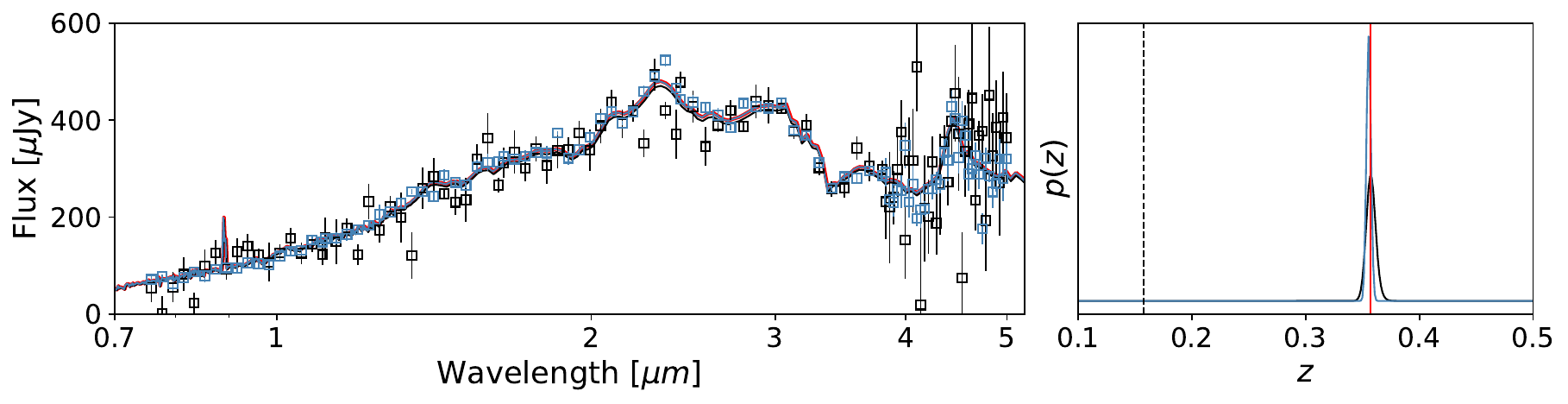}
        \caption{Detailed examination of the impact of blending on spectral
        energy distributions (SEDs) and photometric redshift estimates for two
        example sources. \textit{Image Panels:} Comparison of the simulation and
        extraction steps. For each source, the upper row displays the standard
        pipeline results (``Blended''), while the lower row shows the control run
        where the source was simulated in isolation (``Single''). The columns
        follow the same format described in Figure~\ref{fig:photo_example} (Legacy
        Survey input, High-Res model, SPHEREx-Res model, Mock image, Tractor model,
        and Residual). \textit{Bottom Panels:} (Left) Comparison of the recovered
        spectrophotometry. The red line shows the ground truth input SED. Black squares/line
        represent the ``Blended'' case, while blue squares/line show the ``Single''
        case. (Right) The corresponding redshift probability density functions ($p
        (z)$). The black curve represents the blended case, and the blue curve represents
        the single-source case. The solid red vertical line marks the true
        redshift ($z_{\text{input}}$), while the dashed black vertical line indicates
        the redshift of the nearest blending neighbor ($z_{\text{input, neigh}}$).}
        \label{fig:photoz_pdf}
    \end{figure*}

    Figure~\ref{fig:photoz_pdf} presents two representative examples of this
    comparison. In the single-source control case (blue), the recovered SEDs
    closely match the input, resulting in sharp, well-calibrated PDFs centered
    on the true redshift. In the blended case (black), however, flux contamination
    from neighbors distorts the measured photometry.

    This contamination typically does not produce a secondary peak at the neighbor's
    redshift. Instead, the blending primarily acts as correlated noise or a continuum
    offset that broadens the PDF or systematically shifts its peak away from the
    truth. This degradation of spectra confirms that catastrophic errors in
    crowded fields are primarily driven by complex SED distortion rather than
    simple confusion between distinct source components.

    \bibliographystyle{aasjournalv7}
    \bibliography{ms}
\end{document}